\documentclass[12pt,preprint]{aastex}

\newcommand{\kms}{km~s$^{-1}$}
\newcommand{\subsun}{\mbox{$_{\odot}$}}
\newcommand{\teff}{$T_{\rm{eff}}$}
\newcommand{\grav}{log($g$)}
\newcommand{\etal}{{\it et al.\/}}
\newcommand{\eqw}{$W_{\lambda}$}

\newcommand{\nemp}{8}

\begin{document}

\title{New Extremely Metal-Poor Stars in the Galactic Halo\altaffilmark{1}}

\author{Judith G. Cohen\altaffilmark{2}, Norbert Christlieb\altaffilmark{4}, 
Andrew McWilliam\altaffilmark{3},  
Stephen Shectman\altaffilmark{3}, Ian Thompson\altaffilmark{3}, 
Jorge Melendez\altaffilmark{5},
Lutz Wisotzki\altaffilmark{6} \& Dieter Reimers\altaffilmark{7}  }

\altaffiltext{1}{Based in part on observations obtained at the
W.M. Keck Observatory, which is operated jointly by the California 
Institute of Technology, the University of California, and the
National Aeronautics and Space Administration.}

\altaffiltext{2}{Palomar Observatory, Mail Stop 105-24,
California Institute of Technology, Pasadena, Ca., 91125, 
jlc@astro.caltech.edu, aswenson@caltech.edu}

\altaffiltext{3}{Carnegie Observatories of Washington, 813 Santa
Barbara Street, Pasadena, Ca. 91101, andy, ian, shec@ociw.edu}

\altaffiltext{4}{Current address: Department of
   Astronomy and Space Physics, Uppsala University, Box 515,
   75120 Uppsala, Sweden, formerly at Hamburger Sternwarte, Universit\"at
Hamburg, Gojenbergsweg 112, D-21029 Hamburg, Germany, norbert@astro.uu.se}

\altaffiltext{5}{Palomar Observatory, Mail Stop 105-24,
California Institute of Technology, Pasadena, Ca., 91125,
Current address: Australian National University, Australia,
jorge@mso.anu.edu.au}

\altaffiltext{6}{Astrophysical Institute Potsdam, An der Sternwarte 16,
D-14482 Potsdam, Germany, lwisotzki@aip.de}

\altaffiltext{7}{Hamburger Sternwarte, Universit\"at
Hamburg, Gojenbergsweg 112, D-21029 Hamburg, Germany,
dreimers@hs.uni-hamburg.de}

\begin{abstract}

We present a detailed abundance analysis based on high resolution
and high signal-to-noise spectra of eight extremely metal poor (EMP)
stars with [Fe/H] $\lesssim -3.5$~dex, four of which are new.
Only stars with $4900 <$ \teff\ $< 5650$~K are included.

Two stars of the eight are outliers in each of several
abundance ratios.
The most metal poor star in this sample, HE1424$-$0241,
has [Fe/H] $\sim -4$~dex
and is thus among the most metal poor stars known in the Galaxy.
It has highly anomalous abundance ratios unlike those of any other
known EMP giant, with very low Si, Ca and Ti relative to Fe,
and enhanced Mn and Co, again relative to Fe. 
Only (low) upper limits
for C and N can be derived from the non-detection of the CH
and NH molecular bands.
HE0132$-$2429, another sample star, has 
excesses of N and Sc  with respect to Fe.  

The strong outliers in abundance ratios among
the Fe-peak elements in these C-normal stars,
not found at somewhat higher metallicities
([Fe/H] $\sim -3$ dex),  
are definitely real.  They suggest that at such low metallicities
we are beginning to see the  anticipated and
long sought stochastic effects of individual supernova events
contributing to the Fe-peak material within a single star.
With spectra reaching well into the near-UV we are able to
probe the behvaior of copper abundances in such extreme EMP stars.   

A detailed comparison of the results of the analysis procedures
adopted by our 0Z project compared to those of the First Stars VLT
Large Project finds a systematic difference for [Fe/H] of
$\sim$0.3 dex, our values always being higher.

\end{abstract}


\section{Introduction}

Extremely metal poor stars provide important clues to the chemical
history of our Galaxy, the role and type of early SN, the
mode of star formation in the proto-Milky Way, and the formation
of the Galactic halo.  The number of extremely metal poor (EMP)
stars known is summarized by \cite{beers05}.  They
compiled a list of the key properties of
the 12 stars identified up to that time with [Fe/H] 
$\lesssim -3.5$~dex\footnote{The 
standard nomenclature is adopted; the abundance of
element $X$ is given by $\epsilon(X) = N(X)/N(H)$ on a scale where
$N(H) = 10^{12}$ H atoms.  Then
[X/H] = log$_{10}$[N(X)/N(H)] $-$ log$_{10}$[N(X)/N(H)]\subsun, and similarly
for [X/Fe].}, 7 of which are EMP giants and subgiants within
the range of \teff\ considered here.

Our 0Z project has the goal of increasing the sample of such stars
through data mining of
the  Hamburg/ESO Survey (HES) \citep{wis00}.  This is an 
objective prism survey from which it is
possible to efficiently select QSOs \citep{wis00} as well
as a variety of interesting
stellar objects, among them extremely metal poor (EMP) stars \citep{christlieb03}.

The 0Z project has been systematically searching the 
database of the HES for this purpose over the past five years.
We present in \S\ref{section_sample} a sample of new EMP giants 
with \teff\ $< 6000$~K
and [Fe/H] $\lesssim -3.5$~dex which substantially increases the
number of such stars known.  Details of the analysis are described
in \S\ref{section_analysis}, while the radial velocity data
are discussed in \S\ref{section_vr}.  The abundance ratios for
$\sim$20 elements in each of the sample EMP giants are described
in \S\ref{section_abund}.  In the following section (\S\ref{section_comp})
we check  the consistency of
the analyses and procedures adopted by our 0Z project
with those of the First Stars VLT Large Project using UVES
\citep{cayrel04} and offer some comments on the implication of 
our results on the frequency of carbon-enhanced stars.
The penultimate section presents a discussion of the implications
of our results
for early SN  and for nucleosynthesis in the forming Galactic halo.
A summary of the key results follows.
Two appendices which discuss details of a comparison
of our work with other large spectroscopic and photometric surveys
of EMP stars complete this paper.

\section{Sample of Stars \label{section_sample} }

In pursuit of our goal of exploiting the HES to identify new
EMP stars, we began with a list of candidates selected from the HES
database over the 50\% of its area on the sky to which we have
access.
Our 0Z project has 
now taken moderate resolution spectra of more than 600 candidates
with the Double Spectrograph \citep{oke82} on the Hale 5-m telescope 
at Palomar Mountain, and more than 1100 at the
6.5m Clay and Baade Telescopes at the Las Campanas Observatory.
These spectra have been processed
with the algorithm of \cite{beers99}, which uses a measure
of the strength
of H$\delta$ and of the 3933~\AA\ line
of Ca~II, to determine rough metallicities, denoted [Fe/H](HES).
\cite{cohen05} provides a brief description of the vetting process,
see also \cite{beers05}.
Those stars of special interest with follow-up spectra from Palomar, 
including all those with 
[Fe/H](HES) $< -2.9$ dex,
have been observed with HIRES \citep{vogt94} at the Keck~I telescope
over the past
four years; a total of $\sim$90 such stars have been observed with HIRES to date.
In this paper we present analyses of those stars from the Palomar
sample with \teff\ $< 6000$~K which
turned out to be genuine EMP stars with [Fe/H] $\lesssim -3.5$ dex  as 
determined from our high-resolution, high SNR spectra, and which have
not already been published in our earlier papers.  A future paper will
deal with the most extreme metal-poor stars found in the Las Campanas 
sample.  The limitation on \teff\ ensures that internal comparisons
of abundance ratios within the sample will be as reliable as possible.

Five new EMP stars are presented here.  One turned out to be a rediscovery
of a HK Survey star.  This was not realized for a long time
due to the 32'' difference in the
coordinates of CS22949$-$037 from its  discovery by the HK Survey
\citep{beers85,beers92},
and those of HE2323$-$0256, found in the HES.  It appears
that the HK Survey coordinates are sometimes in error by such a
large amount, as the updated coordinates for this star
given by \cite{cayrel04} are within 1.5'' of those from the HES of
HE2323$-$0256.

We also include a new analysis based on better spectra
of the only genuine EMP giant described previously in our published papers,
HE0132--2429, part of the 
Keck Pilot Project \citep{cohen02,carretta02}.
We do not consider here the three EMP dwarfs whose analyses we have
published, 
BS16545--0089 and HE1346--0427
from \cite{cohen04}, nor HE0218--2738, a double lined
spectroscopic binary \citep{cohen02,carretta02}. 
There are many fewer absorption lines detected in these hotter stars 
and we wish
to restrict ourselves to a narrow range in \teff\ to ensure accurate
comparisons within our sample.

We add to our sample the only star from the 
Hamburg/ESO r-process enhanced star
survey (HERES) \citep{barklem05,christlieb04b} which they believed to have
[Fe/H] $< -3.5$ dex, HE1300+0157.  We also add the star BS16467--062,
which is included in the VLT/UVES First Stars program
\citep{cayrel04}.  A detailed analysis
for the latter, found in the HK Survey, 
was presented by \cite{francois03}, which was superseded
by that of \cite{cayrel04}.
Since BS16467--062 and HE2323$-$0256 are part of the First Stars sample,
several analyses have been published for each of these stars.  
In the present work they serve
as comparison  objects to determine the consistency of the absolute
iron abundance and relative abundances of various
elements as deduced by our group versus those of the First Stars VLT large program.

The sample stars are listed in Table~\ref{table_sample}, which 
gives their J2000 coordinates, new optical photometry,
when available, and other relevant data.

\section{Stellar Parameters \label{section_param} }

We use the procedures described in \cite{cohen02} and used in all
subsequent work by our 0Z project published to date. 
Our \teff\ determinations are based on broad band colors
$V-I, V-J$ and $V-K$.
The IR photometry is taken from 2MASS \citep{2mass1,2mass2}.
We have obtained new photometry at $V$ and $I$ for almost
all of the stars discussed here.  We use 
ANDICAM images taken for this purpose over the past two years via a service
observing queue on
the 1.3m telescope at CTIO operated
by the SMARTS consortium. ANDICAM is a dual channel camera
constructed by the Ohio State University instrument 
group\footnote{See http://www.astronomy.ohio-state.edu/ANDICAM and
http://www.astro.yale.edu/smarts.}.  Our ANDICAM program requires
photometric conditions, and additional standard star fields,
charged to our ANDICAM allocation through NOAO, are always taken for us.
Appendix~A compares our photometry with
that of \cite{beers07} and with that of the Sloan Digital Sky Survey
\citep{york00}.

We derive surface gravities through combining these \teff\ with
an appropriate
12 Gyr isochrone from the grid of \cite{yi01}.

The resulting stellar parameters, which have been
derived with no reference to the spectra themselves,
are given in Table~\ref{table_sample}. The random uncertainties
in \teff\ from photometric errors  (see Appendix A)
are 100~K.  This ignores systematic errors
which may be present.  The adopted uncertainties in \grav\ 
are, following the discussion in \cite{cohen02}, 
100~$d$[\grav]/$d$\teff\ evaluated along the RGB
for stars in this extremely low metallicity range, 0.2~dex at \teff = 5000~K
and 0.15~dex at 5500~K.

\section{Observations}

The EMP  stars in our sample were observed with HIRES at Keck 
during various runs over
the past 4 years.  Details of the best available spectra for each
of the new stars
are listed in Table~\ref{table_spectra}. 
Six of these stars were observed with high SNR spectra from HIRES
at the Keck Observatory after the recent
detector upgrade, which provides complete spectral coverage
from 3180 to 5990~\AA\ in a single exposure.  Two (HE1356$-$0622 and
HE1347$-$1025\footnote{We are grateful to W.~Sargent for obtaining this
spectrum.}) have only relatively short exposures with the new detector, used
primarily to measure the strength of the NaD lines.  Their only high
SNR spectra were taken before the upgrade and hence have 
spectral coverage restricted to 3840 to 5330\,{\AA} with no gaps 
between orders
for $\lambda < 5000$~\AA, and only small gaps thereafter. 
The slit width (either 0.86 arcsec, corresponding to a spectral
resolution of 46,000, or 1.1 arcsec, which corresponds to
a spectral resolution of 35,000) used for each spectrum
is indicated in this table as well.
Each exposure was broken up into 1200 sec segments to expedite
removal of cosmic rays.  The goal was to achieve  
a SNR of 100 per spectral
resolution element in the continuum at
4500\,{\AA}; a few spectra have slightly
lower SNR.   This SNR calculation utilizes only
Poisson statistics, ignoring issues of cosmic ray removal,
night sky subtraction, flattening, etc.   The observations
were carried out with the slit length aligned to
the parallactic angle.  

The processing of the spectra was done with 
MAKEE\footnote{MAKEE was developed
by T.A. Barlow specifically for reduction of Keck HIRES data.  It is
freely available on the world wide web at the
Keck Observatory home page, 
http://www2.keck.hawaii.edu/inst/hires/data\_reduction.html.} and Figaro 
\citep{shortridge93} 
scripts, and follows closely that described by \cite{cohen06}.
The equivalent widths were measured as described in \cite{cohen04}.
Table~\ref{table_eqw} lists the atomic parameters adopted
for each line and their equivalent widths measured in the spectra of
each of the \nemp\ EMP stars.

\section{Analysis \label{section_analysis} }

The analysis is identical to that of \cite{cohen04} with several
important additions.  In particular we use the model
stellar atmosphere grid of \cite{kurucz93}
and a current
version of the LTE spectral synthesis program MOOG \citep{moog},
which treats scattering as LTE absorption.
The improved HIRES spectra now reach into the near-UV, making the
NH and redder part of the OH bands accessible.
We use the molecular line list of \cite{kurucz94}, 
augmented with the strongest atomic features, to analyze the NH band.
We use the line list of \cite{gillis01} for OH.
Our nominal Solar CNO abundances are 8.59, 7.93, and 8.83 dex respectively.
These are close to those of \cite{grevesse98},
but somewhat larger than  the values obtained using 3D model atmospheres 
by \cite{asplund04} and \cite{asplund05}.  We prefer not
to attempt 3D corrections until a full grid of model 3D
atmospheres or of corrections to CNO abundances derived from the
molecular bands from 3D to 1D models becomes
available.  For
CH and for NH we have
adjusted the scale of our $gf$ values so as to 
reproduce the Solar spectrum,
taken from \cite{wallace98}.
We adopt dissociation potentials of  3.47 and 4.39~eV \citep{huber79} for CH
and OH respectively.  For NH we adopt 3.40~eV based on the
theoretical calculations of \cite{bauschlicher87} and the
laboratory spectroscopy of \cite{ervin87}.
Our analysis assumes classical plane
parallel stellar atmospheres and LTE, both for atomic
and for molecular features.

In view of the inclusion of the near-UV in many of these spectra,
a few UV lines of key atomic species have been added to the master line list.
These include, for example, three Fe~II lines near 3270~\AA\ which
are stronger than any in the optical band.  This is important as
often 
only the two strongest Fe~II lines  in the optical are detected
in even high quality spectra of such extremely metal poor giants;
the remaining optical Fe~II lines are too weak.  For the most extreme EMP
dwarfs, none of the optical Fe~II lines can be detected. 
Inclusion of the UV lines
strengthens the determination of the ionization equilibrium
between Fe~I, with its multitude of detected lines,
and Fe~II. Lines of species with no
detectable optical features, such as V~II and Mn~II, have also been
added.  The resonance lines of Cu~I near 3250~\AA\ were added as well,
as the usual Cu~I lines  seen in stars with [Fe/H] $\sim -2.5$~dex, 
including those at 5105 and 5782~\AA,
become undetectable at the extremely low metallicities of the stars
studied here.  In such cases, the $gf$ values were adopted from
Version 3.1.0 of the NIST Atomic Spectra Database 
(phsics.nist.gov/PhysRefData/ASD/index.html).
The HFS patterns for the UV Mn~II lines
are given by \cite{holt99}, but these lines in the sample stars are 
mostly so weak
that the corrections are negligible. Those for the UV Cu~I lines
were downloaded from R.~Kurucz's web site; his primary source
was \cite{biehl76}.  The  isotope ratio $^{63}$Cu/$^{65}$Cu
was assumed to be the solar value. 

Where possible, we have checked the consistency of
the scale of the transition probabilities for a given
species between the rarely used UV lines and the commonly used
optical ones by comparing
the derived abundances of lines for the same species as a function
of wavelength for a small number of stars with HIRES spectra taken
as part of our 0Z project with somewhat higher [Fe/H], including 
as HD~122563, with weak or no detected CH or NH.
The results of this check were satisfactory
for Fe~II.  However, even with HFS included,  
log[$\epsilon$(Mn)] deduced from the 
4030~\AA\ triplet of Mn~I (the strongest lines in the optical region,
two lines of which are sufficiently unblended to use in an abundance analysis)
appear to be $\sim$0.3 dex lower than those found from the 
redder optical lines and from the UV resonance lines of Mn~II.  In the
solar spectrum, as well as for HD~122563, the
nominal Solar Mn abundance of \cite{anders89} is recovered only for lines
with $4783 \leq \lambda \leq 6022$~\AA, as well as from the
Mn~II UV lines.  
A similar problem with the 4030~\AA\ Mn~I triplet was noted by
\cite{cayrel04}.  
\cite{bihain04} has carried out such a consistency check for Cu~I.  They
compare their [Cu/Fe] determinations based on the near-UV resonance lines
with those of \cite{mishenina02} derived from the weak optical lines. 
They find good agreement, i.e. a mean difference
for 16 stars of $-0.04\pm0.04$~dex.

Following \cite{cohen04}, we adopt a non-LTE correction
for Al~I, for which we only detect the resonance doublet at 3950~\AA, 
of +0.60 dex based on the
work of \cite{bau96} and \cite{bau97}.  We adopt a non-LTE correction for Na~I,
for which we can only observe the two D lines, of 
$-0.20$~dex based on the calculations of
\cite{takeda03}. No other non-LTE corrections have been applied.

Our abundances for the CNO elements are based on molecular bands of
CH, NH and OH respectively.
We use 1D model atmospheres to synthesize the molecular features,
ignoring any 3D effects, although \cite{collet06} suggest that
these may be very large.  They claim that  
CNO abundances may be overestimated by $\sim$0.8 dex as
compared to a 1D analysis when molecular bands are used in EMP stars.

\begin{figure}
\epsscale{0.9}
\plotone{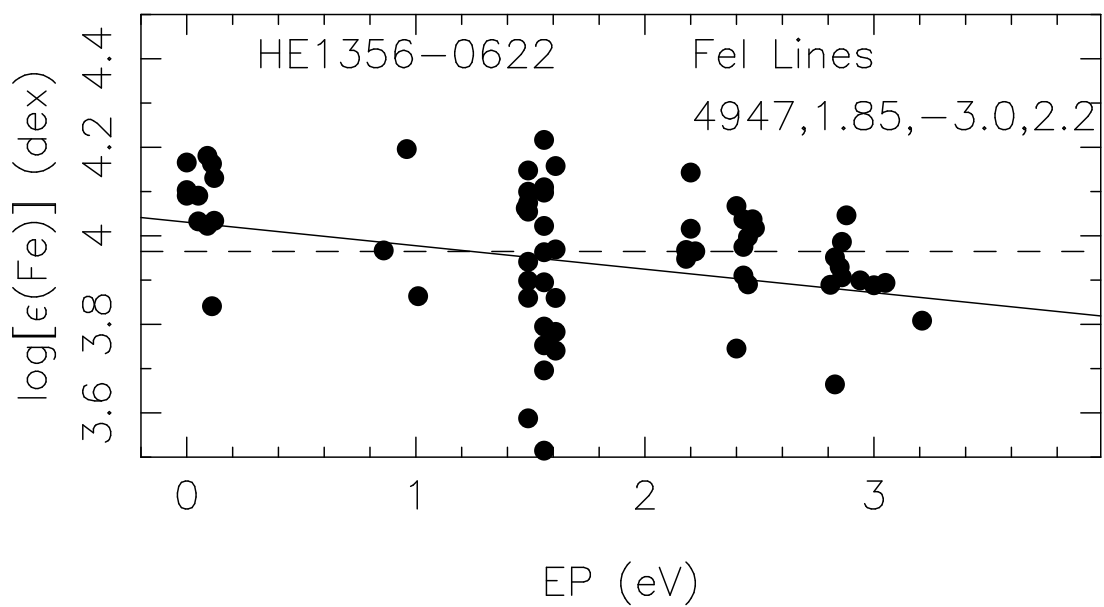}
\caption[]{The iron abundance derived from each of the 63 Fe~I lines
seen in the spectrum of the EMP star HE1356$-$0622 is shown
as a function of $\chi$.  The nominal
set of stellar parameters we derive is used.  The
solid line is the linear fit, with slope $-0.053$~dex/eV with a modest
correlation coefficient of $-0.36$.
However, only the 0~eV lines are discrepant.
The dashed line indicates
the mean log[$\epsilon$(Fe)] derived from the 52 lines with
$\chi > 0.2$~eV.
\label{figure_ep}}
\end{figure}

Table~\ref{table_slopes} gives the slope of a linear fit
to the abundances determined from the set of Fe~I lines
as a function of $\chi$ (the excitation potential of the lower level),
\eqw, and $\lambda$, which are most sensitive to \teff, $v_t$,
and the wavelength dependence of any missing major source
of continuous opacity, respectively.  There are $\sim$ 40 to 60 Fe~I lines detected in
each star, with $\chi$ ranging from 0 to 3~eV.  The correlation coefficients 
$cc(\lambda)$
of the fits with $\lambda$ are, for all except one of the stars,
between 0.11 and $-$0.20, indicating that these fits are not
statistically significant.  The $cc(W_{\lambda} / \lambda$) for the fit
with \eqw\ are within the same range for most of the stars.
The slopes with $\chi$ appear at first sight to be statistically significant
with $\mid cc(\chi) \mid ~ > ~ 0.4$ for one of the 
8 stars; the $cc(\chi)$ are predominantly negative.  
They reach as low as $\sim -0.09$~dex/eV for two of the stars.
If these slopes were
valid, they would 
suggest that for these two stars, \teff\ needs to be decreased by $\sim$300~K
to achieve excitation equilibrium for Fe~I.  However, a careful scrutiny
of the behavior of the derived Fe abundance from individual
Fe~I lines indicates that the problem lies largely in the 0~eV lines; 
for those of higher excitation (the majority of the lines), the
deduced abundance shows no statistically significant
dependence on $\chi$.   A typical example of this is shown
in Fig.~\ref{figure_ep}.

Our plots of abundance versus reduced equivalent width indicate that
the Fe~I overabundances for lines with $\chi ~ < ~ 0.2$~eV do not
appear to arise from
an inappropriate choice of microturbulent velocity parameters.  
The effect could be due
to systematic errors for $gf$ values of low excitation lines, or
may result from resonance scattering \citep*[e.g. see][]{asplund_araa}.
In resonance scattering
the source function, S$_{\nu}$, is reduced to below the local 
Planck function,
thus leading to a stronger line in non-LTE.  Resonance scattering is
seen in the Na~D lines of metal-poor stars \citep*[e.g.][]{andrievsky07},
the OI triplet at
7774~\AA\ in the Sun and may have affected the abundances from the 
Ca~I 4226~\AA\ resonance line 
of \cite{mcwilliam95a}.  If resonance scattering is the cause of the small abundance
enhancement seen in the 0~eV Fe~I lines,
the effect should be more pronounced in the weakest low excitation lines;
thus a plot of \eqw\ versus abundance enhancement should show a positive 
correlation.  Proof
that resonance scattering is the cause of the  apparent overabundances requires a 
non-LTE abundance 
calculation for Fe in our stars, which is beyond the scope of this paper.

\subsection{Radial Velocities \label{section_vr} }

The radial velocities were determined using the procedure
described in \cite{cohen04}, updated 
to improve the long term stability of $v_r$ measurements;
they are given in Table~\ref{table_spectra}.
Each individual measured $v_r$ from the spectrum of a star
taken with the upgraded HIRES detector
for the SNR typical of the present set of spectra taken on
a given run now has an internal
uncertainty of $< 0.2$ \kms, with possible long term systematic drifts
of comparable size; this was not true of the $v_r$ published
in earlier papers of the 0Z project.

For six of our sample EMP stars we have HIRES spectra taken 
between 2 and 5 years apart.  The older spectra have
been re-reduced using our improved procedures and codes to
measure more accurate $v_r$.   R. Cayrel has generously
provided the $v_r$ measured by the First Stars project
from their UVES spectra
for the two stars discussed here which are in common.

Table~\ref{table_vr} compiles the available high precision
radial velocities for these EMP stars from HIRES
and from UVES.  Two of the stars
with  $v_r$ measurements from multiple HIRES spectra show
$v_r$ variations, HE0132--2429 at more than 10$\sigma$ and
HE1012--1540 at the 4$\sigma$ level.  In addition,
BS16467$-$062 may also be a $v_r$ variable, with a difference
exceeding 5$\sigma$ between the two UVES measurements from
M.~Spite \& R.~Cayrel and that from our HIRES spectrum, but
this should be verified.  Such $v_r$ variations presumably
result from orbital motion in a binary system.

\section{Abundances \label{section_abund} }

Tables~\ref{table_abunda} and \ref{table_abundb} give the 
derived abundances for each
detected species in each of the \nemp\ EMP giants. 
We divide the stars into
two groups: low C stars with the G band of CH
barely detectable, if present at all, and  three stars which appear
to have C enhanced. 
The first group includes
five of the 8 stars,
HE0132$-$2429, HE1347$-$1025, HE1356$-$0622, HE1424$-$0241 and BS16467$-$062.
The second group includes three stars, two of which
(HE1300+0157 and HE2323$-$0256) are in the present
analyses hovering just at
the border line of being C-rich, defined by
\cite{beers05} as [C/Fe] $> +1.0$ dex.
The third star,
HE1012$-$1540, is a highly C-enhanced EMP star.
The fraction of C-rich stars in our sample of EMP giants with this definition
is a minimum of 1/8 and a maximum of 3/8,
depending on which side of the boundary of [C/Fe] = +1.0~dex
the two borderline stars fall.

\subsection{The Low Carbon Stars \label{section_low_c} }

We first consider the 5 stars with low C.
These stars span a relatively small total range in \teff\ of 420~K, from 
4950 to 5370~K.  Their spectra show very weak metallic absorption
lines and  lack strong molecular
features, so measuring \eqw\ for them is straightforward; 
the uncertainties in the \eqw\ are thus low, particularly
for the three stars with high SNR spectra taken with the updated
HIRES detector.  A breakdown of
the abundance errors resulting from uncertainties in \teff,
\grav, and $v_t$ can be found in Table~6 of \cite{ramirez03}.
Given our uncertainties in the determination of the stellar parameters,
the dominant contribution to the uncertainties in the abundance ratios [X/Fe]
is that of \teff.  The contribution from errors in $v_t$, \eqw\ and the assumed
metallicity of the model atmosphere are small for most
elements due to the use of mostly weak lines and to the stars being so metal poor. 
(Exceptions are Ca~I, Mg~I and Ni~I, where the expected contributions
to the uncertainty in [X/Fe]
from \teff\ and from \grav\ are each $\leq 0.05$~dex, and
Sc~II, La~II, Ba~II and Eu~II, where they are each $\leq 0.08$~dex
in absolute value.)
We therefore expect the abundance ratios when compared among
this group of stars to have small errors, $\pm0.15$ dex. 
Systematic
errors of comparable size may exist for
the CNO elements as these are derived from hydride
bands and accurate values for many molecular parameters
are required for their analysis.
This estimate is too conservative, at least for
comparisons internal to the 0Z project, for neutral species with at least
three detected absorption lines which have a temperature dependence
similar to that of Fe~I, an example of which is Cr~I, when enough
lines are detected in a star.

\begin{figure}
\epsscale{0.9}
\plotone{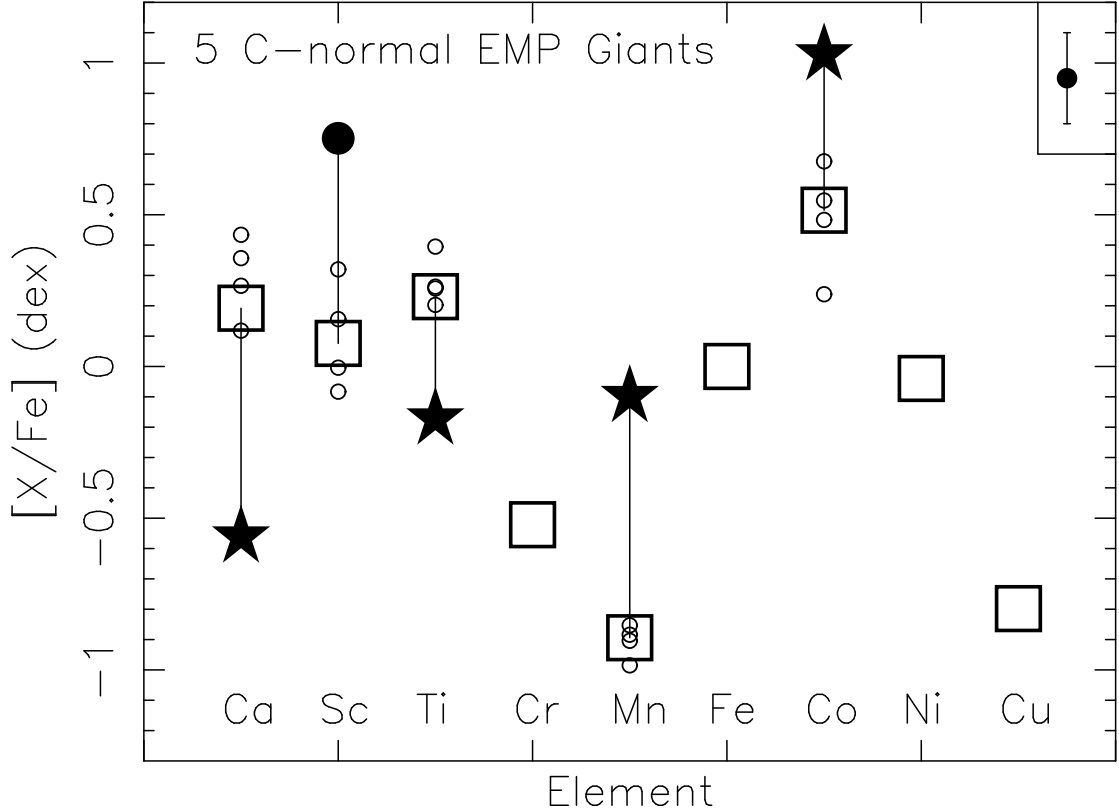}
\caption[]{The median of [X/Fe] for 9 elements from Ca to Cu is shown
for the 5 C-normal stars as a box.  Any ratio [X/Fe] which differs
from the median by more than 0.3 dex for HE1424$-$0241 is shown as a large star;
the same for HE0132$-$2429 is shown as a large filled circle.  If there is
an outlier for a particular species, the abundance ratios for the
remaining four C-normal EMP stars in our sample are shown as small open
circles.
A typical error bar for each ratio [X/Fe] in a star is shown at the upper
right.
\label{figure_cnormal_heavy}}
\end{figure}

The abundance ratios in these stars among the heavy elements
are shown in Fig.~\ref{figure_cnormal_heavy} for 9 species in the
range Ca to Cu.  The median is indicated, and stars with
[X/Fe] which deviate from the median by more than 0.3 dex are shown
individually; these are always either HE0132$-$2429 or HE1424$-$0241.
If there is an outlier for a particular species, the individual values
of [X/Fe] for the remaining four stars in the sample are shown as
well.  Thus this figure demonstrates graphically how deviant
the outliers really are.

The medians show (see Fig.~\ref{figure_cnormal_heavy})
a small excesses for Ca and for Ti with respect to Fe.   
Cr, Mn and Cu are deficient relative to Fe, while
Co is strongly enhanced relative to Fe.
Ni appears to be tied to Fe so that [Ni/Fe] $\sim 0.0$~dex.  
These are in agreement with the usual trend seen among extreme
EMP stars as found in our previous work \citep{cohen04} and in the
First Stars project \citep{cayrel04}.

\begin{figure}
\epsscale{0.9}
\plotone{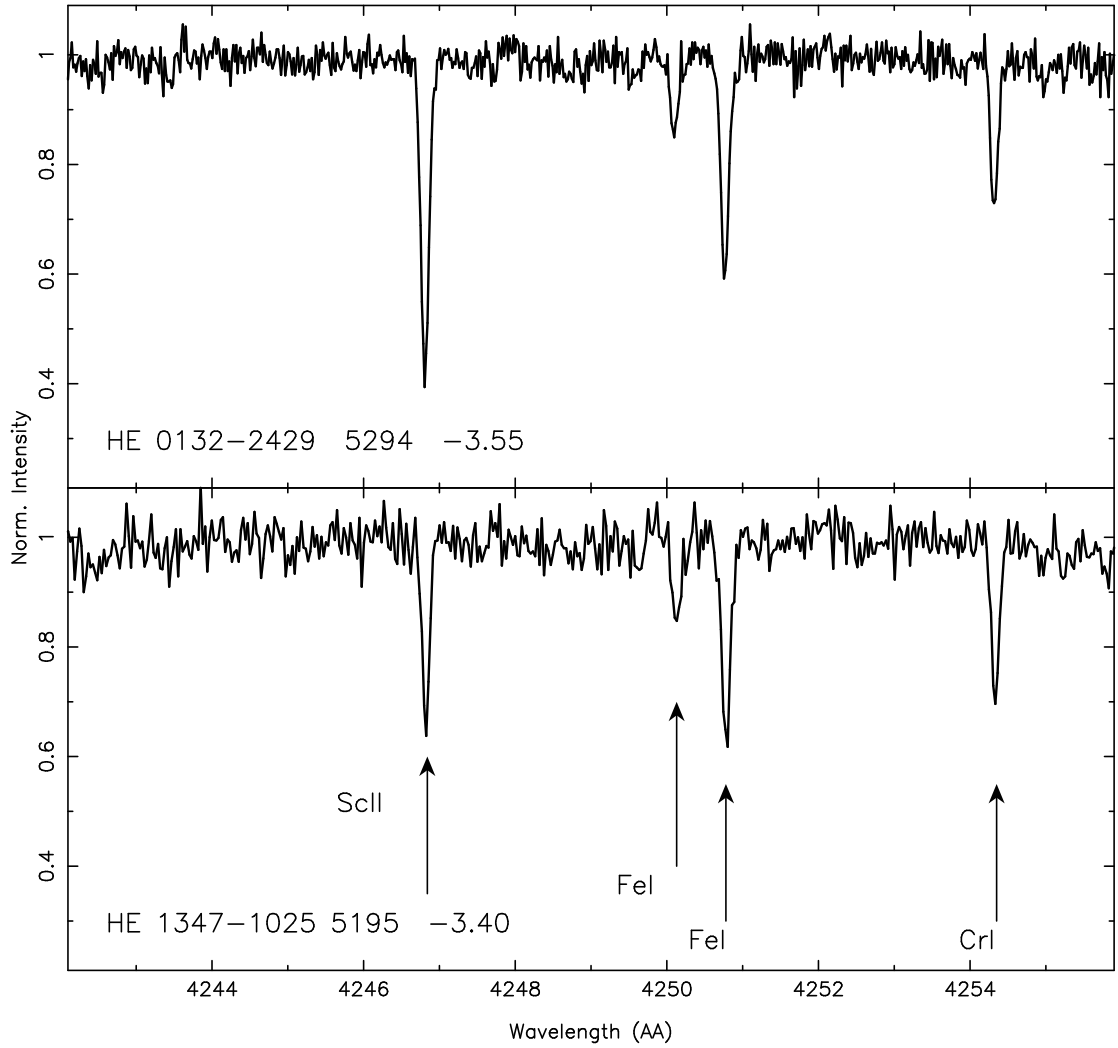}
\caption[]{The region of the Sc~II line at 4246~\AA\ is shown in HE0132$-$2429 and
in HE~1347$-$1025. 
[Fe/H] differs for these two giants by only 0.15 dex, but the ScII line is
much stronger in the former.
\label{figure_sc2_4246}}
\end{figure}

\begin{figure}
\epsscale{0.9}
\plotone{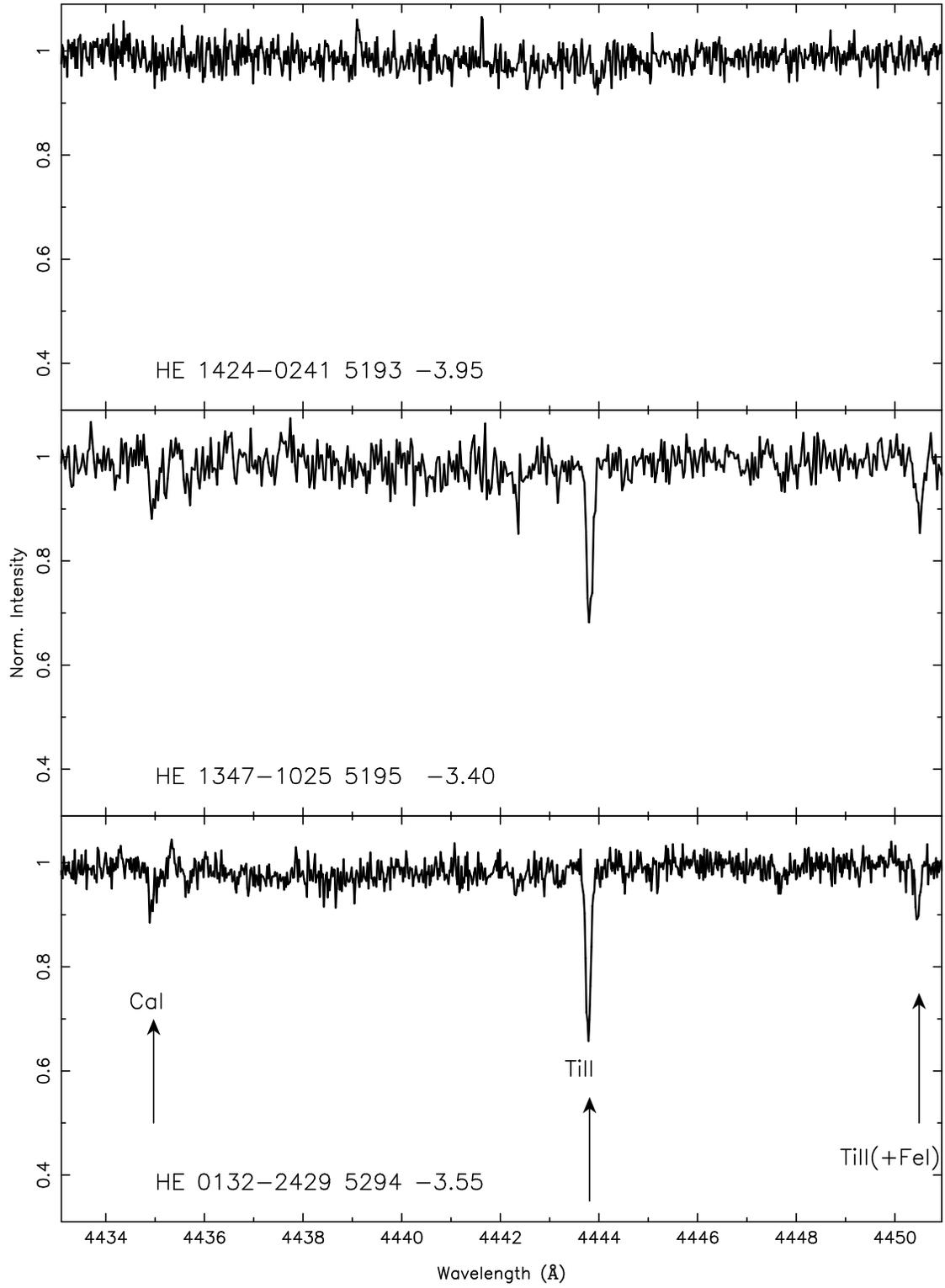}
\caption[]{The region of the Ti~II line at 4444~\AA\ is shown in 
three of the C-normal EMP giants of our sample.
The abnormally low [TiII/Fe] ratio in HE1424$-$0241 is apparent.
\label{figure_ti2_4444}}
\end{figure}

\begin{figure}
\epsscale{0.9}
\plotone{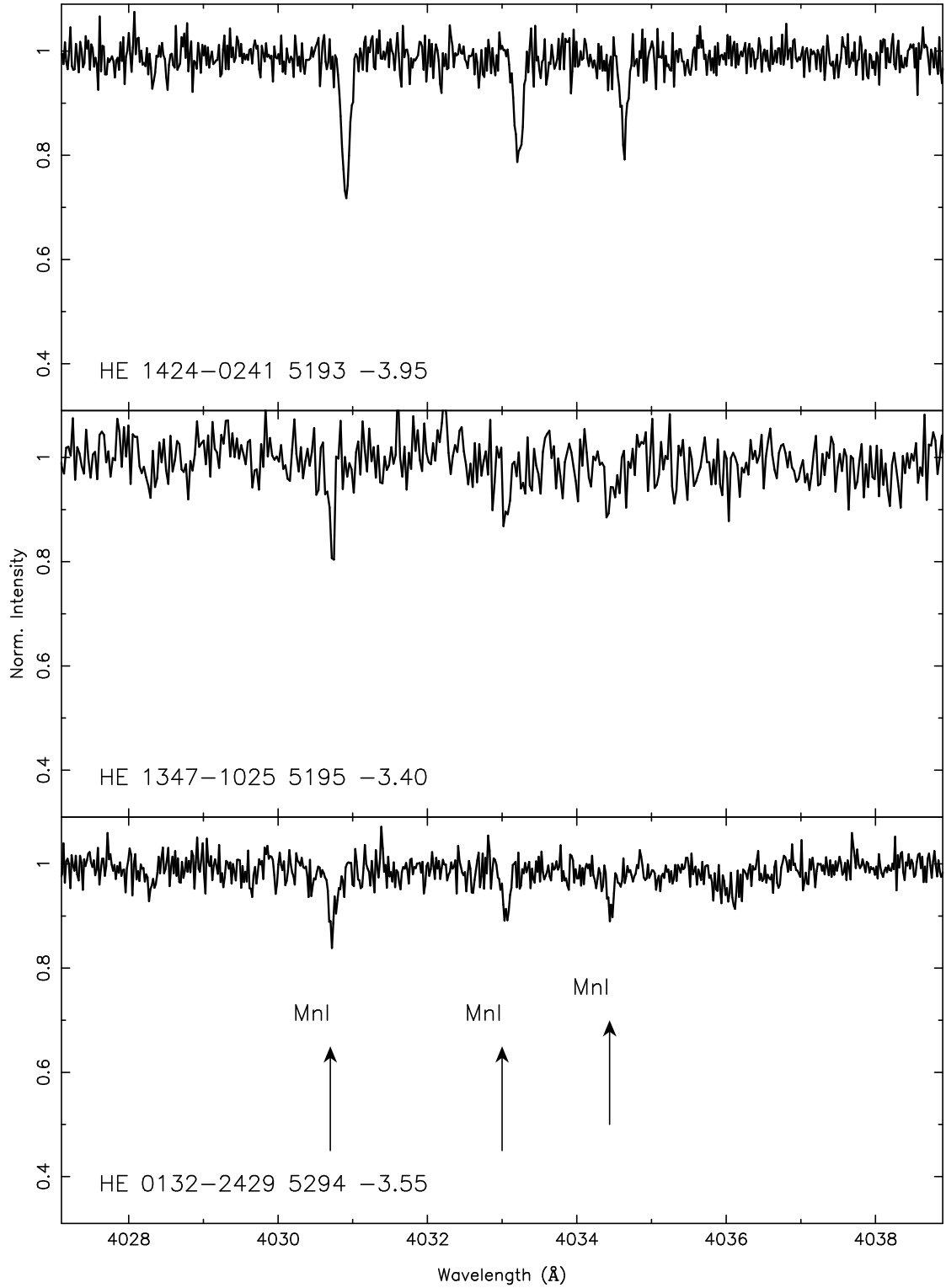}
\caption[]{The region of the Mn~I triplet at 4030~\AA\ is shown in 
three of the C-normal EMP giants of our sample.
The extremely high [Mn/Fe] ratio in HE1424$-$0241 as compared to the
other two stars whose spectra are displayed is apparent.
\label{figure_mn_4030}}
\end{figure}

The surprise is the outliers.  Five  species with outliers are marked 
in Fig.~\ref{figure_cnormal_heavy}, all of which arise in only two
stars, HE0132$-$2429 and HE1424$-$0241.
Spectral regions illustrating lines of three of these cases, contrasting
the outlier star with a star close to the median value of [X/Fe], are shown
in Figs.~\ref{figure_sc2_4246}, 
\ref{figure_ti2_4444} and \ref{figure_mn_4030}.  The
stars displayed in each figure have been chosen to have \teff\ as
close to each other as possible. These figures
demonstrate the reality of the outlier in [X/Fe] for each of Sc~II,
Ti~II and Mn~I.  The spectra of the each outlier star for each of these
five cases have been checked twice.  Two independent HIRES spectra
exist for several of these stars. There is no question that
the outliers in each of [Ca, Sc, Ti, Mn, Co/Fe] are real. 

\begin{figure}
\epsscale{0.9}
\plotone{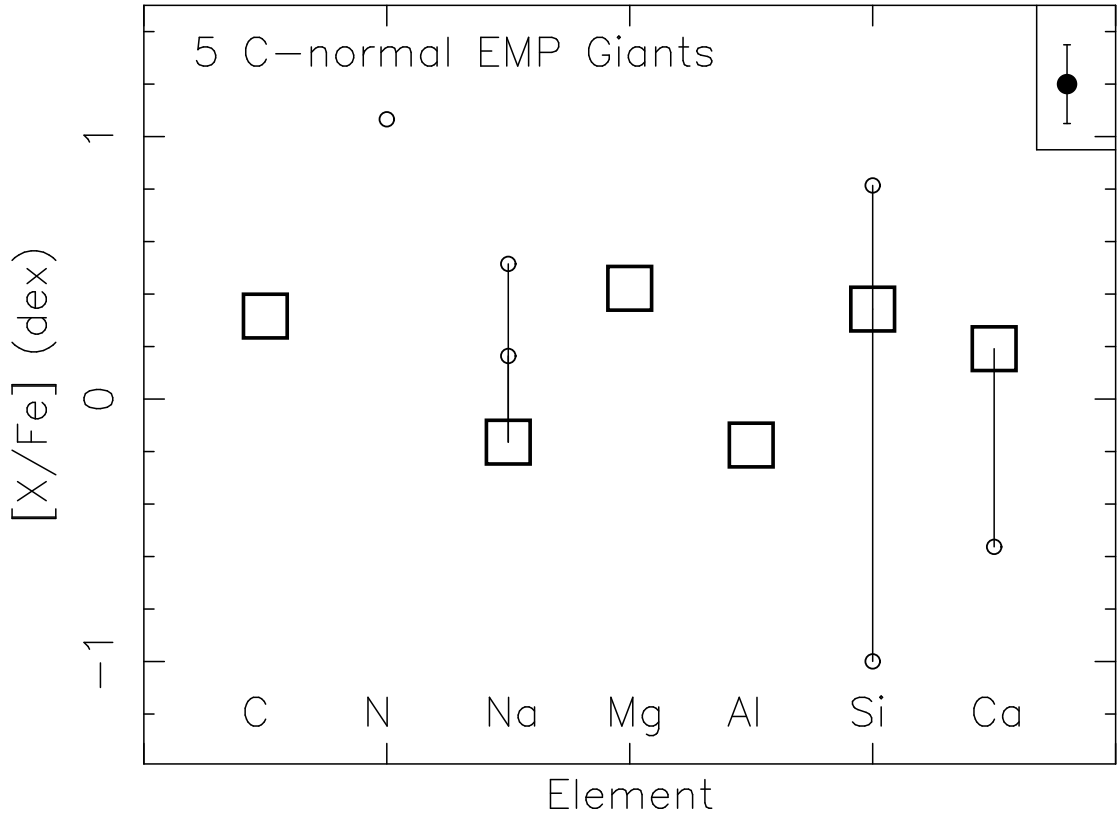}
\caption[]{The median of [X/Fe] for 7 elements from C to Ca is shown
for the 5 C-normal stars as a box.  Any ratio [X/Fe] which differs
from the median by more than 0.3 dex is shown as a small open circle.
Upper limits are excluded.
Only one of these stars has a detectable NH band.
A typical error bar for each ratio [X/Fe] 
in a star is shown at the upper right.
\label{figure_cnormal_light}}
\end{figure}

\begin{figure}
\epsscale{0.9}
\plotone{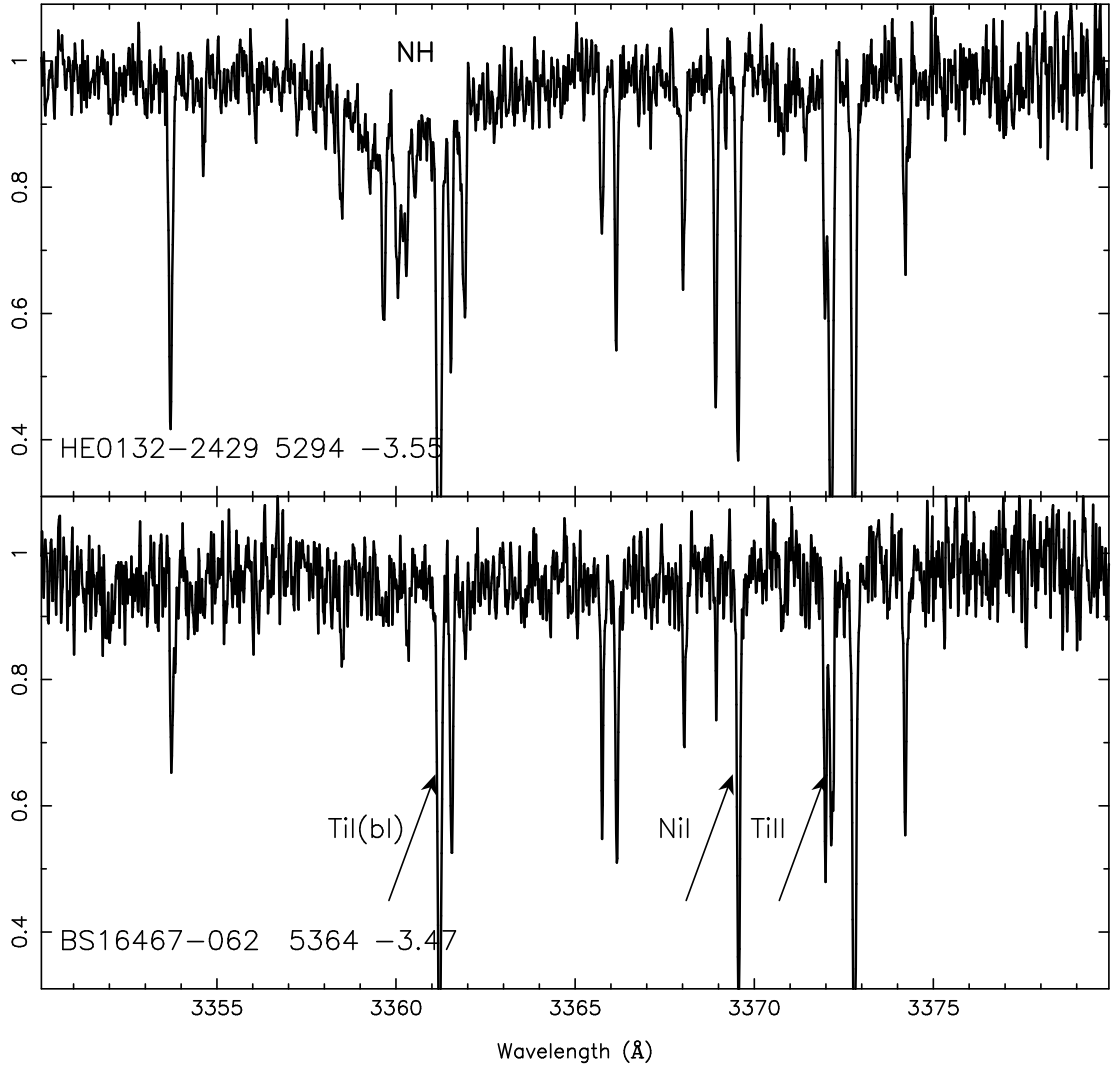}
\caption[]{The region of the NH band at 3360~\AA\ is shown for two
C-normal stars; there is a huge difference in their N abundances.
The \teff\ and [Fe/H] for each star are given following the
object name in the text within each panel, and
the X axis displays rest frame wavelengths in this figure
as well as in the next 4 figures.
\label{figure_nh3360}}
\end{figure}

Fig.~\ref{figure_cnormal_light} shows the behavior of [X/Fe] for
7 light elements in the range C to Ca for
the five C-normal stars.  Upper limits are ignored, and only one of
these stars has a detectable NH band.  Again there are outliers.
One might expect outliers among those elements
(C, N, O, Na, Mg, and Al)  where mixing
of proton-burning material has already been demonstrated to occur
among luminous EMP giants by 
\cite{spite05} and \cite{spite06}, while \cite{andrievsky07}
present slight modifications in the details
due to non-LTE effects without altering the overall picture.  
Large variations in [N/Fe] 
are clearly present
among this small sample of C-normal EMP giants, as illustrated in
Fig.~\ref{figure_nh3360}, where it is shown that HE0132$-$2429
has a strong enhancement for [N/Fe].

\begin{figure}
\epsscale{0.9}
\plotone{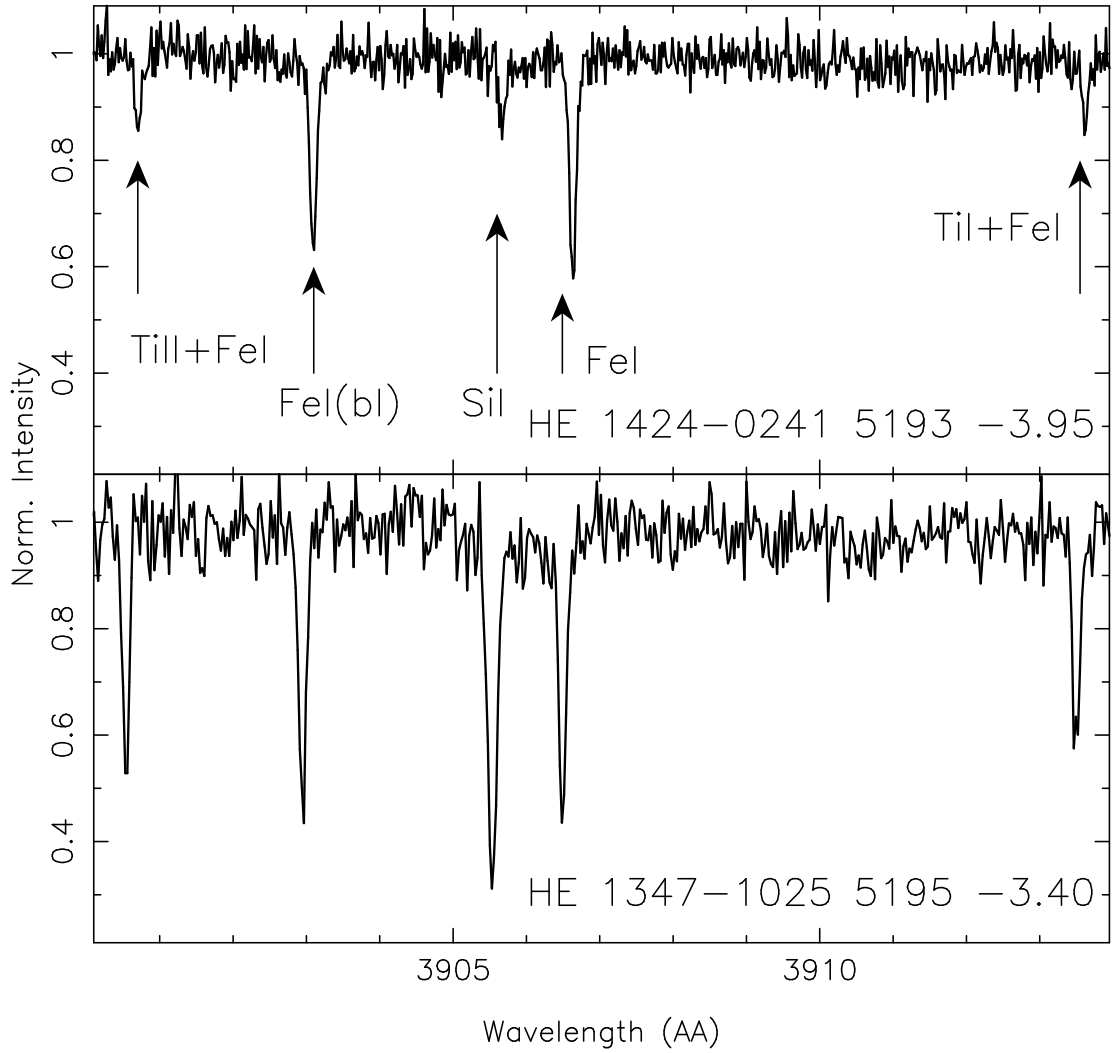}
\caption[]{The region of the Si~I line at 3905~\AA\ is shown in HE~1424$-$0241 and
in HE~1347$-$1025; the \teff\ of both stars is the same. Although the Fe-abundance
is roughly 4 times higher in the latter star,  the ratio of line strengths
clearly demonstrates that [Si/Fe] is abnormally low
in the former star.
\label{figure_si3905}}
\end{figure}

The most peculiar star in this sample of EMP giants, HE1424$-$0241, is 
also the most metal poor, with [Fe/H] $\sim -4$~dex.
The very low [Si/Fe] and other abundance anomalies found  in
a preliminary analysis of this star were
briefly  reported in
\cite{cohen07}. HE1424$-$0241 has a ratio of [Si/Fe] which is 1.2 dex 
below that of any other star in the present sample,
a result which is
completely unexpected.  Fig.~\ref{figure_si3905} shows
the spectral region of the 3905~\AA\ Si~I line
to demonstrate the obvious reality
of this very discrepant abundance ratio. 
This star also has anomalously low [Ca/Fe] and moderately low [Ti/Fe],
accompanied by unusually high [Mn/Fe] (by 0.6 dex) and [Co/Fe].
The G band of CH and the NH band at 3360~\AA\ are not detected in the
HIRES spectra of this star, implying fairly low upper limits for
C and for N.

The EMP giant HE0132$-$2429 has [Sc/Fe] higher than any other star in the
sample of C-normal stars by 0.4 dex, accompanied by very high [N/Fe],
and [N/C]$ > 0$.  HE1356$-$0622 shows 
an apparent small excess for [Na/Fe] and for [Si/Fe].
The modest anomalies in this star are small enough that their reality
is dubious.

The plots comparing
spectral regions around key absorption lines presented here
reinforce our claim that
most if not all of the discrepant points in Figs.~\ref{figure_cnormal_heavy}
and \ref{figure_cnormal_light} are
unquestionably real, and not the result of observational error
nor of uncertainties in the analysis.  They are not consistent
with a dependence on condensation temperature nor on first
ionization potential.
Table~\ref{table_abunda} presents
a summary of the abundance ratios found among the five C-normal stars.

\begin{figure}
\epsscale{0.9}
\plotone{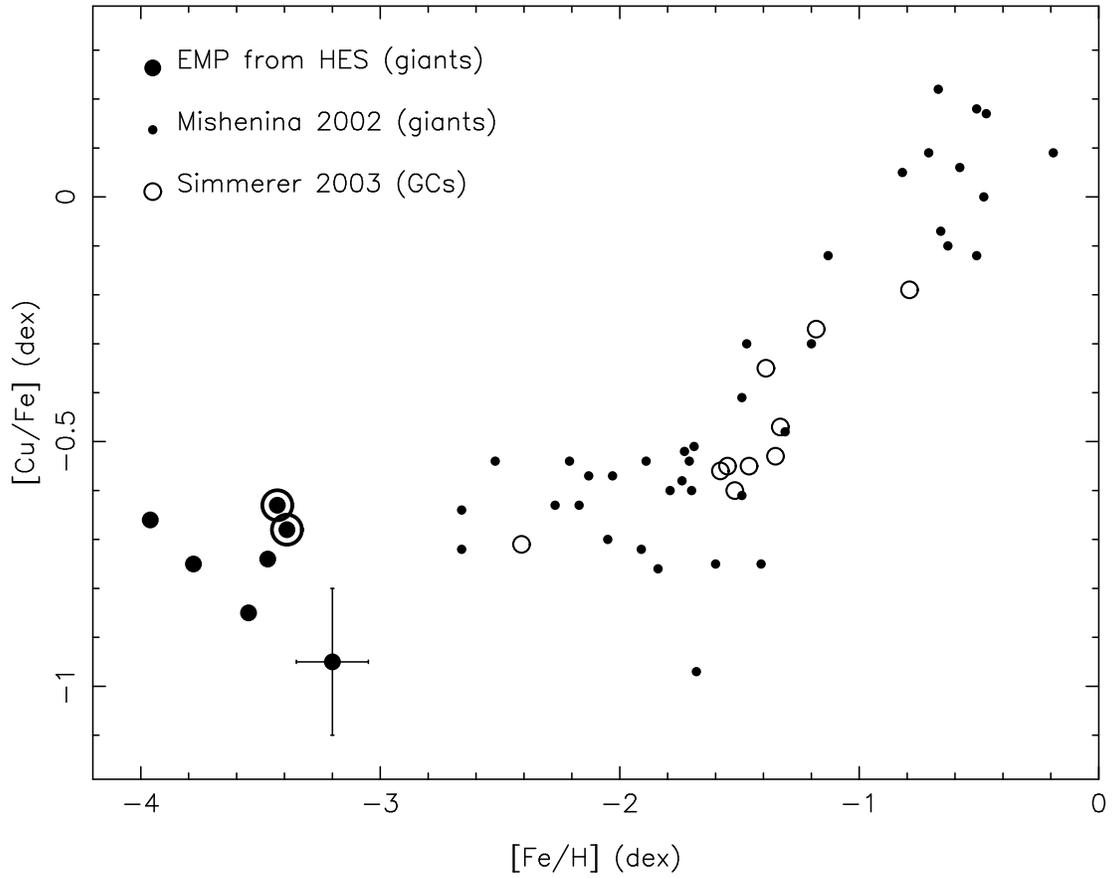}
\caption[]{[Cu/Fe] is shown as a function of [Fe/H] for giants, combining
data for the extreme EMP stars presented here with that of \cite{mishenina02}
and \cite{simmerer03}.  The C-rich stars from the HES
are circled.
\label{figure_copper}}
\end{figure}

We have presented in Tables~\ref{table_abunda} and \ref{table_abundb}
the first Cu abundances for extreme EMP stars, made possible
by the high efficiency of HIRES in the near-UV, so that we can reach
Cu~I resonance lines near 3250~\AA.  Fig.~\ref{figure_copper}
displays [Cu/Fe] as a function of [Fe/H] for our sample of EMP giants
(including one whose analysis has not yet been published).
Earlier results using the weak optical Cu~I lines by \cite{mishenina02}
(for the giants in their sample) and by \cite{simmerer03} who compiled
the means for Galactic globular clusters are shown in this figure.
A steady decrease of [Cu/Fe] as the Fe-metallicity
decreases was found in previous work.  Our new results
demonstrate that [Cu/Fe] reaches a plateau at  low
Fe-metallicity below [Fe/H] $-2.0$~dex which continues through
the extreme EMP stars, as might be expected if 
Cu is formed primarily in massive stars.

\begin{figure}
\epsscale{0.9}
\plotone{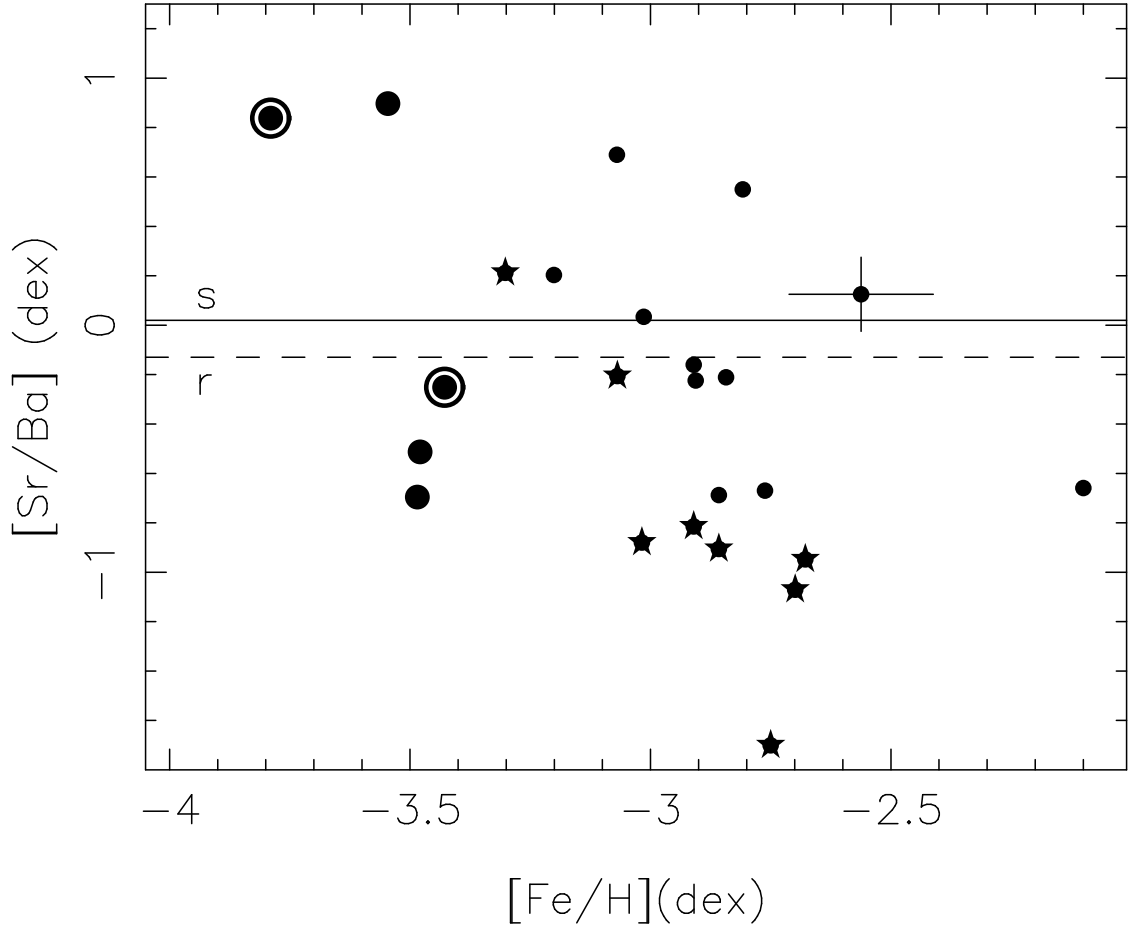}
\caption[]{[Sr/Ba] from singly ionized lines of both elements is shown
as a function of [Fe/H] for the entire sample of stars from the HES
with \teff $< 6000$~K.  Only stars with a secure detection of an
absorption feature of at least one of these two species are shown.
The stars denote carbon stars with detected
bands of C$_2$.  The large filled circles are the EMP giants in the
present sample; those with apparent C-enhancements
are circled.  A typical error bar is shown for a single star.
%
%
\label{figure_sr_ba}}
\end{figure}

The behavior of the heavy element ratio [Sr/Ba] is presented in
Fig.~\ref{figure_sr_ba}.  Here the values from all the stars in
our sample of candidate EMP giants from the HES from our published
and unpublished work are indicated as well to provide guidance
as to the typical behavior. Stars from the present sample often have
very weak lines of both of these elements. If no detected line
has \eqw $< 10$~m\AA, we consider the abundance $\epsilon$(Sr) or 
$\epsilon$(Ba)
for that star to be an upper limit. 

The abundance ratio
$\epsilon$(Sr)/$\epsilon$(Ba) ranges over more than a factor of 100,
in agreement with \cite{mcwilliam95b} and \cite{mcwilliam98}.
The fractions predicted to arise from pure $r$ or $s$ process nucleosynthesis
for the Sun,
taken from \cite{simmerer04}, are indicated in Fig.~\ref{figure_sr_ba}
by the dashed and solid horizontal lines.  
CS22892$-$052, the prototype for the rare extreme
$r$-process stars, shows [Sr/Ba] $-0.4$~dex
\citep*[see, e.g.][]{sneden03}, a value somewhat below the
$r$-process line indicated in Fig.~\ref{figure_sr_ba}.

The presence of numerous EMP stars with [Sr/Ba] larger than that from
either the standard $s$ or $r$-process demonstrates that another
process must exist which produces the light neutron capture elements,
and in particular Sr, in EMP stars, without producing those in
the second peak (i.e. Ba), as was originally suggested by
\cite{mcwilliam95b}.  Early calculations by \cite{prantzos90} have been updated and
augmented by \cite{travaglio04}, who emphasize the many nucleosynthetic
processes that can produce Sr, Y and Zr, and who suggest again
that a secondary source of Sr from an as yet unidentified 
nucleosynthetic site is required. 

Supernova calculations by \cite{woosley92} found production of elements 
significantly 
heavier than the iron-peak, up to A$\sim$100, occurs for high neutron excess 
material (greater than $\eta$$\sim$0.05) during the alpha-rich freezeout.  
They also suggested that the alpha-rich
freezeout might merge, naturally, into an $r$-process.

\cite{chieffi04} computed supernova yields for a range of masses and metallicities,
including charged particle reactions up to Mo.  In their models they found that 
elements heavier
Zn could only be produced for metallicities greater than 
Z/Z$_{\odot}$=10$^{-3}$, essentially
confirming the neutron-excess sensitivity found by \cite{woosley92}
for the production
of elements up to A$\sim$100.  This metallicity limit excluded the supernovae 
considered by \cite{chieffi04} as sources for the enhancements of elements up to
A$\sim$100 seen in EMP stars \citep*[e.g. as found by][]{mcwilliam98}.

\cite{nomoto06} explored theoretical supernova and hypernova yields; 
while they included the
alpha-rich freezeout in their calculations, they only considered species 
up to A=74.  Their
hypernovae were characterized with kinetic energies more than 10 times 
that of normal
core-collapse supernovae.  It was found that in the complete Si-burning 
region of hypernovae
elements produced by the alpha-rich freezeout are enhanced.  
Thus, we speculate that some form
of alpha-rich freezeout, perhaps from hypernova explosions, with 
metallicities lower than the low
limit determined by \cite{chieffi04}, may yet provide an explanation for the 
extra source of A$\sim$100 elements seen in some EMP stars 
(also known as a ``second r-process'').

Most of the
stars with [Sr/Ba] significantly less than that of the $r$-process  (and of the
$s$-process as well) are carbon stars with highly enhanced Ba
from the $s$-process running at low Fe-metallicity
\citep*[see, e.g.][]{busso99}.
Some  of the EMP stars studied here fall somewhat below the
$r$-process line, but not by more than 2$\sigma$.  
 
All of the EMP stars in the present sample, including the three C-rich stars,
have [Ba/Fe] $< -0.2$~dex.
With such weak lined stars, and no excess of Ba in any of them,
no other heavy elements beyond the Fe-peak besides Ba and Sr could be 
detected in any of the \nemp\ EMP stars studied here.

\subsection{The Stars with Higher C \label{section_crich} }

\begin{figure}
\epsscale{0.9}
\plotone{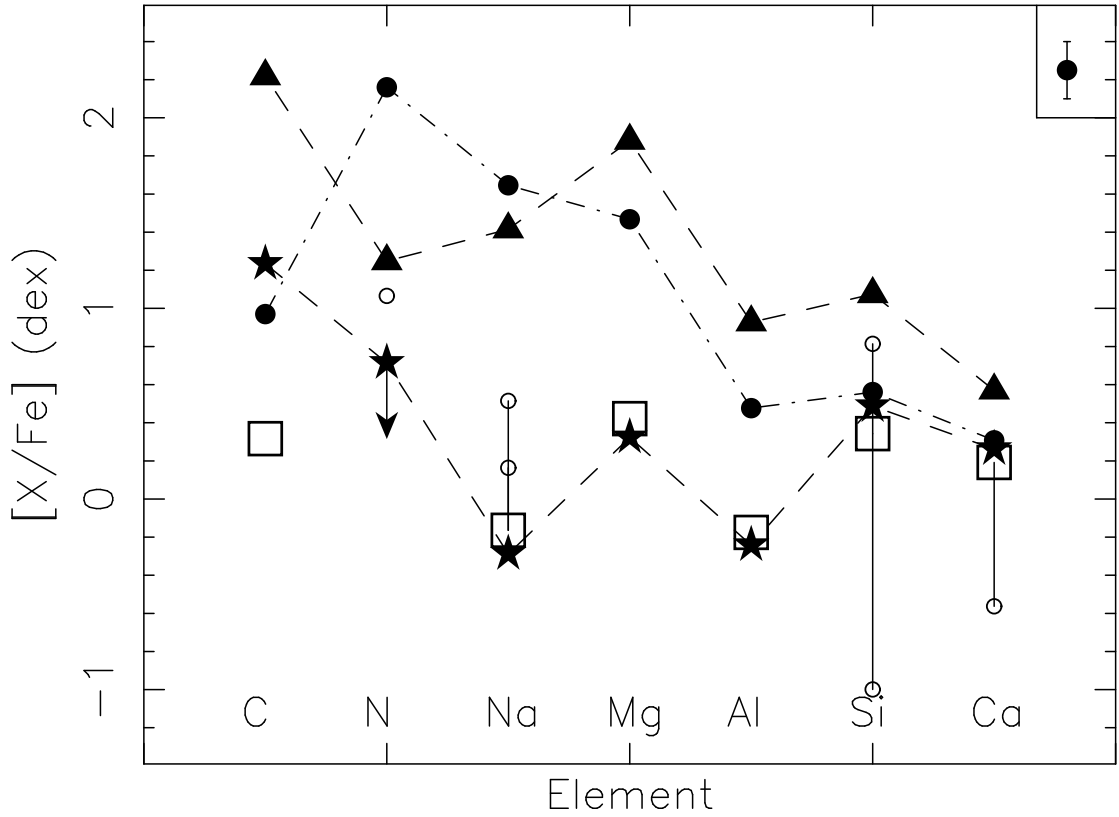}
\caption[]{The same as Fig.~\ref{figure_cnormal_light}.
with [X/Fe] for each of the three C-enhanced stars shown as well,
indicated by filled symbols.  The abundance ratios for each of the C-rich stars
are connected by dashed or dot-dashed lines.  Upper limits are excluded
for the C-normal stars; the single upper limit (for N) 
which occurs in one C-rich star is indicated.  A typical error bar for 
each ratio [X/Fe] in a star is shown at the upper right.
\label{figure_crich_light}}
\end{figure}

There are three stars with the G band of CH obviously much
stronger than the five C-normal stars discussed above.
Two of the then 
have larger enhancements of C than of N, but one 
(HE2323$-$0256, CS22949$-$037)) has [N/Fe] $>$ [C/Fe], as shown
in Fig.~\ref{figure_crich_light}.
In two of the three stars, including the highly N-enhanced star,
Na and Mg are also enhanced, while the enhancement of Al is
more modest.  Such enhancements have been seen among other very metal poor
carbon stars, for example HE0336+0113 from our 0Z survey,
with [Mg/Fe] +1.0 dex \citep{cohen05}, the EMP giant CS29498$-$043,
%
%
studied by \cite{aoki04} with  highly enhanced Na and Mg as well, 
and the very extreme EMP dwarf
CS22958$-$042 (\teff\ 6250~K, [Fe/H] $-2.85$~dex) 
with [Na/Fe] +2.8 dex analyzed by \cite{sivarani06}.
%
%
On the other hand,
HE1300+0157, which has the smallest C+N-enhancement of the three stars,
[C/Fe] $\sim 1.2$~dex, has only an upper limit for N (from NH), with
highly enhanced O (from OH).  It
shows normal abundances for all other detected elements.

\begin{figure}
\epsscale{0.9}
\plotone{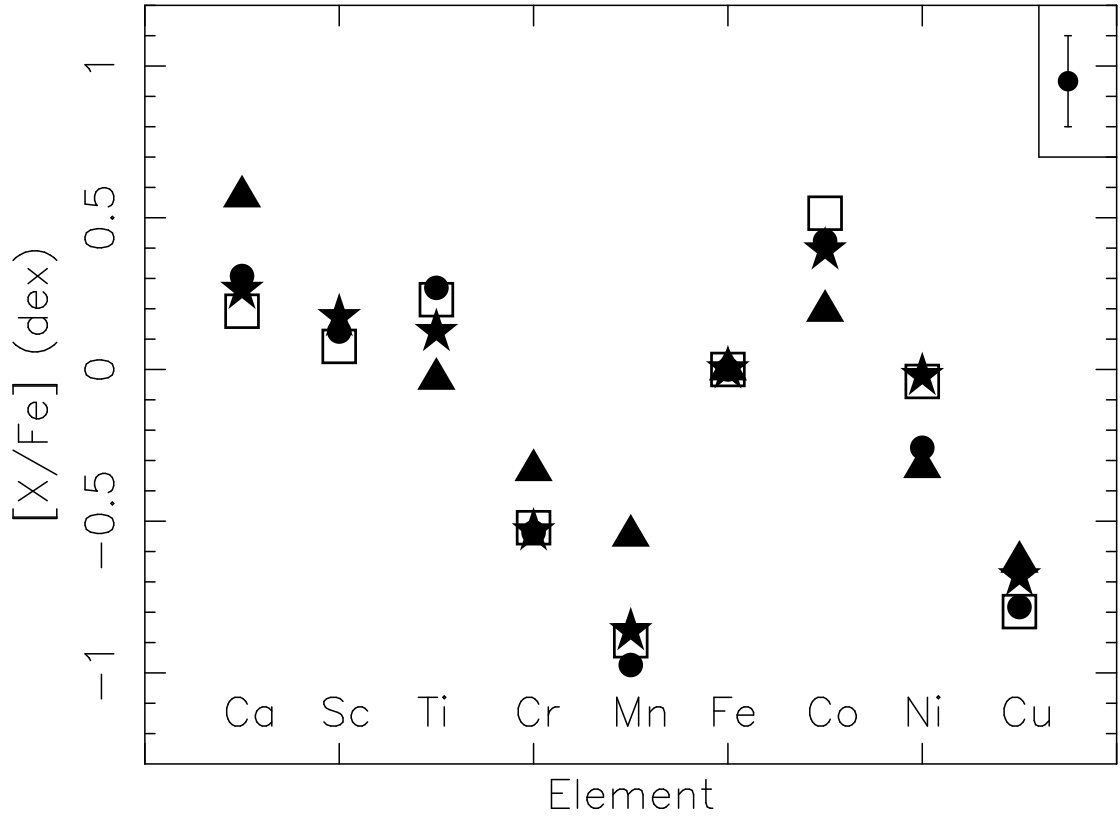}
\caption[]{[X/Fe] is shown for each of the three C-enhanced EMP stars
for 9 elements from Ca to Cu. The median for the five C-normal stars
is indicated by a box.  A typical error bar for each ratio [X/Fe] 
in a star is shown at the upper right.
\label{figure_crich_heavy}}
\end{figure}

Each of the three C-rich
EMP stars shows good agreement for the abundance ratios [X/Fe]
for the elements with detected features from Ca through Cu
with each other and with the median from the five C-normal stars.  This is
illustrated in Fig.~\ref{figure_crich_heavy}.   The C-enhancement, even
when extreme, does not affect the relative abundances of elements
in this range, which includes the Fe-peak, as was suggested earlier
by, e.g. \cite{cohen06}.

\section{Comparison with the Abundance Analyses of \cite{cayrel04}
\label{section_comp} }

The 0Z project and the First Stars project of \cite{cayrel04}
using UVES at the VLT are two large independent
efforts to determine the chemical abundance ratios in EMP stars
and use those to draw inferences on the properties of the
early Galaxy, the first supernovae, etc. 
Our sample largely consists of new EMP stars we have found through
painstaking, time consuming searches of the HES database coupled
with the expenditure of very large and generous allocations of telescope time.
We have observed a few stars from thee sample of the First Stars Survey at the VLT
\citep{cayrel04} and have analyzed them independently.  We compare 
our results with those of \citep{cayrel04} to determine the 
consistency of the absolute
iron abundance and relative abundances of various
elements as deduced by our group versus those of the First Stars VLT large program.

\begin{figure}
\epsscale{0.9}
\plotone{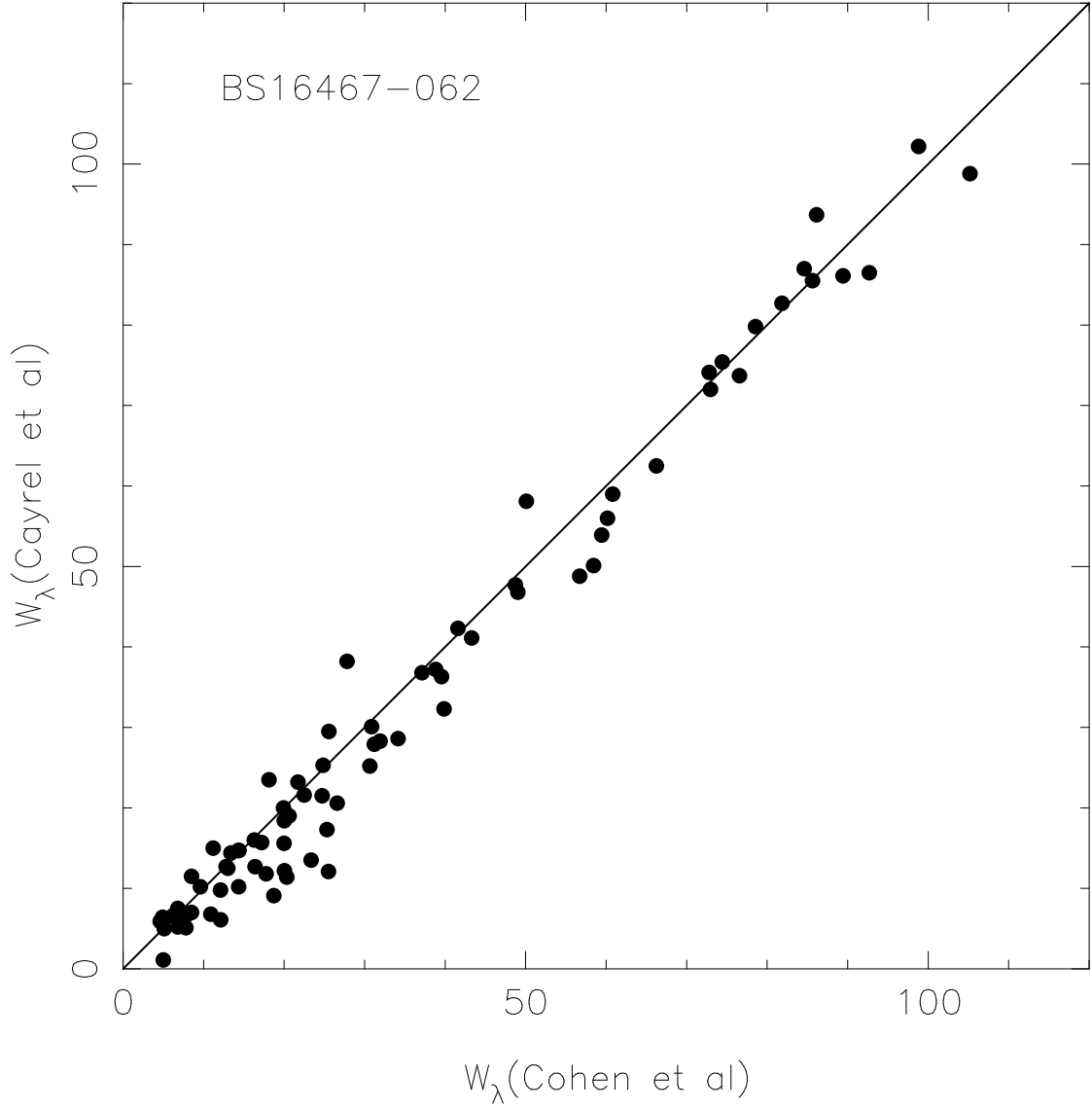}
\caption[]{The \eqw\ we measure from our Keck/HIRES spectra for the
EMP star BS16467--062 are compared to those of the First Stars VLT/UVES
program data from \cite{cayrel04}.
\label{figure_eqw_bs16467}}
\end{figure}

\begin{figure}
\epsscale{0.9}
\plotone{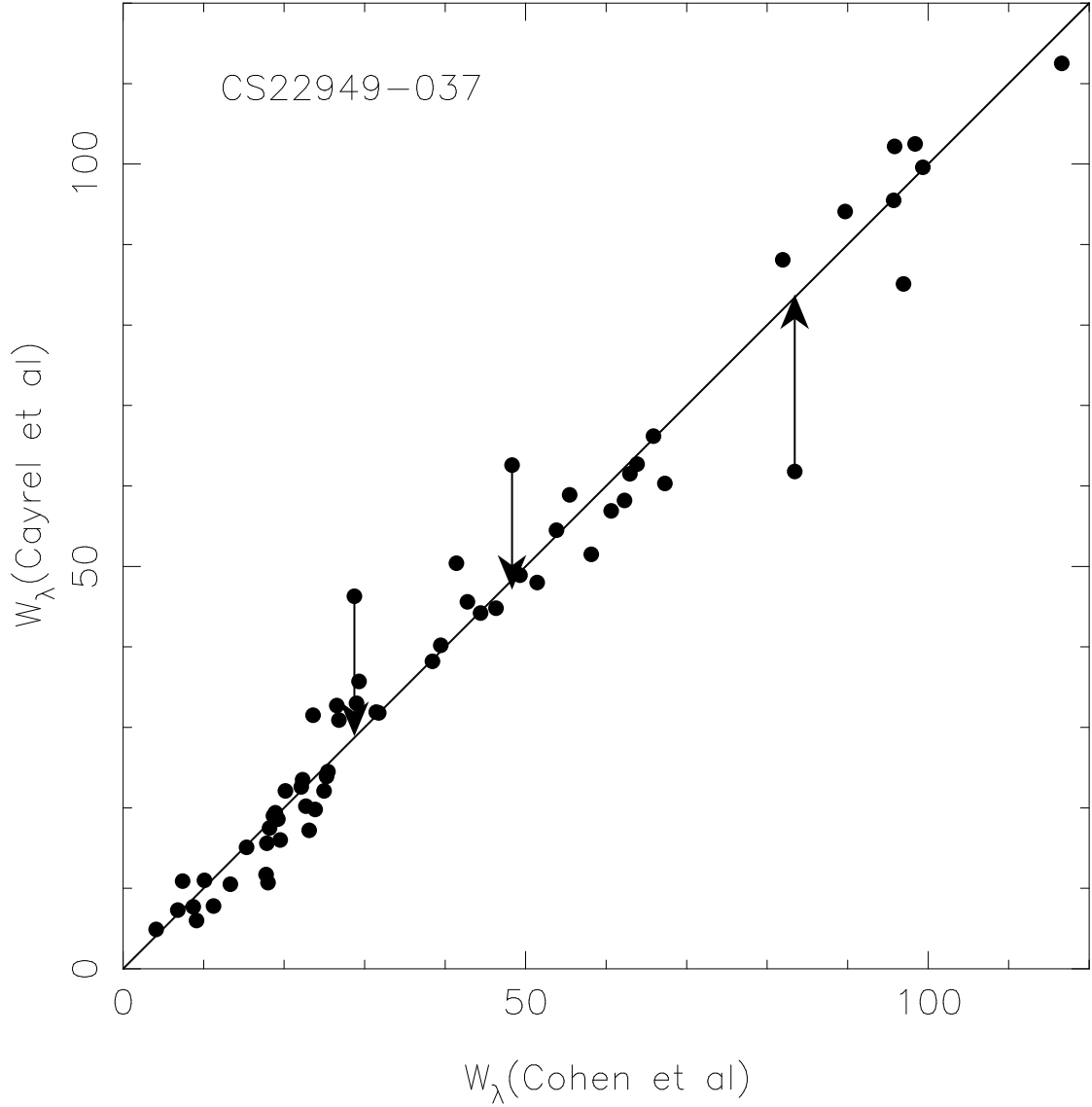}
\caption[]{The same as Fig.~\ref{figure_eqw_bs16467} for
the star HE2323--0256 (a.k.a. CS22949--037).  The arrows denote
corrections to the published \eqw\ of \cite{cayrel04}
(M.~Spite, private communication, June 2007).
\label{figure_eqw_he2323}}
\end{figure}

We begin by testing the measured equivalent widths
for the
two stars in common, HE2323$-$0256 (a.k.a. CS22949$-$037) and
BS16467$-$062, the latter of which we added to our sample specifically
for this purpose.  There are 79 lines in common for
BS16467$-$062, with a mean difference in \eqw\ of 1.8~m\AA\
and $\sigma$ of the differences of 4.4~m\AA.  This extremely
good agreement is shown in Fig.~\ref{figure_eqw_bs16467}.
The agreement in measured \eqw\ for HE2323--0256 (a.k.a CS22949--037)
is not quite as good (see Fig.~\ref{figure_eqw_he2323}); the
dispersion of the differences in 
measured \eqw\ for the 66 lines in
common is 9.5~m\AA, with a mean difference of only 0.3~m\AA.
Most of the dispersion arises from 5 discrepant lines, as is shown
in Fig.~\ref{figure_eqw_he2323}. M.~Spite advises (private
communication, June 2007), on behalf of
the First Stars Project, that their published \eqw\ for these 5
lines are not correct, and that the correct values from their UVES spectra are
much closer to those given here in Table~\ref{table_eqw}.  She further
advises that she believes that their problems with \eqw\ are restricted to this
particular star.

We next examine the scale of the transition probabilities adopted by
each group. For 43 Fe~I lines in common in the spectrum of 
BS16467$-$062, the mean difference in log($gf$) is only
0.004 dex, with $\sigma$ for the differences of 0.06 dex.
The scale of the $gf$ values for all lines of species 
in common with \cite{cayrel04}
have been compared.
The maximum scale difference for the lines of a given species
was only 0.04 dex (occurring
for Fe~II), with the largest dispersion about the mean
for the lines in common reaching 0.07~dex (for Ti~II).
Thus we find that the  parameters adopted for atomic lines
by the two groups are in very good agreement, and specifically 
for Fe~I are identical in the mean to within $\pm0.01$~dex.

We have adopted the \cite{schlegel98} reddening map, which 
has a small but non-zero reddening at the Galactic pole, while
the First Stars project appears to be using the older \cite{burstein82}
values based on 21 cm HI surveys.  This map has zero 
reddening at the Galactic pole.  Hence we have systematically
larger reddening values for each star than does the First
Stars project.

A detailed discussion of the differences in the stellar
parameters between us and the First Stars project
and the differences in abundance, both absolute
(i.e. [Fe/H]) and ratios with
respect to Fe ([X/Fe]), is given in  Appendix~B.  Overall
for a particular star with measured \eqw\ for a set of detected absorption features
and values of observed optical and 2MASS colors,
the [Fe/H] value derived by the First Stars project
as described in \cite{cayrel04} will be systematically $\sim$0.3~dex lower
than that for the same star as analyzed by the 0Z project.
It is interesting to note that the sample of
stars analyzed by \cite{aoki05} also included two stars in common
with the First Stars
project sample. \cite{aoki05} derived [Fe/H] values higher than those
of \cite{cayrel04} by $\sim 0.2$~dex.

Table~1 of \cite{cohen07} compares the mean for [X/Fe] for C-normal
EMP giants between our 0Z survey and the First Stars survey of
\cite{cayrel04}.  A detailed comparison for a small number of 
individual stars in common is given in Appendix~B.
We find much better agreement of the abundance ratios [X/Fe]
between the two large surveys
than for absolute Fe-metallicities [Fe/H]. This is as expected since
many of the error terms in the absolute iron abundance
[Fe/H] largely cancel out in an abundance ratio [X/Fe].  
If we ignore C deduced from an analysis of the CH band
and N inferred from the UV NH band, 
we find
differences in [X/Fe] for 11 or 12 elements in a star 
ranging from $-0.07$ to +0.05 dex
when we adopt their equivalent widths but use our stellar parameters.
Somewhat larger differences, $\pm$0.15 dex with $\sigma = 0.10$~dex, 
occur when we analyze
our own HIRES spectra with our own choice of stellar parameters, in
part because of the errors in the \eqw\ of \cite{cayrel04}.
As indicated above, only a maximum of $\pm0.04$~dex
of these differences arise from differences in the scale of the
transition probabilities adopted by each of
these large surveys.

\subsection{Comparison with the HERES Sample \label{section_heres}}

Analyses of 253 stars from the Hamburg/ESO $r$-process enhanced
star survey \citep[HERES,][]{christlieb04b} was presented by
\cite{barklem05}.
This survey  relied on modest SNR
high resolution spectra of candidate EMP giants from the HES.
 We have observed with HIRES at the Keck~I
telescope the most Fe-poor star found
in that survey, HE1300+0157, as a comparison object.
A very detailed abundance analysis based on a high quality
Subaru/HDS spectrum for this star was recently presented by \cite{frebel07}.

Both HERES and \cite{frebel07} utilize the 
relations between broad band colors and \teff\ for giants of
\cite{alonso99} evaluated at [Fe/H] $-2.0$~dex, as these
relations have not been calibrated adequately at still lower
metallicities.  As was discussed in \cite{cohen02}, while
the MARCS and ATLAS9 \teff\ color-relations are in very good
agreement, they disagree with those of \cite{alonso99}.  In this
regime of \teff, \grav\ and [Fe/H], the difference
in deduced \teff\ for a fixed $V-K$ color is $\sim$200~K, with
the value derived from the \cite{alonso99} relations being
cooler.  The \teff\ adopted by \cite{barklem05}, \cite{frebel07},
and that we derive for HE1300+0157
are 5411, 5450, and 5630~K respectively.
This difference in \teff\ corresponds to a difference in
[Fe/H] of $\sim$0.35~dex, with  HE1300+0157 having the
higher Fe-metallicity of $-3.4$~dex in our analysis instead
of the value they obtained, $-3.7$~dex.

We first consider the comparison for the star HE1300+0157 if we
adopt \teff\ and \grav\ from \cite{barklem05}, but 
analyze with our own codes and atomic parameters 
our own set of \eqw\ from our HIRES/Keck spectra.
Overall, with this assumption, the agreement between the results 
presented here
based on high SNR Keck/HIRES spectra and those of HERES 
is very good given
the lower SNR of their spectra and the automatic analysis
codes employed in the analysis of \cite{barklem05}.
For the 10 elements in common, the agreement in log$\epsilon$(X) 
is in all cases within
the errors assigned by HERES for absolute abundances (their ``errA'' 
values, ranging from $\sim$0.2 to 0.3 dex), and in almost all
cases is within the smaller relative errors they assigned for star-to-star
comparison within HERES.  The techniques employed by HERES are certainly
more than adequate to find interesting EMP stars and determine the general
nature of their chemical inventory.

We detected several additional elements beyond those that HERES
could reach.  We suggest, as did \cite{frebel07},  that
the HERES claimed
detection of Y using their automatic
abundance analysis code  is almost surely incorrect;
they found [Y/Fe] +0.56 dex.
They could not detect Sr; we find a marginal detection of 
a single Sr~II line which yields [Sr/Fe] $=~-1.55$~dex.
Given this very low Sr abundance, any accessible
optical Y line would be expected to be undetectable.

We now compare  our derived [Fe/H] and abundance
ratios [X/Fe] for HE1300+0157 with those of the very
detailed and careful analysis by
\cite{frebel07}.   Their \teff\ is 180~K cooler than ours, hence
they derive [Fe/H]  0.34 dex lower than we do.  However, the difference
in abundance ratios should be smaller assuming the same stellar parameters
are adopted.  

For the CNO elements, we note that there is agreement to within
0.2 dex for the [C/Fe] and [O/Fe] abundance ratios in this
star between our analysis and that of  \cite{frebel07},
while both fail to detect the NH band and only have an
upper for [N/Fe].  Our [C/Fe] is identical to  that derived by
\cite{lucatello06}, who analyzed the HERES spectra for the CNO
elements.  Since their [Fe/H] for this star was only $-2.9$~dex,
this implies a difference in $\epsilon$(C) between the
value they derive and that of either of the two
high dispersion analyses of about a factor of 3 (0.5~dex),
with the  value of \cite{lucatello06} for HE1300+0157 being too large.  

We have carried out the  comparison for this star
adopting first the stellar
parameters of \cite{frebel07}, then those we have derived.
In both cases
we use our own set of \eqw\ measured from our HIRES spectra.
The results  are given in
Table~\ref{table_he1300}.  Our measured  \eqw\ for Fe~I lines show no
systematic difference with those of \cite{frebel07}.
Similarly $\Delta$(Fe) = [Fe~I/H] -- [Fe~II/H] has the same
sign (positive) in both analyses, but our ionization equilibrium
is slightly better than theirs ($\Delta$(Fe) = 0.08 vs 0.15~dex)
for the identical stellar parameters.  However,
log[${\epsilon}(Fe~I)$] is slightly higher in our analysis (by 0.13~dex) 
with same stellar parameters; the origin of this offset is not clear.
  
Adopting their (cooler) \teff, 
we find that of the 17 elements in common (ignoring upper limits),
log[$\epsilon(X$)] differs by less than 0.10~dex for 7 of them,
but disagrees by more than 0.15~dex for 6 species. 
Adopting our hotter \teff\ raises [Fe/H] substantially.  But
the values of [X/H] are only slightly altered, 
$\mid\Delta {\rm{[X/H]}} \mid ~ < ~ 0.15$~dex
for most elements in our analysis.  The largest change in abundance
ratio is seen for C, which is derived from the CH molecular band;
with the higher \teff\ [C/Fe] increases by $\sim$0.2~dex.  The
same holds for O (from the OH band).

\section{Discussion}

Our 0Z survey and the First Stars survey at the VLT
(see, e.g. Cohen \etal\ 2004, Cayrel \etal\ 2004),
following in the footsteps of many earlier studies, including,
for example, \cite{mcwilliam95a} and \cite{mcwilliam95b}, have established
over the past five years the general behavior of abundances
among EMP stars, with substantial samples of stars analyzed
with [Fe/H] $< -2.5$~dex.
If one ignores the light elements which might be affected by
proton burning, 
definite trends of [X/Fe] with [Fe/H] have been established 
beginning with Ca and extending through the Fe-peak
which hold
down to [Fe/H] $\sim -4$~dex.  The
data available to date show that there is a scatter about these
trends which for most stars with normal carbon abundances 
and for most elements
is not larger than the observational uncertainties.
A substantial theoretical effort has gone into calculations of
nucleosynthesis yields in core collapse SN directed towards understanding the
behavior of these ``typical'' VMP and EMP stars.  The work of
\cite{chieffi04}, \cite{kobayashi} and
\cite{tominaga07} are examples of recent 
computations for grids of metal-poor stars over a range of initial mass and
chemical composition.  These models have been tuned to  reproduce the 
previously observed trends of abundance ratios among
``typical'' EMP stars.

In comparing the properties of our sample of EMP giants with the
predictions of such calculations, it behooves us to recall the 
enormous difficulty of these calculations and the many parameters whose
values must be calculated from theory, assumed, or inferred from the data and which
substantially affect the resulting predicted nucleosynthesis yields.  
Among the most crucial of these
factors are the explosion energy, the mass cut, the neutron excess,
and  previous mass loss in earlier evolutionary stages.

There are three stars in the present sample of EMP giants which we consider as
``typical''.  They have normal or low carbon.
Their abundance ratios follow the patterns
previously delineated  by our work and that of the First Stars project.
However, there are also two C-normal stars which are definite outliers.
Some abundance ratios in these two stars are
definitely anomalous.  The small \teff\ range of our sample discussed
here ensures that intercomparisons within the 0Z project set of 
abundance analyses are valid and that differences exceeding
0.3~dex are real. Furthermore, in Table~1 of \cite{cohen07} we compared our
mean abundance ratios with those of \cite{cayrel04} for
those ``typical'' giants studied by each group.  In our
case this included our published and unpublished abundance analyses,
and for the First Stars project we relied upon the 
fits tabulated in \cite{cayrel04}.  The agreement was very good,
within 0.10~dex, with one
exception (Mg).  Although we
emphasize again that we have shown here that our [Fe/H] scale is 
systematically 0.3 dex higher than that of \cite{cayrel04}, the consequences
for abundance ratios of differences in the details of the analyses
between these two surveys are much smaller.
We are therefore
confident that any outliers  found are real and are not the result
of problems or uncertainties in our measurements or analyses.

HE1424$-$0241, with [Fe/H] $\sim -4$~dex,
is the  most extreme outlier we have found.  Its peculiarities were
briefly described in \cite{cohen07}.   This extreme EMP giant
has a very low abundance of Si, and moderately
low Ca and Ti, with respect to Fe.  Si, Ca and Ti are produced
primarily via explosive $\alpha$-burning.  But Mg/Fe, where Mg is produced
largely by hydrostatic $\alpha$-burning, is normal in this star.
Mn and Co are enhanced with respect to Fe.  Only (low) upper limits
for C and for N could be determined.
Older calculations of nucleosynthetic yields by \cite{woosley95}
come close to reproducing at least some of this behavior with
ejecta from
SN biased towards the lower end of the relevant mass range,
but more current grids of SNII nucleosynthesis fail to reproduce
the very unusual chemical inventory seen in this extreme EMP star.
We defer to
our theoretical colleagues to try to find an explanation for
this very peculiar star.

The anomalies seen in HE1424$-$0241 are unique and,
as far as we are aware,
are not seen in any other EMP giant studied to date.
This is illustrated in Fig.~\ref{figure_si} and
\ref{figure_ca}, which show [Si/Fe] and [Ca/Fe] as a function
of [Fe/H] for all the C-normal giants analyzed by our 0Z project to
date, all those of the First Stars project \citep{cayrel04},
and those from several other sources noted in the figure legends.
In both cases, HE1424$-$0241 has the lowest value by far
of the relevant abundance ratio.

There is one star, CS22966$-$043,
studied by
\cite{preston00} and again by \cite{ivans03} which has
abundance ratios somewhat similar
to those of  HE1424$-$0241.  CS22966$-$043 has [Fe/H] $-1.9$~dex, with 
[Si/Fe] $-1.0$~dex
\citep{ivans03} and [Ca/Fe] $-0.2$~dex, with [Cr/Fe] and [Mn/Fe]
somewhat high and [Sr/Fe] somewhat low for its Fe-metallicity.  CS22966$-$043,
however,
has \teff\ = 7200~K.  It is a SX Phe variable and a binary.
It shows rotation, with $v_{rot} {\rm{sin}}(i) = 20$~\kms ~\citep{preston96}, 
not uncommon among
stars in the \cite{preston00} sample of blue metal poor stars.
It may be a blue straggler, the outcome today of extensive past mass transfer
within the binary system.
\cite{ivans03} attribute its anomalies
to local differences in the chemical history within 
different regions of the Galactic halo,
presumably arising from accretion of one or more dwarf satellite galaxies.
We assume, perhaps incorrectly, that whatever may be causing its anomalies
is not directly relevant to those of the EMP giant HE1424$-$0241.

The second of the two outliers is HE0132$-$2429, with [Fe/H] $-3.55$~dex.
The spectrum of this star indicates moderately high [C/Fe],
very high [N/Fe], and high [Sc/Fe].
We suggest that this is the result of a major contribution
to its chemical inventory from a SNII with a higher
than typical mass.  \cite{limongi06} reproduce the
general nature of these anomalies with a massive 
SNII with $M \sim 60 M$\subsun\ (see their
Fig.~5). 
Although this calculation was carried out for
solar metallicity, we take it as 
applying at least partially to extremely metal poor SNII.  We
could not locate any
published SNII models with nucleosynthetic yields
for a large set of isotopes for such massive extremely metal poor 
stars; the published grids typically end at 40$M$\subsun, with some
studies, for example \cite{umeda02}, then jumping to treat
very massive pair instability
SN with $M \sim 150M$\subsun, omitting the mass range of ``normal'' SNII
with $M > 50M$\subsun.  The very recent calculations of
\cite{tominaga07} for nucleosynthesis in Pop~III SNII explosions
end at $50M$\subsun.

The peculiarities in  abundance ratios found
in HE0132$-$2429 are reminiscent of those found in the most Fe-poor star known,
HE1327$-$2326 \citep{frebel05}, which also has N highly enhanced 
(with C enhanced as well,
but not by as much) and high Sr relative to Fe.  The Sc~II lines
are too weak to be detected in such an extreme star even if the same
anomaly were present for this element as well.

HE1356$-$0622 is a modest outlier in [Na/Fe] and
in [Si/Fe], being high in both cases by perhaps $\sim$0.5 dex.
However, [Na/Fe] shows a definite range among EMP giants
\citep{cayrel04}, which \cite{spite06}  subsequently explained
as mixing of material processed through proton-burning
along the RGB, thus enhancing Na by $\sim$0.5~dex
\citep*[see also ][for a discussion
of non-LTE effects.]{andrievsky07}.  Since
HE1356$-$0622 is one of the coolest stars in our sample
of EMP giants, it presumably is among those with the highest
luminosity and thus has a high probability of being a mixed star.
The apparent anomaly in [Si/Fe] is due primarily to the 
extremely large deficiency of that ratio in HE1424$-$0241, which
couples with our adopted definition of ``anomalous''
via a median over our
small sample of C-normal EMP giants. 
After careful consideration, we find that HE1356$-$0622 is probably
a mixed EMP giant and has no  statistically significant
anomalies in its abundance ratios for the set of elements
we have detected.

The three more C-rich stars in the present sample of EMP giants
all obey patterns previously seen for such stars (see, e.g. Cohen \etal\ 2005
or Aoki \etal\ 2006).  Two of them show strong enhancements of the light elements
up to and including Al; one does not.

There are two stars in our sample of EMP giants with 
[N/C] $> 0$, one of which is very highly N-enhanced.
CS 22949-037 (a.k.a HE2323$-$0256), is a N-rich star, 
with [N/Fe] +2.16 dex, while [C/Fe] is is +0.97 dex.  HE0132$-$2429
is a milder case of a star with [N/C] $> 0$.   Several
similar stars are known, although EMP
C-rich stars with [N/C] $< 0$ are much more common than
EMP giants with  [N/C] $> 0$.  If we assume that the C-rich EMP giants 
are the result
of mass transfer across a binary system when the former primary
was an AGB star, then
predictions of nucleosynthesis 
in AGB stars (Lattanzio 1992, Herwig 2004, and references therein)
suggest that hot bottom burning in intermediate mass
AGB stars (3 to 6 $M$\subsun) leads to strong N-enhancements

\cite{johnson06} discuss the predicted frequency of N-enhanced stars
with [N/C] $> 0$ as a function of mass of the AGB star contributing.
They suggest that N-rich stars represent the contribution from
the upper mass limit of such stars near $\sim$6~M\subsun, and 
note that
their observed frequency appears to be
quite low compared to that expected for a normal mass distribution
of AGB stars.
They speculate that factors
which decrease the efficiency of mass transfer in binary systems
with large mass ratios may be responsible for the apparent lack of
N-enhanced stars.

To summarize the situation as we view it,
HE1424$-$0241 and HE0132$-$2429, both analyzed here, are the only
EMP giants known to us which show peculiar abundance ratios 
among the Fe-peak elements.  One EMP dwarf, HE2344$-$2800
with [Fe/H] $\sim -2.7$~dex, first
studied in the Keck Pilot Project, was a suspected outlier, with 
[Cr/Fe] $\sim 0.3$~dex higher than typical EMP dwarfs, a large
sample of which were studied in \cite{cohen04}.  Analysis of
a new HIRES spectrum of this EMP dwarf taken in Oct. 2004 confirms the
excess in [Cr/Fe],
and suggests an excess in [Mn/Fe] of $\sim 0.5$~dex as well.
Many EMP stars show
unexpectedly high CNO abundances, which are likely due to
intrinsic production followed
by mixing (for luminous giants only)
or pollution from a former AGB binary companion.
A smaller number of stars, including two of the three C-rich stars
studied here,
show large enhancements of the light elements
Na, Mg and Al as well.
All such $\alpha$-enhanced stars
with the exception of BS16964$-$002, very recently discovered by \cite{aoki07},
are C-rich; Aoki's new star is, however, highly O-rich.  
A few stars, such as CS22952$-$015 \citep{mcwilliam95a} and
CS22169$-$035 \citep{cayrel04} 
show small (at least compared to those of the present sample) deficiencies 
of the $\alpha$-elements, with normal C.

With better spectra and analyses, and the larger sample
of known EMP stars enabled in part by our searching 
for such in the Hamburg/ESO Survey,
we are now able to discern the impact on the chemical inventory of
a star from contributions
by individual SNII among extreme EMP stars.  At slightly higher Fe-metallicity,
we see abundance ratios which show slow trends as functions of [Fe/H] 
with low dispersions about the mean trends which presumably arise
from summing the ejecta of SNII over a stellar population with 
a normal (i.e. Salpeter or similar)  initial mass function.
These trends can often be reproduced in detail by theoretical models
of Galactic chemical evolution containing the most recent
nucleosynthetic yields such as those of
\cite{prantzos06} or \cite{matteucci07}.
It is now up to the theorists who model SNII explosions to 
try to develop a set of nucleosynthesis yields
which will lead to the variety of chemical inventories we have seen in the
EMP stellar population in the Milky Way, particularly among
the lowest [Fe/H] stars known in the Galaxy, and especially
for the very anomalous extreme EMP star HE1424$-$0241.

Turning to the elements beyond the Fe-peak,
another by now well established observational fact is the decoupling between the
production of the Fe-peak elements and the heavy neutron capture elements.
The ratios of [Sr/Fe] and [Ba/Fe] 
among the EMP stars analyzed by our 0Z project, including those discussed here,
show a very wide range among both the
C-normal and C-rich EMP stars (see Fig.~\ref{figure_sr_ba}). 
It is interesting to note that no star in the present
sample, neither C-normal nor C-rich, has [Ba/Fe] $> -0.2$~dex.  Among
more metal-rich C-rich stars, the fraction of stars with enhancements
of the $s$-process elements is large, exceeding 75\% \citep[see, e.g.][]{cohen06}.

\subsection{Implications for C/Fe ratio and Frequency of C-enhanced stars}

The fraction of C-enhanced stars among EMP stars  is a very
contentious issue.  Adopting the definition of
C-enhanced stars of \cite{beers05} as those with
[C/Fe] $> 1.0$~dex, recent values for 
this fraction from several independent survey
for stars with [Fe/H] $< -2.0$~dex
cover the range from  $>21~\pm$0.2\% \citep{lucatello06}
to 9${\pm}$2\% \citep{frebel07}, with  
our 0Z survey yielding a preliminary value of 14${\pm}$4\% \citep{cohen05}. 

The samples are in each case reasonably large, but there is a fairly 
large range in the deduced frequency of C-rich
EMP stars.  This suggests that differences in the analysis
are contributing to this problem.
It is not surprising in the context of the previous discussion
(see, for example, that of \S\ref{section_heres})
that the fraction of C-rich stars calculated from a set of different independent
analyses of large samples would result in different estimates from
different surveys.
Our [Fe/H] values are systematically
higher than those of \cite{cayrel04} by $\sim$0.3 dex, and the
differences in 
C and N abundances determined from molecular bands
between the two surveys  show a larger
dispersion than do the abundances based on atomic absorption lines.
Stars near the boundary
of the C-enhanced class 
could easily be shifted into or out of  the
C-rich class as a result of small systematic differences
between the various ongoing large projects.  This 
would affect such frequency calculations, whatever
the specific abundance characteristic of interest might be.

Since there are many stars near the boundary of the C-enhanced
class as defined above,
we suggest that a substantial part of the  variation in the deduced
fraction of C-enhanced EMP stars arises from such differences
in the details of the analysis.  It is thus extremely important
for people engaged in this type of work to publish the full
details of their analyses, as we did in \cite{cohen06},
and to analyze a few stars in common with
other major groups working in this area.  Among the details
that must be described, in addition to those discussed above, is the issue
of the way one handles the recent upheaval in the Solar CNO abundances
through the work of \cite{asplund04} and \cite{asplund05} with 3D models.

The modified definition of  the cutoff for enhanced [C/Fe] suggested by
\cite{aoki07} offers the advantage of a cleaner cut with fewer stars
near the boundary between C-normal and C-enhanced classes.  It is preferable
to the definition we are using, which is that of \cite{beers05}.
With the latter definition, two of the three stars are just at the
boundary, while with the newer definition, all three
of the C-rich stars discussed here would clearly be considered
C-rich.

\section{Summary}

We have presented detailed abundance analyses of five extremely
metal poor giants which are newly discovered
from our datamining of the Hamburg/ESO Survey. One of these
turned out to be a rediscovery
of a star found in the HK Survey with an unusually large error
in its published coordinates.
We include here a
new analysis based on better spectra of HE0132$-$2429, part of
the Keck Pilot Project (see Cohen et al 2002 and Carretta et al 2002).
We also analyze new high resolution and high signal-to-noise
ratio spectra of the only EMP giant found in the HERES project
\citep{barklem05} and of an EMP giant from the First Stars project
to use as a calibration object for comparison of the two projects,
for a total sample of 8 EMP giants.

The high quality of our HIRES spectra and our discovery of many more EMP
stars, including some close to $-4$~dex, makes it possible to search
for, to find, and to confirm outliers which have
anomalous abundance ratios [X/Fe] among the Fe-peak elements,
where such have not been detected previously.
The lowest metallicity star in our sample,  HE1424$-$0241,
with [Fe/H] $-3.95$ dex, has only upper limits for the C and N
abundances, based on our non-detection of the G band of CH and
of the 3360~\AA\ band of NH.  This star 
shows highly anomalous abundance
ratios, with extremely low [Si/Fe] ($<-1$~dex) and very
low [Ca/Fe] ($-0.6$ dex) and [Ti/Fe] ($-0.18$ dex), while
[Mg/Fe] is normal. Mg is produced in hydrostatic $\alpha$-burning,
while the other three elements are made in explosive $\alpha$-burning.
In essentially all other EMP
stars, these three abundance ratios are positive. These deficits 
in HE1424$-$0241 are accompanied
by strong excesses for the odd atomic number elements, so that
[Mn/Fe] and [Co/Fe] are significantly larger than is typical
of all other EMP giants.  We
speculate that the parcel of gas from which this star 
formed in the early Milky Way contained ejecta from only
a few SNII, and was deficient in ejecta from core collapse SN
whose progenitors had masses at the upper end of the relevant range.
The nucleosynthesis was such that the explosive $\alpha$-elements
(Si, Ca and Ti) were not produced by the SNII at typical rates, while the
hydrostatic $\alpha$-element Mg, formed during the course of
normal stellar evolution even in zero metallicity stars, was
produced at normal rates.

A second outlier, HE0132$-2429$, shows  enhanced
Sc relative to Fe, with [N/C] $>0$.
We suggest that this chemical inventory of this star had 
the opposite bias, namely a
larger contribution from SNII toward the upper end of
the progenitor mass range near 60$M$\subsun.   The remaining
three C-normal stars have abundance ratios typical
of slightly more metal rich EMP stars; they do not
show any detectable anomalies in their chemical inventory.

No other EMP giant known to us shows peculiar abundance ratios 
among the Fe-peak elements.  One EMP dwarf,
HE2344$-$2800, first
studied in the Keck Pilot Project, and now revisited  with a better HIRES spectrum,
appears to have [Cr/Fe] $\sim 0.3$~dex and [Mn/Fe] of $\sim 0.5$~dex
higher than typical EMP dwarfs, a large
sample of which were studied in \cite{cohen04}.  C and N
have only upper limits in this hot dwarf, where the molecular
bands are, even for normal [C/Fe] ratios, 
very weak and difficult to detect.  Many EMP stars show
unexpectedly high CNO abundances, which are likely due to
intrinsic production followed
by mixing (for luminous giants only)
or pollution from a former AGB binary companion. 
A smaller number of stars, including two of the three C-rich stars
studied here,
show large enhancements of the light elements
Na, Mg and Al as well.
All such $\alpha$-enhanced stars
with the exception of BS16964$-$002, very recently discovered by \cite{aoki07},
are C-rich, and this star is highly O-rich.

The behavior of the C-rich stars contains no surprises.  Two in our 
present sample
have large enhancements of the light elements through Al; the third
shows normal abundance ratios for Na, Mg and Al with respect to Fe.
Two of the sample stars have [N/C] $> 0$, which is not common
among C-rich EMP giants.  It
is generally believed that C-rich EMP stars are  the result
of mass transfer within a binary system when the former primary 
was an AGB star.  If this is true,
nucleosynthesis calculations by \cite{lattanzio92}, \cite{herwig04}
and references therein suggest that to produce such a large excess of N
the former primary must have a mass towards the upper
end of the range for AGB stars.

We present the first determination of [Cu/Fe] for EMP giants
below [Fe/H] $-3$~dex based on the rarely measured
UV resonance lines of Cu~I near 3250~\AA.
We find that the plateau level which was
suggested for [Cu/Fe] for dwarfs with [Fe/H] $<-2$~dex
by \cite{bihain04} continues to even lower metallicities.

The  heavy neutron capture elements are low in all eight EMP stars
in our sample, with
[Ba/Fe] $< -0.2$~dex.  This is rare among more Fe-rich C-stars,
which often have strong $s$-process enhancements. 
The ratio $\epsilon$(Ba)/ $\epsilon$(Sr)
varies by more than a factor of 100 among the stars studied here,
and suggests again that another nucleosynthesis mechanism
that preferentially produces the light neutron capture elements such as Sr is
required.

A careful comparison of the procedures and results 
of the detailed abundance analyses carried out by our 0Z project
with those of the First Stars project \citep{cayrel04} demonstrates
that there is
a systematic offset between the deduced [Fe/H] values; ours
being on average 0.3~dex higher. 
Thus the most metal poor star we found, HE1424$-$0241,
with [Fe/H] $-3.95$ dex based on our 0Z project analysis, would
translates roughly into $-4.2$ dex if it were analyzed by
the First Stars project of \cite{cayrel04}. 
For elements which display
measurable atomic absorption features, the differences 
between these two projects results in systematic
changes in abundance ratios with respect to Fe which are much smaller.  

Inter-comparison
between surveys of abundances  [X/Fe] derived from molecular bands,
typically used to determine CNO abundances, is much less
common and more difficult to carry out.  Systematic differences
between surveys for CNO abundances from
CH, NH, CN, CO  or OH bands may be common and may be larger than for elements
where atomic lines can be utilized.    Any such systematic
differences between surveys may
affect the deduced fraction of C-rich stars found in a survey and
may contribute to the wide range in published values
for this parameter.

\section{Appendix A: Comparison of Photometry \label{appendix_phot} }

We have observed $\sim$100 EMP candidates with ANDICAM in queue mode over the past
three years.  We calibrate to the Johnson-Kron-Cousins
photometric system using standard star fields
from \cite{landolt92}.  The observer at CTIO running the queue only carries
out our program if the night is believed to be photometric.  Data from
nights which at sunset were believed to be photometric, but subsequently
the observer changed his opinion of the sky conditions, were discarded.
The zero points of our photometric
calibration are based on two sets of images of standard
star fields per night in almost all cases, and never more than two.  Most
stars were observed only on one night; about 1/3 were observed 
on two nights, and a few on three nights.

We assess whether our assigned photometric errors for ANDICAM photometry
of the HES candidate EMP stars are valid by comparison of stars from
our sample which are included in
much larger, and hopefully better calibrated, photometric surveys.  
This is a key issue
since we use this photometry, together with $J, H, K_S$ from
2MASS, to determine stellar parameters.  

We compare
our values with those from the SDSS \citep{york00} DR4 release
\citep{sdss_dr4}, using the
transformations of \cite{sdss_trans} for Johnson $V$
and Kron-Cousins $I$.
There are 26 stars in common.  The mean difference in $V$ and in $I$ 
between that we measure using ANDICAM and that from the SDSS database
appropriately transformed is less than 0.01 mag.  The dispersion
about the mean is 0.06 mag for each of $V$ and $I$, 
somewhat larger than one might expect
for our nominal errors of $\pm 0.03$~mag for each combined with 
the nominal uncertainties of $\pm 0.02$~mag for the SDSS.  This suggests
that our assessment of the  uncertainty in our ANDICAM photometry may be
somewhat underestimated.

The SDSS is a very large area survey with a very extensive
photometric calibration effort, and our ANDICAM measurements agree
well in the mean with the SDSS values, suitably transformed.
This suggests that the photometric calibration of our ANDICAM data
has no systematic errors.

The recent large photometric survey of EMP candidate stars 
of \cite{beers07} includes 17 stars from our ANDICAM sample.
\cite{beers07} include data taken at many sites with many different
instruments on many different runs.
For these 17 stars, our $V$ is
fainter by 0.05~mag on average, with $\sigma$ of the differences
being 0.05~mag.  Our $V-I$ colors are identical to within $\pm0.01$ mag
on average with those of \cite{beers07} for the 12 stars with
such colors in the survey of \cite{beers07}, with $\sigma$ of the difference
being 0.03~mag. The (identical) systematic difference in $V$ and in 
$I$ may arise from calibration difficulties across a survey with
such a wide variety of data sources.

HE1424$-$0241, the most metal-poor star discussed here,
has $V$(ANDICAM,SDSS,Beers) = 15.47, 15.36, 15.32~mag respectively,
and is among the three stars with the largest deviation in $V$
for both of these comparisons.  Its $I$ mag is the same
(differing by only 0.015 mag) for
the SDSS and for the ANDICAM data, while the \cite{beers07} photometry
for this star has $V-I$ identical to the ANDICAM result, but $V$
brighter by 0.15~mag.
If we assume the SDSS $V,I$ to be correct, then
we have underestimated \teff\ for this star by $\sim$130~K,
the nominal uncertainty we adopt for this key
stellar parameter is 100~K.

\section{Appendix B: Details of the Comparison with the First Stars Project
of \cite{cayrel04} \label{appendix_cayrel} }

In this appendix we provide additional details,
beyond the discussion given in \S\ref{section_comp}, of the comparison
of the analyses we have carried out for EMP giants with
those of the First Stars project as given in \cite{cayrel04}.

We selected a representative sample of stars
from \cite{cayrel04}
to cover the range of stellar parameters of interest here
([Fe/H] below $-3$~dex and \grav\ as expected for RGB stars).
Three values of \teff\ determined for each of these stars are 
shown in Table~\ref{table_teff_cayrel}.
The first is
the value we would derive using the codes and procedures of the 0Z project,
including reddening from the map of \cite{schlegel98}. 
The second is that we would derive 
if we used $E(B-V)$ from \cite{cayrel04} instead.  The third is that
adopted by the First Stars project, taken from
Table~4 of \cite{cayrel04}.  In all cases, the $V$~mag
given in Table~2  of \cite{cayrel04} was used.  \teff\ for the second and
third cases are given in the table as differences from that
of our 0Z project.

At first sight the differences between the \teff\ adopted by each project
for the representative sample of stars
given in the last column of Table~\ref{table_teff_cayrel} are
satisfactory considering the quoted
accuracies of \teff\ by us of $\pm100$~K and
by the First Stars project of $\pm$80~K.  
However, we stress that this is a systematic,
not a random, effect.  
Table~\ref{table_teff_cayrel} shows, by comparing the first and second
values of \teff\ given for each star, that  the
somewhat larger values of $E(B-V)$ that we adopt result in
our \teff\ being higher than that of the First Stars project 
by up to $\sim$100~K. 
This difference depends both on the reddening difference
for the particular star (i.e. where the star is in the sky)
and on \teff\ for the star, as hotter stars have smaller values
of $\Delta(V-K,V-J)/\Delta($\teff).

In addition, for a specified set of stellar colors and a fixed choice
of reddening, comparing the  second and third choices for \teff,
we find that our \teff\ are higher by up to 100~K.  This must
arise from some systematic difference in the transformation within
the grid of colors (specifically $V-J$ and $V-K$) and the stellar parameters
for EMP giants.  It should be noted that the \teff\ given by
\cite{cayrel04} are based on the \cite{alonso99} relations while
ours are based on those of the MARCS and ATLAS model atmosphere grids.
This issue was discussed in \cite{cohen02}, where it was demonstrated
that the MARCS and ATLAS9 \teff\ color-relations are in very good
agreement, but that they disagree with those of \cite{alonso99}.
We also note that the \cite{alonso99} color relations
have not been calibrated adequately at the very low values of [Fe/H]
considered here; a theoretical calibration is easier to achieve at present
than an observational one for EMP stars.

The result of adding these two effects is that our \teff\ would be
from 30 to 165~K hotter than those adopted by \cite{cayrel04} for the
same input set of observed colors, with the hotter stars in the sample
(\teff $\sim~5200$~K) having larger differences than the
cooler ones; the coolest stars in our sample have \teff$~\sim~4900$~K.
This of course will result in a systematic offset such that the
[Fe/H] value deduced for a star with a particular set of \eqw\ will
be higher 
as determined by the 0Z project than as determined by the First Stars
project \citep{cayrel04}.   This is accompanied by
a smaller effect
on the deduced abundance ratios, as we will see below.


Since \cite{cayrel04} use ionization equilibrium
to determine \grav, given a \teff, while we use evolutionary
tracks, there is some scatter of the difference of \grav\
between the two projects for a  given star.  In most cases the
difference is small; the largest difference among the stars
compared in Table~\ref{table_teff_cayrel} 
is 0.4 dex for the same input parameters.

We next isolate the differences introduced
into the abundance ratios by the different abundance analysis codes
and model atmosphere grids
between our 0Z project and the First Stars Project.  Recall
that we use model atmospheres from ATLAS \citep{kurucz93} 
while the First Stars project uses OSMARCS models
\citep{gustaffson02}.  

We first note that changing to the latest and currently
preferred Kurucz models which have
a somewhat different treatment of convection with no overshooting
described in \cite{castelli}
and an improved opacity distribution function
(which are labeled as ``ODFNEW'' models) leads to
very small changes (not exceeding 0.01 dex) in the derived
log[{$\epsilon$}(X)] for all species considered here.  The 
predicted model colors
$V-J$ and $V-K$ differ
by less than 0.02 mag in this range of [Fe/H] and
\teff, introducing a change 
of less than 50~K in deduced \teff.
Since these
stars are so metal-poor, the details of the treatment of the
line opacity used in computing the model atmosphere do not matter.

For this test, we assume, for a given star, the  same 
stellar parameters as \cite{cayrel04} derived.  We then carry out an abundance
analysis using our codes and our model atmospheres appropriate
to those input stellar parameters, and compare the resulting
[Fe/H] values and abundance ratios [X/Fe]. 
First we compare the deduced values of log[$\epsilon$(X)] for three
stars.  The first case we examined
is the bright EMP giant CD $-38$ 245.  We use the \eqw\
and $gf$ values from \cite{cayrel04}; we do not have any HIRES
spectra of this star. For the other two stars, BS16467$-$062
and CS22949$-$037 (a.k.a. HE2323$-$0256) we have independent
HIRES spectra; we use our own measured \eqw\ and our own adopted
atomic parameters.

The comparison for the three stars is shown in Fig.~\ref{figure_cayrel_diff}.
This figure shows our absolute abundances
being systematically about 0.15~dex larger for all species.
This difference is similar to that
found by \cite{luck06} (see their Fig.~1), who have carried out a similar
test for stellar parameters appropriate to Cepheid variables.
This systematic offset  would be largely eliminated
if we were to use a stellar atmosphere which was $\sim$90~K lower than
the \teff\ adopted by the First Stars project.
The agreement for the C and N abundances, with differences as large
as 0.4 dex, is not as good.  This is not surprising, as we have shown
in \S\ref{section_comp}
that the $gf$ values for atomic absorptoin lines
adopted by both projects are in very good
agreement.
The  C and N abundances, on the other hand, are derived from molecular bands and
many more parameters enter, none of which have been compared between
the two projects.

The offset from equality shown in Fig.~\ref{figure_cayrel_diff} does
not affect the deduced abundance ratios [X/Fe] derived
from our work and the First Stars project as the offset is approximately
constant for all species considered.
However the
First Stars project adopts  log$\epsilon$(Fe) = 7.50~dex
for the Sun, while we use a value 0.05 dex smaller.   Since, as we have
shown earlier, the $gf$ value scales are identical,
we will therefore see
a systematic difference of 0.05~dex in all values of [X/Fe] and in
[Fe/H], with our values being higher (i.e. more metal-rich).  This is
in addition to the offset of 0.15~dex due to differences in the
model atmosphere grids and analysis codes.

For CD$-$38~245 
both analyses give very good ionization
equilibrium, so it matters little whether one uses
Fe~I or Fe~II as the reference.  Ideally, since log$\epsilon($FeII)
is less sensitive to \teff, it would be better to use that
in calculating [Fe/H] and [X/Fe], but
there are few Fe~II lines detected in these EMP stars, and many
of those detected are very weak with somewhat unreliable \eqw.
The differences in [X/Fe] for the 12 species in common (ignoring
the reference species, Fe~I) range from $-0.07$ to +0.05 dex,
with a mean of $-0.02$ dex and $\sigma$ of 0.04 dex, which we consider
very good agreement.

The difference in the deduced abundance ratios between our 0Z project
and the First Stars project for HE2323$-$0256, ignoring
N (deduced from the NH band at 3360~\AA\  in both cases),
ranges from $-0.18$ to +0.11 dex for the 11 species in common,
with a mean of $-0.07$ dex and $\sigma$ 0.10 dex. 
(That for [N/Fe] is 0.38 dex, with our value being lower.)
Part of this may arise in the problem with the \eqw\ for this
star; incorrect values were published by \cite{cayrel04} for
at least 7\% of the  lines in common with our 0Z values
given here in Table~\ref{table_eqw} (M.~Spite, 
private communication, July 2007).
The ionization equilibrium of Fe is good in both analyses.

BS~16467--062, the third case we checked, gives similar results.
Here, with the stellar parameters set by
the First Stars project, the difference in log$\epsilon$(Fe~I)
is 0.16~dex, with our value again being higher.
The ionization equilibrium in our solution for this
set of stellar parameters is 
log($\epsilon$)(Fe~I) -- log($\epsilon$)(Fe~II) +0.04 dex;
for the First Stars project derived +0.12~dex.

But the true comparison is what happens when we use the stellar
parameters  
derived with our own codes and procedures and the reddening
values we adopt (i.e. our systematically higher \teff\ as compared to
those of the First Star project). 
For BS~16467--062,
if we compare log$\epsilon$(Fe~I) as published by \cite{cayrel04}
versus that given in Table~\ref{table_abunda} and \ref{table_abundb}, 
the difference in [Fe/H]
0.30 dex, with our value being higher.  The higher \teff\ we adopt
(164~K higher than that of the First Stars project)
based on our higher reddening and on our \teff\ scale is somewhat
compensated by the difference in adopted \grav. 
The difference in adopted \teff\ for CS22949-037 (a.k.a. HE2323$-$0256)
between the two projects
is small only because our $V$~mag from ANDICAM photometry
is 0.05~mag fainter than that adopted by \cite{cayrel04}.  We thus
find a difference of +0.18~dex in the final deduced [Fe/H], our value
being higher.  This, given the essentially identical \teff, reflects
just the difference in analysis details discussed above.

In summary, it appears that the different
analysis codes, and stellar atmospheres grids adopted lead to [Fe/H] values
from our 0Z project being systematically 0.15$\pm0.03$~dex higher than
those of the First Stars project.  
The difference becomes somewhat larger,
$\sim$0.25 dex,  when the hotter
stellar parameters determined
from the independent codes, procedures, and adopted reddening map
of our 0Z project are used instead of those adopted by
\cite{cayrel04}.  There is an additional
contribution of 0.05 dex to the difference in derived
[Fe/H] from the two projects which arises
from the difference in the adopted Solar Fe abundance.
Overall, the [Fe/H] value derived by the First Stars project
as described in \cite{cayrel04} will be systematically $\sim$0.3~dex lower
than that for the same star, the same set of \eqw,
and the same set of observed stellar
photometry as analyzed by the 0Z project.

\acknowledgements

We are grateful to the many people  
who have worked to make the Keck Telescope and HIRES  
a reality and to operate and maintain the Keck Observatory. 
The authors wish to extend special thanks to those of Hawaiian ancestry
on whose sacred mountain we are privileged to be guests. 
Without their generous hospitality, none of the observations presented
herein would have been possible.
Some of the data presented herein were obtained at the Palomar
Observatory.
This publication makes use of data from the Two Micron All-Sky Survey,
which is a joint project of the University of Massachusetts and the 
Infrared Processing and Analysis Center, funded by the 
National Aeronautics and Space Administration and the
National Science Foundation.
J.G.C. is grateful to NSF grant AST-0507219  for partial support.
      N.C. is a Research Fellow of the Royal Swedish Academy of
      Sciences supported by a grant from the Knut and Alice
      Wallenberg Foundation. He also acknowledges financial
      support from Deutsche Forschungsgemeinschaft through grants
      Ch~214/3 and Re~353/44.  We thank W.~Huang for help accessing
      the SDSS database.

{}

\clearpage

\begin{deluxetable}{lclr rrrr}
\tabletypesize{\footnotesize}
\tablenum{1}
\tablewidth{0pt}
\tablecaption{New EMP Stars With $T_{eff} < 6000$~K From the 0Z Project
\label{table_sample}}
\tablehead{
\colhead{ID} & \colhead{Coords.} & 
\colhead{V\tablenotemark{a}} &  \colhead{I\tablenotemark{a}} &
\colhead{E(B$-$V)\tablenotemark{b}} & \colhead{\teff} &
\colhead{\grav} & \colhead{$v_t$}  \\
\colhead{} & \colhead{(J2000)} &
\colhead{(mag)} &   \colhead{(mag)} &
\colhead{(mag)} & \colhead{(K)} & \colhead{(dex)} &
\colhead{(\kms)} }
\startdata 
HE0132$-$2429\tablenotemark{f} & 01 34 58.8  $-24$ 24 18 & 14.82 & \nodata &
    0.012 & 5294 & 2.75 & 1.8 \\
HE1012$-$1540 & 10 14 53.5  $-$15 55 54 & 14.04 & 13.21 & 0.070 &
    5620 & 3.40  & 1.6   \\
HE1347$-$1025 &  13 50 22.4 $-$10 40 19 & 15.06 & 14.16 & 0.058 &
    5195 & 2.50 &  1.8 \\
HE1356$-$0622 &  13 59 30.3  $-$06 36 35 &  14.31& 13 36  &0.030 &
   4947 & 1.85 & 2.2 \\
HE1424$-$0241 &  14 26 40.3  $-$02 54 28 &  15.47 & 14.54 &  0.064 &
 5193 & 2.50 & 1.8 \\
HE2323$-$0256\tablenotemark{c} &  23 26 29.8  $-$02 39 58 &  14.41 & 13.40 & 0.051 &
 4915 & 1.70 &  2.0 \\  
From Other Surveys: \\
HE1300+0157\tablenotemark{d}  & 13 02 56.3  +01 41 51 &  14.11 & 13.39 &  0.022 & 
  5632  & 3.37 &  1.3 \\  
BS16467$-$062\tablenotemark{e}  &  13 42 00.6  +17 48 48 & 
  14.09\tablenotemark{e} &   \nodata &  0.018 &
  5364 & 2.95 & 1.6 \\
\enddata
\tablenotetext{a}{Our photometry from ANDICAM images.}
\tablenotetext{b}{Based on the reddening map of \cite{schlegel98}. }
\tablenotetext{c}{Rediscovery of CS22949$-$037 from the HK Survey.}
\tablenotetext{d}{From HERES \citep{christlieb04b,barklem05}.}
\tablenotetext{e}{Star from the HK Survey included in the First Stars \citep{cayrel04}
sample, $V$ mag from this source.}
\tablenotetext{f}{Star from the Keck Pilot Project, $V$ from
\cite{cohen02}.}
\end{deluxetable}

\begin{deluxetable}{lllrr r}
\tabletypesize{\footnotesize}
\tablenum{2}
\tablewidth{0pt}
\tablecaption{Details of the HIRES Observations
\label{table_spectra}}
\tablehead{
\colhead{ID} & \colhead{Exp.Time} & \colhead{Julian Date} &
\colhead{SNR\tablenotemark{a}} & \colhead{Slit Width} &
\colhead{$v_r$\tablenotemark{b}} \\
\colhead{} & \colhead{(sec)} &
\colhead{($-$2453000.00)} &   \colhead{} & \colhead{('')} &
\colhead{(\kms)} }
\startdata 
HE0132$-$2429 & 7200 & 289.89 & 95 &  0.86 & +289.2 \\
HE1012$-$1540 &  3600 & 489.77 &    109 &  0.86 &  +226.2 \\
HE1300+0157  & 3600 & 843.87 &  $>$110  & 1.1 &  +73.4  \\
HE1347$-$1025  &  3600 & 149.80 & 80  & 1.1 & +48.6 \\
HE1356$-$0622 &  3600 & 149.89 & 105 &   1.1 & +93.5 \\
HE1424$-$0241  &  6000 & 844.96 & 90  & 1.1 &    +60.4  \\
HE2323$-$0256  &   7200 & 312.76 &  100  &  0.86 &   $-125.9$ \\ 
BS16467$-$062 &   3600  & 490.02 &  100  & 0.86 &   $-91.7$ \\  
\enddata
\tablenotetext{a}{SNR per spectral resolution element in the continuum at
4500\,{\AA}.}
\tablenotetext{b}{Heliocentric $v_r$.}
\end{deluxetable}

\clearpage

\begin{deluxetable}{lllll}
\tablenum{3}
\tablewidth{0pt}
\tablecaption{Radial Velocities for Stars with Multiple
High Resolution Observations
\label{table_vr}}
\tablehead{
\colhead{ID} & \colhead{Julian Date\tablenotemark{a}} & 
   \colhead{Julian Date\tablenotemark{a}} &
   \colhead{Julian Date\tablenotemark{a}} &
   \colhead{Julian Date\tablenotemark{a}} \\
\colhead{} & \colhead{$v_r$(\kms)\tablenotemark{b}} 
    & \colhead{$v_r$(\kms)\tablenotemark{b}}
    & \colhead{$v_r$(\kms)\tablenotemark{b}}
    & \colhead{$v_r$(\kms)\tablenotemark{b}} } 
\startdata 
HE0132$-$2429 & 1811.96 &  3243.00 & 3289.89 \\
~ & +296.0 (0.1) &  +289.5 (0.2) & +289.2 (0.1) \\
HE1012$-$1540 &  2396.82 &  3489.77  \\
~  &   +226.3 (0.4) &  +224.8 (0.1) \\
HE1300+0157 & 2830.50\tablenotemark{c} & 3157.74\tablenotemark{d} & 
   3432.01\tablenotemark{d} &   3843.87   \\
~ & +73.6 (2.0) & 74.6 (0.6) & 74.6 (0.6) & +73.4 (0.1) \\
HE1346$-$1025 & 3149.80 & 4198.92 \\
~  & +48.6 (0.5) & +49.4 (0.1) \\
HE1356$-$0622 & 3149.89 & 3989.72 \\
~  & +93.5 (0.4) & +94.1 (0.2) \\
HE1424$-$0241  &  3152.90 & 3844.96 \\
~   & +58.8 (0.6) &     +60.4 (0.1)  \\
HE2323$-$0256  & 1764.73\tablenotemark{e} & 2158.73\tablenotemark{e} & 
    2544.90  & 3312.76  \\
~ & $-$125.6 (0.1)\tablenotemark{e} & $-$125.6 (0.1)\tablenotemark{e} &
     $-125.9$ (0.1) &  $-125.9$ (0.1) \\ 
BS16467$-$062 & 2064.55\tablenotemark{e} & 2095.45\tablenotemark{e} & 
     3490.02 \\
~  & $-$90.6 (0.1) &  $-$90.5 (0.1) &   $-91.7$ (0.1) \\  
\enddata
\tablenotetext{a}{Julian date  $-$ 2450000.00.}
\tablenotetext{b}{Heliocentric $v_r$ and its 1$\sigma$ uncertainty.}
\tablenotetext{c}{$v_r$ from HERES/UVES \citep{barklem05}.}
\tablenotetext{d}{$v_r$ from Subaru/HDS spectra of \cite{frebel07}.}
\tablenotetext{e}{$v_r$ provided by M.Spite \& R.Cayrel (from UVES spectra).}
\end{deluxetable}


%

\clearpage

%
%
%
\begin{deluxetable}{l  crrr  rrrr  rrrr }
\tablenum{4}
\tabletypesize{\tiny}
\rotate
\tablewidth{0pt}
\tablecaption{$W_{\lambda}$ for the Sample EMP Stars From the HES \label{table_eqw}}
\tablehead{
\colhead{$\lambda$} & \colhead{Species} & \colhead{EP} &
\colhead{log($gf$)} &
\colhead{HE0132$-$2429} & \colhead{HE1012$-$1540} &
\colhead{HE1300+0157} & \colhead{HE1356$-$0622} &
\colhead{HE1347$-$1025} & \colhead{HE1424$-$0241} & 
\colhead{BS16467$-$062}  & \colhead{HE2323$-$0256}  \\
\colhead{($\AA$)} & \colhead{} &
\colhead{(eV)}    &   \colhead{}  & \colhead{(m$\AA$)} 
}
\startdata 
 3189.30 & OH &   1.03 & $-$1.990 &    $\leq$21.2 & \nodata & \nodata & \nodata & \nodata & \nodata & \nodata & \nodata  \\
 3255.50 & OH &   1.30 & $-$1.940 &     $\leq$14.3 & \nodata & \nodata & \nodata & \nodata & \nodata & \nodata & 40.0  \\
 5889.95 & Na~I  &   0.00 &   0.110 &    62.8 &   168.8 &    48.2 &   132.8 &  64.9 &    65.7 &    50.1 &   192.7  \\
 5895.92 & Na~I  &   0.00 & $-$0.190 &    50.4 &   146.7 &    26.8 &   117.3 & 52.2 &    39.0 &    27.8 &   159.2  \\
 3829.36 & Mg~I  &   2.71 & $-$0.210 &    93.3 &   259.0 &    84.2 & \nodata & \nodata &    79.0 &    81.8 &   139.7  \\
 4057.52 & Mg~I  &   4.34 & $-$1.200 & \nodata &    22.0 & \nodata &     9.0 & \nodata & \nodata & \nodata &    22.6  \\
 4167.28 & Mg~I  &   4.34 & $-$1.000 & \nodata & \nodata & \nodata &    10.5 & \nodata & \nodata &     6.8 &    28.7  \\
 4703.00 & Mg~I  &   4.34 & $-$0.670 &     9.0 &    72.2 & \nodata &    24.4 &    16.0 & \nodata & \nodata &    47.5  \\
 5172.70 & Mg~I  &   2.71 & $-$0.380 &   101.3 &   301.2 &    92.6 &   128.2 &   110.5 &    89.7 &    92.7 &   160.6  \\
 5183.62 & Mg~I  &   2.72 & $-$0.160 &   114.3 &   391.5 &   103.7 &   142.2 &   120.8 &   104.7 &   105.2 &   179.9  \\
 5528.40 & Mg~I  &   4.34 & $-$0.480 &    12.1 &    59.3 & \nodata & \nodata & \nodata & \nodata & \nodata &    48.3  \\
 3944.01 & Al~I  &   0.00 & $-$0.640 &    39.1 & \nodata & \nodata &    61.5 &    58.5 &    23.0 &    34.2 &   123.8  \\
 3961.52 & Al~I  &   0.00 & $-$0.340 &    42.7 &   103.0 &    36.4 &    76.8 &    62.5 &    27.1 &    37.1 &    89.7  \\
 3905.53 & Si~I  &   1.91 & $-$1.040 &   110.5 &   133.7 &    93.5 &   141.7 &   116.5 &    16.0 &    93.4 &   150.7  \\
 4102.94 & Si~I  &   1.91 & $-$3.140 & \nodata & \nodata & \nodata &    41.3 &     8.0 & \nodata & \nodata &    13.7  \\
 4226.74 & Ca~I  &   0.00 &   0.240 &   105.7 & \nodata &    94.1 &   134.0 &   111.1 &    58.7 &    98.8 &   116.6  \\
 4289.37 & Ca~I  &   1.88 & $-$0.300 & \nodata & \nodata & \nodata &    16.9 &     9.8 & \nodata & \nodata & \nodata  \\
 4302.54 & Ca~I  &   1.90 &   0.280 &    22.6 & \nodata & \nodata &    39.5 &    32.4 & \nodata &    12.0 & \nodata  \\
 4318.66 & Ca~I  &   1.90 & $-$0.210 & \nodata & \nodata & \nodata &    16.3 & \nodata & \nodata & \nodata & \nodata  \\
 4425.44 & Ca~I  &   1.88 & $-$0.360 & \nodata & \nodata & \nodata &    14.0 &     9.0 & \nodata & \nodata & \nodata  \\
 4435.69 & Ca~I  &   1.89 & $-$0.520 & \nodata &     7.3 & \nodata &    13.0 & \nodata & \nodata & \nodata & \nodata  \\
 4454.79 & Ca~I  &   1.90 &   0.260 &    17.4 &    21.3 &    15.9 &    36.2 &    26.7 & \nodata &    17.2 &    82.0  \\
 3736.90 & Ca~II &   3.15 & $-$0.148 &    75.0 &    50.9 &    69.7 & \nodata & \nodata &    44.0 &    70.8 & \nodata  \\
 4246.82 & Sc~II &   0.32 &   0.242 &    79.3 & \nodata &    30.9 &    87.6 &    43.3 &    20.5 &    43.3 &    63.8  \\
 4314.08 & Sc~II &   0.62 & $-$0.100 &    43.1 & \nodata & \nodata &    49.5 &    20.0 & \nodata &    13.4 & \nodata  \\
 4320.73 & Sc~II &   0.60 & $-$0.260 &    33.6 & \nodata & \nodata &    36.8 & \nodata & \nodata &    11.1 & \nodata  \\
 4670.41 & Sc~II &   1.36 & $-$0.580 & \nodata & \nodata & \nodata &     7.5 & \nodata & \nodata & \nodata & \nodata  \\
 3958.22 & Ti~I  &   0.05 & $-$0.160 &    15.3 & \nodata & \nodata &    18.9 & \nodata & \nodata & \nodata &  $<$13.5  \\
 3998.64 & Ti~I  &   0.05 & $-$0.050 &    13.5 & \nodata & \nodata &    24.7 &  $<$25.0 & \nodata &     9.6 & \nodata  \\
 4533.25 & Ti~I  &   0.85 &   0.480 & \nodata & \nodata & \nodata &    14.0 &  $<$16.0 & \nodata & \nodata & \nodata  \\
 4534.78 & Ti~I  &   0.84 &   0.280 & \nodata & \nodata & \nodata &    11.0 & \nodata & \nodata & \nodata & \nodata  \\
 4981.74 & Ti~I  &   0.85 &   0.500 &    12.0 & \nodata & \nodata &    18.0 &  $<$12.0 & \nodata &     4.9 & \nodata  \\
 4999.51 & Ti~I  &   0.83 &   0.250 &  \nodata &    $<$6.0 &  \nodata &  \nodata &   $<$10.0 &  \nodata &  \nodata &  \nodata  \\
 3900.54 & Ti~II &   1.13 & $-$0.450 &    52.2 &    13.6 &    27.9 &    76.5 &    54.1 &    14.0 &    41.6 &    61.1  \\
 3987.61 & Ti~II &   0.61 & $-$2.730 & \nodata & \nodata & \nodata &     5.0 & \nodata & \nodata & \nodata & \nodata  \\
 4012.39 & Ti~II &   0.57 & $-$1.610 &    24.4 & \nodata & \nodata &    46.8 &    23.7 &     8.5 &    20.3 &    31.4  \\
 4028.35 & Ti~II &   1.89 & $-$0.870 &     5.3 & \nodata & \nodata &    14.0 & \nodata & \nodata & \nodata & \nodata  \\
 4300.05 & Ti~II &   1.18 & $-$0.490 &    42.6 & \nodata & \nodata &    67.2 &    46.6 &     7.5 &    30.9 & \nodata  \\
 4301.93 & Ti~II &   1.16 & $-$1.200 &    17.9 & \nodata & \nodata &    37.6 & \nodata & \nodata &    10.3 & \nodata  \\
 4312.86 & Ti~II &   1.18 & $-$1.160 &    23.4 & \nodata & \nodata &    41.0 &    23.7 & \nodata &     8.5 & \nodata  \\
 4395.03 & Ti~II &   1.08 & $-$0.510 &    49.5 & \nodata &    24.4 &    76.4 &    56.7 &    12.0 &    39.5 &    60.6  \\
 4399.77 & Ti~II &   1.24 & $-$1.290 &    19.9 & \nodata & \nodata &    31.6 &    17.5 & \nodata &    12.1 &    20.2  \\
 4417.72 & Ti~II &   1.16 & $-$1.160 &    23.0 & \nodata & \nodata &    38.6 &    31.3 & \nodata &    12.1 &    25.4  \\
 4443.81 & Ti~II &   1.08 & $-$0.700 &    44.0 &    11.8 &    20.3 &    69.8 &    51.1 & \nodata &    39.9 &    58.2  \\
 4468.51 & Ti~II &   1.13 & $-$0.600 &    46.8 &    19.9 &    22.1 &    74.3 &    50.3 &    10.7 &    31.2 &    53.8  \\
 4501.28 & Ti~II &   1.12 & $-$0.760 &    41.7 &    11.2 &    20.7 &    66.8 &    49.9 &    12.2 &    31.9 &    51.4  \\
 4533.97 & Ti~II &   1.24 & $-$0.640 &    37.4 &    14.0 &    15.5 &    63.5 &    42.2 &     8.3 &    24.8 &    49.3  \\
 4563.77 & Ti~II &   1.22 & $-$0.820 &    30.1 &    10.6 &    13.4 &    56.1 &    32.3 &     5.1 &    24.7 &    39.5  \\
 4571.98 & Ti~II &   1.57 & $-$0.340 &    33.4 &    13.7 &    17.3 &    58.5 &    28.9 & \nodata &    19.7 &    42.8  \\
 4589.95 & Ti~II &   1.24 & $-$1.650 & \nodata & \nodata & \nodata &    15.5 & \nodata & \nodata & \nodata & \nodata  \\
 4111.77 & V~I   &   0.30 &   0.408 &  \nodata &  \nodata &  \nodata &  \nodata & 
      $<$5.6 &  \nodata &  \nodata &  \nodata  \\
 3545.19 & V~II  &   1.10 & $-$0.390 & \nodata & \nodata & \nodata & \nodata & \nodata &  $<$10.9 &    11.6 &    11.8  \\
 3592.03 & V~II  &   1.10 & $-$0.370 &    12.4 & \nodata & \nodata & \nodata & \nodata & \nodata &     8.2 &    11.5  \\
 4254.33 & Cr~I  &   0.00 & $-$0.110 &    31.5 & \nodata &    21.5 &    58.9 &    40.6 &    23.6 &    30.6 &    38.4  \\
 4274.79 & Cr~I  &   0.00 & $-$0.230 &    29.2 & \nodata &    16.8 &    47.3 &    32.5 &    19.7 &    22.5 &    29.3  \\
 4289.72 & Cr~I  &   0.00 & $-$0.361 &    22.8 & \nodata & \nodata &    42.3 &    26.5 &    13.4 &    20.0 &    31.8  \\
 5206.04 & Cr~I  &   0.94 &   0.030 &    11.9 &     8.4 &     4.7 &    23.2 &    14.0 &     5.8 &     6.2 &     7.4  \\
 5208.43 & Cr~I  &   0.94 &   0.158 &    14.3 &     8.5 &     7.6 & \nodata & \nodata &     7.7 &    12.8 &   18.2  \\
 4030.75 & Mn~I  &   0.00 & $-$0.470 &    17.3 &    20.0 &    10.8 &    36.6 &    26.0 &    34.7 &    17.4 &    22.1  \\
 4033.06 & Mn~I  &   0.00 & $-$0.620 &    12.3 &    15.0 &     8.5 &    30.0 &    18.0 &    26.3 &    14.4 &    17.9  \\
 3441.99 & Mn~II &   1.78 & $-$0.273 &    28.0 &    10.0 &    16.2 & \nodata & \nodata &    48.4 &    32.0 & \nodata  \\
 3460.32 & Mn~II &   1.81 & $-$0.540 &    23.5 & \nodata &    14.3 & \nodata & \nodata &    37.9 &    20.4 &    32.3  \\
 3488.68 & Mn~II &   1.85 & $-$0.860 &    15.9 & \nodata & \nodata & \nodata & \nodata &    29.8 &    13.0 & \nodata  \\
 3865.52 & Fe~I  &   1.01 & $-$0.980 &    63.8 &    46.5 &    48.8 &    88.1 &    74.2 &    50.2 &    66.2 &    72.8  \\
 3886.29 & Fe~I  &   0.05 & $-$1.080 & \nodata & \nodata & \nodata & \nodata & \nodata &    90.5 & \nodata &  \nodata  \\
 3887.06 & Fe~I  &   0.91 & $-$1.140 & \nodata & \nodata & \nodata & \nodata & \nodata &    40.5 & \nodata &  \nodata \\
 3895.67 & Fe~I  &   0.11 & $-$1.670 &    79.4 & \nodata &    69.2 &   101.9 &    83.4 &    69.4 &    78.6 &    96.4  \\
 3899.72 & Fe~I  &   0.09 & $-$1.530 &    83.7 &    70.2 &    72.9 &   115.0 &    92.1 &    73.5 &    84.6 &    95.8  \\
 3902.96 & Fe~I  &   1.56 & $-$0.470 &    55.9 &    48.0 &    45.9 &    80.6 &    70.8 &    41.4 &    56.1 &    59.7  \\
 3906.49 & Fe~I  &   0.11 & $-$2.240 &    60.7 &    36.0 &    41.8 &    92.4 &    73.9 &    47.5 &    60.7 &    68.3  \\
 3920.27 & Fe~I  &   0.12 & $-$1.750 &    83.5 &    59.9 &    66.4 &   106.4 &    82.9 &    70.2 & \nodata &    95.7  \\
 3922.92 & Fe~I  &   0.05 & $-$1.650 &    90.5 &    58.1 &    74.1 &   113.3 &    95.4 &    79.4 &    85.6 &    98.4  \\
 3930.31 & Fe~I  &   0.09 & $-$1.590 & \nodata &    75.0 & \nodata & \nodata & \nodata &    75.7 & \nodata &  \nodata  \\
 3949.96 & Fe~I  &   2.18 & $-$1.160 &     8.6 & \nodata & \nodata &    18.2 &     9.2 &     8.5 &     8.6 &    12.6  \\
 4005.24 & Fe~I  &   1.56 & $-$0.610 &    54.9 &    45.8 &    41.3 &    80.3 &    68.1 &    39.5 &    56.7 &    63.0  \\
 4045.81 & Fe~I  &   1.49 &   0.280 &    90.0 &    94.1 &    79.5 &   121.5 &   106.7 &    78.1 &    89.4 &    99.3  \\
 4063.59 & Fe~I  &   1.56 &   0.060 & \nodata &    69.1 & \nodata &   101.3 &    82.9 & \nodata & \nodata & \nodata  \\
 4071.74 & Fe~I  &   1.61 & $-$0.020 & \nodata &    68.3 & \nodata &    99.6 &    83.8 & \nodata & \nodata & \nodata  \\
 4118.55 & Fe~I  &   3.57 &   0.140 & \nodata & \nodata & \nodata & \nodata &    18.0 & \nodata & \nodata & \nodata  \\
 4132.06 & Fe~I  &   1.61 & $-$0.820 &    49.0 &    37.6 &    36.0 &    81.7 &    68.2 &    40.5 &    48.7 &    62.3  \\
 4143.87 & Fe~I  &   1.56 & $-$0.620 &    57.8 &    42.6 &    47.7 &    84.9 &    66.8 &    49.6 &    60.2 &    65.9  \\
 4147.67 & Fe~I  &   1.49 & $-$2.100 &    13.6 & \nodata & \nodata &    19.0 &     8.0 & \nodata &     6.4 &     8.7  \\
 4172.76 & Fe~I  &   0.96 & $-$3.070 & \nodata & \nodata & \nodata & \nodata &     7.1 & \nodata & \nodata &     6.0  \\
 4181.75 & Fe~I  &   2.83 & $-$0.370 &    12.2 & \nodata &     7.3 &    21.5 &    17.1 & \nodata &    14.4 &    13.3  \\
 4187.05 & Fe~I  &   2.45 & $-$0.550 & \nodata & \nodata & \nodata &    35.4 &    19.1 &    16.1 &    14.4 &    19.2  \\
 4187.81 & Fe~I  &   2.43 & $-$0.550 & \nodata & \nodata &    12.6 &    38.8 &    27.9 &    13.2 &    20.0 &    25.0  \\
 4198.33 & Fe~I  &   2.40 & $-$0.720 & \nodata & \nodata & \nodata &    33.4 &    20.0 & \nodata &    12.4 &    20.2  \\
 4199.10 & Fe~I  &   3.05 &   0.160 &    14.0 & \nodata &    14.5 &    31.8 &    22.3 & \nodata &    16.3 &    25.3  \\
 4202.04 & Fe~I  &   1.49 & $-$0.710 &    59.1 & \nodata &    45.9 &    80.6 &    65.0 &    47.2 &    59.5 &    67.3  \\
 4216.19 & Fe~I  &   0.00 & $-$3.360 &    18.2 & \nodata & \nodata &    41.2 &    19.9 &     8.2 &    20.1 &    23.5  \\
 4222.22 & Fe~I  &   2.45 & $-$0.970 & \nodata & \nodata & \nodata & \nodata & \nodata & \nodata &     7.8 & \nodata  \\
 4227.44 & Fe~I  &   3.33 &   0.270 &    15.1 & \nodata &    11.9 & \nodata & \nodata & \nodata &    11.2 &    15.3  \\
 4233.61 & Fe~I  &   2.48 & $-$0.600 &    17.7 & \nodata & \nodata &    32.4 &    22.0 & \nodata &    14.3 &    18.7  \\
 4235.95 & Fe~I  &   2.43 & $-$0.340 &    26.4 & \nodata & \nodata &    44.2 &    32.2 &    11.6 &    39.4 & \nodata  \\
 4250.13 & Fe~I  &   2.47 & $-$0.410 &    16.6 & \nodata &    15.9 &    44.6 &    25.2 & \nodata &    20.6 &    22.1  \\
 4250.80 & Fe~I  &   1.56 & $-$0.380 &    54.3 & \nodata &    39.5 &    76.0 &    53.3 &    39.4 &    49.4 &    64.0  \\
 4260.49 & Fe~I  &   2.40 &   0.140 &    41.0 & \nodata &    29.0 &    64.9 &    48.3 &    28.2 &    41.6 &    44.4  \\
 4271.16 & Fe~I  &   2.45 & $-$0.350 &    30.1 & \nodata &    15.0 &    41.5 &    29.3 &    11.8 &    18.1 &    26.5  \\
 4271.77 & Fe~I  &   1.49 & $-$0.160 &    74.9 & \nodata &    61.5 &    94.3 &    83.4 &    60.6 &    74.4 &    81.9  \\
 4282.41 & Fe~I  &   2.18 & $-$0.780 &    19.9 & \nodata & \nodata &    38.6 & \nodata & \nodata &    20.0 &    19.5  \\
 4294.14 & Fe~I  &   1.49 & $-$0.970 &    51.6 & \nodata & \nodata &    83.1 &    57.4 & \nodata &    46.3 &    66.1  \\
 4299.25 & Fe~I  &   2.43 & $-$0.350 &    25.8 & \nodata & \nodata &    47.8 &    47.3 &    15.0 &    20.4 & \nodata  \\
 4307.91 & Fe~I  &   1.56 & $-$0.070 &    85.0 & \nodata & \nodata & \nodata &    97.6 &    68.9 &    84.2 & \nodata  \\
 4325.77 & Fe~I  &   1.61 &   0.010 &    76.0 & \nodata & \nodata &   102.6 &    80.2 &    60.8 &    72.8 & \nodata  \\
 4337.05 & Fe~I  &   1.56 & $-$1.690 &    21.5 & \nodata & \nodata &    41.9 & \nodata & \nodata &    25.5 &    28.3  \\
 4375.94 & Fe~I  &   0.00 & $-3$.030 &    32.1 & \nodata &    17.4 &    67.7 &    34.6 &    19.7 &    30.8 &    36.9  \\
 4383.56 & Fe~I  &   1.49 &   0.200 &    96.7 & \nodata &    79.2 &   126.2 &    99.0 &    78.9 &    86.1 &    96.9  \\
 4404.76 & Fe~I  &   1.56 & $-$0.140 &    75.1 &    64.3 &    66.0 &   100.9 &    86.2 &    62.5 &    72.9 &    83.4  \\
 4415.13 & Fe~I  &   1.61 & $-$0.610 &    51.9 &    43.9 &    45.3 &    81.1 &    65.3 &    46.5 &    58.5 &    62.0  \\
 4427.32 & Fe~I  &   0.05 & $-$3.040 &    33.6 & \nodata &    15.9 &    59.6 &    39.7 & \nodata &    30.5 &    40.0  \\
 4442.35 & Fe~I  &   2.20 & $-$1.250 &     9.0 & \nodata &     7.2 &    24.0 & \nodata &     6.6 &     8.5 &    16.2  \\
 4447.73 & Fe~I  &   2.22 & $-$1.340 & \nodata & \nodata & \nodata &    13.7 & \nodata & \nodata & \nodata &     9.1  \\
 4459.14 & Fe~I  &   2.18 & $-$1.280 & \nodata & \nodata & \nodata & \nodata & \nodata & \nodata &     5.2 &     8.8  \\
 4461.66 & Fe~I  &   0.09 & $-$3.210 &    22.1 &    17.5 &    16.4 &    52.1 &    30.6 &    10.2 &    26.6 &    29.0  \\
 4489.75 & Fe~I  &   0.12 & $-$3.970 & \nodata & \nodata & \nodata &    12.6 & \nodata & \nodata & \nodata &     5.0  \\
 4494.57 & Fe~I  &   2.20 & $-$1.140 &    19.6 & \nodata &     9.7 &    23.4 &    16.8 &     8.9 &    10.9 &    17.8  \\
 4531.16 & Fe~I  &   1.49 & $-$2.150 &     6.2 & \nodata & \nodata &    17.5 & \nodata & \nodata & \nodata &     5.0  \\
 4592.66 & Fe~I  &   1.56 & $-$2.450 & \nodata & \nodata & \nodata &     9.1 & \nodata & \nodata & \nodata &     5.0  \\
 4602.95 & Fe~I  &   1.49 & $-$2.220 &     7.8 & \nodata & \nodata &    16.2 & \nodata & \nodata & \nodata &     5.0  \\
 4871.33 & Fe~I  &   2.86 & $-$0.360 &    11.9 &    10.6 & \nodata &    24.9 &     9.8 & \nodata &    18.7 &    18.0  \\
 4872.14 & Fe~I  &   2.88 & $-$0.570 &     6.7 &     8.0 & \nodata &    18.0 & \nodata & \nodata &     4.6 &    11.2  \\
 4891.50 & Fe~I  &   2.85 & $-$0.110 &    19.0 &    21.5 &    17.5 &    35.3 & \nodata & \nodata &    23.4 &    23.9  \\
 4919.00 & Fe~I  &   2.86 & $-$0.340 & \nodata &    10.0 & \nodata &    22.5 &    14.0 & \nodata &     8.5 &    10.1  \\
 4920.51 & Fe~I  &   2.83 &   0.150 &    17.4 &    16.5 &    13.1 &    36.2 &    26.0 &     9.9 &    19.9 &    22.3  \\
 4957.61 & Fe~I  &   2.81 &   0.230 &    34.3 &    31.7 &    26.0 &    56.1 &    37.5 &    22.0 & \nodata &    37.7  \\
 5083.34 & Fe~I  &   0.96 & $-$2.960 & \nodata & \nodata & \nodata &    17.6 & \nodata & \nodata & \nodata & \nodata  \\
 5166.28 & Fe~I  &   0.00 & $-$4.200 & \nodata & \nodata & \nodata &    11.5 & \nodata & \nodata & \nodata & \nodata  \\
 5171.61 & Fe~I  &   1.48 & $-$1.790 &    19.2 &    17.9 &    13.2 &    38.2 &    27.4 & \nodata &    13.0 & \nodata  \\
 5192.35 & Fe~I  &   3.00 & $-$0.420 &     6.1 & \nodata & \nodata &    14.4 & \nodata & \nodata & \nodata & \nodata  \\
 5194.95 & Fe~I  &   1.56 & $-$2.090 & \nodata & \nodata & \nodata &    18.0 & \nodata & \nodata &     7.8 & \nodata  \\
 5216.28 & Fe~I  &   1.61 & $-$2.150 & \nodata & \nodata & \nodata &    12.8 & \nodata & \nodata & \nodata &     6.5  \\
 5227.19 & Fe~I  &   1.56 & $-$1.350 &    36.4 &    21.8 &    24.8 &    62.8 &    45.9 & \nodata &    34.6 &    46.6  \\
 5232.95 & Fe~I  &   2.94 & $-$0.100 &    15.5 &    15.0 & \nodata &    30.4 & \nodata &     5.6 &    16.4 &    17.9  \\
 5269.55 & Fe~I  &   0.86 & $-$1.320 &    75.8 &    55.8 &    55.7 &   103.1 &    97.3 &    61.8 &    76.6 &    85.6  \\
 5324.19 & Fe~I  &   3.21 & $-$0.100 & \nodata & \nodata & \nodata &    14.6 & \nodata & \nodata &     6.8 & \nodata  \\
 5405.79 & Fe~I  &   0.99 & $-$1.840 &    39.3 &    25.5 &    25.7 & \nodata & \nodata &    27.5 &    38.8 &    46.3  \\
 5434.53 & Fe~I  &   1.01 & $-$2.130 &    19.7 &     7.0 & \nodata & \nodata & \nodata & \nodata &    21.7 &    23.6  \\
 5506.79 & Fe~I  &   0.99 & $-$2.790 & \nodata & \nodata & \nodata & \nodata & \nodata & \nodata &     6.8 & \nodata  \\
 3255.90 & Fe~II &   0.99 & $-$2.498 &    49.6 &    36.4 &    33.6 & \nodata & \nodata &    50.4 &    54.2 &   71.2  \\
 3277.36 & Fe~II &   0.99 & $-$2.191 &    52.3 &    52.1 &    46.6 & \nodata & \nodata &    53.4 &    58.6 &   74.5  \\
 3281.30 & Fe~II &   1.04 & $-$2.678 &    41.0 &    28.7 &    24.8 & \nodata & \nodata &    34.6 &    45.0 & \nodata  \\
 4178.86 & Fe~II &   2.57 & $-$2.530 & \nodata & \nodata & \nodata &    16.6 &    10.2 & \nodata &     5.1 & \nodata  \\
 4233.17 & Fe~II &   2.57 & $-$2.000 &    13.9 & \nodata &     9.0 &    40.5 & \nodata &    12.0 &    17.8 &    22.7  \\
 4416.82 & Fe~II &   2.77 & $-$2.430 & \nodata & \nodata & \nodata &    10.8 & \nodata & \nodata & \nodata &     4.1  \\
 4508.30 & Fe~II &   2.84 & $-$2.280 &     5.2 & \nodata & \nodata &    10.8 & \nodata & \nodata & \nodata &     6.8  \\
 4555.89 & Fe~II &   2.82 & $-$2.170 & \nodata & \nodata & \nodata &    14.4 &    16.8 & \nodata & \nodata &     6.8  \\
 4583.84 & Fe~II &   2.81 & $-$2.020 &    10.5 & \nodata & \nodata &    30.5 &    15.0 & \nodata &     8.6 &    17.4  \\
 4923.93 & Fe~II &   2.88 & $-$1.320 &    17.5 &     5.2 &    12.6 &    45.8 & \nodata & \nodata &    17.7 &    25.2  \\
 5018.45 & Fe~II &   2.89 & $-$1.220 &    26.6 &     7.8 &    16.5 &    58.6 &    36.6 &    12.6 &    24.6 &    35.1  \\
 3842.05 & Co~I  &   0.92 & $-$0.763 &     9.0 & \nodata & \nodata & \nodata & \nodata &    18.0 & \nodata & \nodata  \\
 3845.46 & Co~I  &   0.92 &   0.009 &    19.9 &    10.3 &    16.5 &    36.8 &    33.4 &    28.0 &    25.6 &    26.8  \\
 3873.11 & Co~I  &   0.43 & $-$0.666 &    29.7 & \nodata &     9.6 &    41.7 &    40.0 &    34.8 &    43.9 &    38.8  \\
 4121.31 & Co~I  &   0.92 & $-$0.315 &    19.8 & \nodata &     9.9 &    25.0 &    18.2 &    26.0 &    25.3 &    23.1  \\
 3807.15 & Ni~I  &   0.42 & $-$1.180 &    41.4 &    16.5 &    29.0 &    58.9 &    45.3 &    42.7 &    49.0 &    41.4  \\
 3858.30 & Ni~I  &   0.42 & $-$0.967 &    56.6 &    29.8 &    41.7 &    79.2 &    56.5 &    53.0 &    60.8 &    55.5  \\
 4401.55 & Ni~I  &   3.19 &   0.084 & \nodata & \nodata & \nodata &     7.0 & \nodata & \nodata & \nodata & \nodata  \\
 3247.53 & Cu~I  &   0.00 & $-$0.060 &    26.0 & \nodata &    22.6 & \nodata & \nodata & \nodata &    30.8 & 
     \nodata\tablenotemark{a}  \\
 3273.95 & Cu~I  &   0.00 & $-$0.360 &    18.4 & 13.0 &    14.4 & \nodata & \nodata &    13.9 &    20.0 &    25.5  \\
 4810.54 & Zn~I  &   4.08 & $-$0.170 &   $<$4.5 & \nodata & \nodata &   $<$5.0 &   $<$6.0 &
      \nodata &   $<$5.0 & \nodata  \\
 4077.71 & Sr~II &   0.00 &   0.170 &    82.9 &    44.4 & \nodata &    27.6 &    37.2 &   $<$5.0 &
      \nodata &   106.1  \\
 4215.52 & Sr~II &   0.00 & $-$0.140 &    75.6 & \nodata &     4.8 &    13.3 &    34.2 &   $<$5.0 &     5.2 &    96.1  \\
 4554.04 & Ba~II &   0.00 &   0.170 &     8.4 &    13.7 &   $<$5.5 &    22.2 &    21.5 &     4.5 &  
      $<$4.3 &    19.7  \\
 4934.16 & Ba~II &   0.00 & $-$0.150 &     5.2 &     8.3 &   $<$5.5 &     7.0 &    12.6 & \nodata &  
      $<$5.0 &    21.7  \\
 5853.70 & Ba~II &   0.60 & $-$1.010 & \nodata & \nodata & \nodata & \nodata & \nodata & \nodata &   $<$2.0 & \nodata  \\
 3774.33 & Y~II  &   0.13 &   0.220 &  \nodata &  \nodata &  \nodata &  \nodata &  \nodata & 
       $<$10.0 &  \nodata &  \nodata  \\
 3950.36 & Y~II  &   0.10 & $-$0.490 &   $<$6.9 & \nodata & \nodata &   $<$9.0 & \nodata & 
       $<$4.5 & \nodata & \nodata  \\
 3819.67 & Eu~II &   0.00 &   0.510 &    $<$8.0 &  \nodata &  \nodata &    $<$7.5 &  \nodata & 
      \nodata &  \nodata &  \nodata  \\
 3971.96 & Eu~II &   0.21 &   0.270 & \nodata & \nodata & \nodata & \nodata & \nodata &   $<$4.5 & 
     \nodata & \nodata  \\
 4129.70 & Eu~II &   0.00 &   0.220 &    $<$8.0 &    $<$8.0 &  \nodata &    $<$9.0 &   $<$15.0 &  \nodata
      &  \nodata &    $<$5.0  \\ 
 3407.80 & Dy~II &   0.00 &   0.180 &  \nodata &  \nodata &  \nodata &  \nodata &  \nodata &  
      $<$16.1 &  \nodata &  \nodata  \\
 3531.71 & Dy~II &   0.00 &   0.770 &  \nodata &  \nodata &  \nodata &  \nodata &  \nodata &  
       $<$8.8 &  \nodata &  \nodata  \\
 4057.81 & Pb~I  &   1.32 &  $-$0.220 &  \nodata &    $<$8.0 &  \nodata &  \nodata &  \nodata & 
       \nodata &  \nodata &  \nodata  \\
%
%
%
\enddata
\tablenotetext{a}{Too blended to use.}
\end{deluxetable}

\clearpage

\begin{deluxetable}{l rrr}
\tablenum{5}
\tablewidth{0pt}
\small
\tablecaption{Fit Fe~I Slopes With EP, Equivalent Width, and Wavelength 
\label{table_slopes}}
\tablehead{\colhead{Star ID} &
\colhead{$\Delta$[X/Fe]/$\Delta$(EP)\tablenotemark{a}} &
\colhead{$\Delta$[X/Fe]/${\Delta}[W_{\lambda}/\lambda]$} &
\colhead{$\Delta$[X/Fe]/${\Delta}\lambda$} \\ 
\colhead{} & \colhead{(dex/eV)} & \colhead{(dex)} & 
\colhead{($10^{-4}$dex/$\AA$)}
}
\startdata
C-normal \\
HE0132$-$2429 & $-0.068$ & $-0.074$ & +0.09 \\
HE1347$-$1025 & $-0.044$ & $-$0.007 & +0.32 \\
HE1356$-$0622 & $-0.053$ & $-0.136$ & +0.33 \\
HE1424$-0241$ &  $-0.041$  & $-0.079$ & $-0.32$ \\
BS16467$-$062 & $-0.091$ &  +0.076 & $-0.05$ \\
C-rich \\
HE1012$-$1540 & +0.014 & $-0.091$ & +0.71 \\
HE1300+0157 & $-0.098$ & +0.022 & +0.11 \\
HE2323$-$0256 & $-0.036$ & +0.005 & $-0.67$ \\
\enddata
\tablenotetext{a}{Typical range of EP is 3 eV.  Often only the 0~eV lines
are discrepant.}
\end{deluxetable}

\clearpage

\thispagestyle{empty}

\begin{deluxetable}{l | rrrr | rrrr | rrrr | rrrr|  rrrr}
\tablenum{6}
\tabletypesize{\tiny}
\rotate
\tablewidth{0pt}
\setlength{\tabcolsep}{.1cm}
\tablecaption{Abundances for the Five C-Normal EMP Stars From the HES \label{table_abunda}}
\tablehead{\colhead{Species} & 
\colhead{} & \multispan{3}{HE0132$-$2429~~~[Fe/H] $-$3.55} &
\colhead{} & \multispan{3}{HE1347$-$1025~~~[Fe/H] $-$3.48} &
\colhead{} & \multispan{3}{HE1356$-$0622~~~[Fe/H] $-$3.49} &
\colhead{} & \multispan{3}{HE1424$-$0241~~~[Fe/H] $-$3.96} & 
\colhead{} & \multispan{3}{BS16467$-$062~~~[Fe/H] $-$3.47}  \\
\colhead{} & 
\colhead{[X/Fe]} & \colhead{log$\epsilon(X)$} & \colhead{No.} & \colhead{$\sigma$} &  
\colhead{[X/Fe]} & \colhead{log$\epsilon(X)$} & \colhead{No.} & \colhead{$\sigma$} &
\colhead{[X/Fe]} & \colhead{log$\epsilon(X)$} & \colhead{No.} & \colhead{$\sigma$} &
\colhead{[X/Fe]} & \colhead{log$\epsilon(X)$} & \colhead{No.} & \colhead{$\sigma$} &
\colhead{[X/Fe]} & \colhead{log$\epsilon(X)$} & \colhead{No.} & \colhead{$\sigma$} \\
\colhead{} & \colhead{(dex)} & \colhead{(dex)} & \colhead{Lines} & \colhead{(dex)} &
\colhead{(dex)} & \colhead{(dex)} & \colhead{Lines} & \colhead{(dex)} &
\colhead{(dex)} & \colhead{(dex)} & \colhead{Lines} & \colhead{(dex)} &
\colhead{(dex)} & \colhead{(dex)} & \colhead{Lines} & \colhead{(dex)} &
\colhead{(dex)} & \colhead{(dex)} & \colhead{Lines} & \colhead{(dex)} 
}
\startdata 
C(CH)  &   0.62 &   5.66   &    1 & \nodata   &   0.15 &   5.26   &    1 & \nodata   &  $\leq -$0.05 &   $\leq$5.06 & 1 & \nodata 
  &   $\leq$0.63 &   $\leq$5.26 &    1 & \nodata   &   0.48 &   5.60   &    1 & \nodata   \\
N(NH)  &   1.07 &   5.45   &    1 & \nodata   &\nodata & \nodata   &  \nodata & \nodata   & \nodata & \nodata   &  \nodata & \nodata   & 
  $\leq$1.13 &   $\leq$5.10 &    1 & \nodata   &   $\leq$0.54 &   $\leq$5.00 &    1 & \nodata   \\
O(OH)  &   $\leq$1.67 &   $\leq$6.95 &    2 &    0.20   & \nodata & \nodata   &  \nodata & \nodata   & \nodata &\nodata   &  \nodata
   & \nodata   & \nodata &\nodata   &  \nodata & \nodata   &   $\leq$1.79 &  $\leq$ 7.15 &    2 &    0.20   \\
Na~I  &  $-$0.31 &   2.46   &    2 &    0.09   & $-$0.42 & 2.42 & 2 & 0.08   &   0.31 &   3.15   &    2 &    0.03   & 
  $-0.04$ &   2.32   &    2 &    0.07   &  $-$0.60 &   2.25   &    2 &    0.07   \\
Mg~I  &   0.40 &   4.39   &    5 &    0.15   &   0.49 &   4.55   &    3 &    0.22   &   0.67 &   4.72   &  5 &    0.32   &   0.45 & 
  4.03   &    3 &    0.12   &   0.31 &   4.37   &    4 &    0.33   \\
Al~I  &  $-$0.19 &   2.74   &    2 &    0.17   &  $-$0.02 &   2.97   &    2 &    0.17   &  $-$0.13 &   2.86   &    2 &    0.05   & 
 $-$0.16 &   2.35   &    2 &    0.15   &  $-$0.28 &   2.72   &    2 &    0.18   \\
Si~I  &   0.57 &   4.57   &    1 & \nodata   &   0.41 &   4.48   &    2 & 0.23   &   0.81 &   4.88   &    2 &    0.16 
  &  $-$1.00 &   2.59   &    1 & \nodata   &   0.27 &   4.35   &    1 & \nodata   \\
Ca~I  &   0.27 &   3.08   &    3 &    0.15   &   0.36 &   3.24   &    5 &    0.18   &   0.43 &   3.31   &    7 &    0.17  
 &  $-$0.56 &   1.84   &    1 & \nodata   &   0.12 &   3.00   &    3 &    0.12   \\
Ca~II &   0.00 &   2.81   &    1 & \nodata   & \nodata &\nodata   &  \nodata & \nodata   & \nodata & \nodata   &  \nodata & \nodata  
 &  $-$0.30 &   2.10   &    1 & \nodata   &  $-0.02$ &   2.87   &    1 & \nodata   \\
Sc~II &   0.75 &   0.31   &    3 &    0.05   &  0.00 &  $-$0.38   &    2 &    0.12   &   0.32 &  $-$0.06   &    4 &    0.17 
  &  $-$0.08 &  $-$0.94   &    1 & \nodata   &   0.16 &  $-$0.22   &    3 &    0.03   \\
Ti~I  &   0.61 &   2.05   &    3 &    0.10   &   $\leq$0.62 &   $\leq$2.13 &    4 &    0.09   &   0.40 &   1.91   &  5 &    0.06 
  & \nodata &\nodata   &  \nodata & \nodata   &   0.28 &   1.80   &    2 &    0.09   \\
Ti~II &   0.39 &   1.84   &   15 &    0.10   &   0.26 &   1.77   &   13 &    0.11   &   0.26 &   1.76   &   17 &    0.07   & 
 $-$0.17 &   0.85   &    8 &    0.17   &   0.20 &   1.72   &   14 &    0.12   \\
V~I   & \nodata & \nodata   &  \nodata & \nodata   &   $\leq$0.69 &   $\leq$1.21 &    1 & \nodata   &\nodata & \nodata   & 
 \nodata & \nodata   &\nodata &\nodata   &  \nodata & \nodata   & \nodata &\nodata   &  \nodata & \nodata   \\
V~II  &   0.36 &   0.81   &    1 & \nodata   & \nodata & \nodata   &  \nodata & \nodata   & \nodata & \nodata   &  \nodata &
 \nodata   &  $\leq$ 0.60 &   $\leq$0.64 &    1 & \nodata   &   0.28 &   0.81   &    2 &    0.14   \\
Cr~I  &  $-$0.43 &   1.69   &    5 &    0.12   &  $-$0.52 &   1.67   &    4 &    0.11   &  $-$0.52 &   1.66   &    4 &    0.13
   &  $-$0.38 &   1.33   &    5 &    0.09   &  $-$0.54 &   1.65   &    5 &    0.10   \\
Mn~I\tablenotemark{b}  &  $-$0.90 &   0.94   &    2 &    0.02   &  $-$0.88 &   1.03   &    2 &    0.04   &  $-$0.98 &   0.92   &    2 &    0.02 
  &  $-$0.10 &   1.33   &    2 &    0.02   &  $-$0.85 &   1.06   &    2 &    0.04   \\
Mn~II &  $-$0.48 &   1.37   &    3 &    0.16   & \nodata & \nodata   &  \nodata & \nodata   & \nodata & \nodata   &  \nodata & 
  \nodata   &   0.19 &   1.62   &    3 &    0.13   &  $-$0.48 &   1.44   &    3 &    0.07   \\
Fe~I  & $-3.55$\tablenotemark{a} &   3.90   &   53 &    0.18   & $-3.48$ & 3.97   &   50 &    0.22   & $-3.49$ &   3.96   &   63 &    0.16   & 
   $-3.96$ &   3.49   &   39 &    0.18   & $-3.47$ &   3.98   &   57 &    0.19   \\
Fe~II &  $-$0.05 &   3.85   &    8 &    0.18   &   0.20 &   4.17   &    4 &    0.19   &   0.13 &   4.10   &    8 &    0.13   & 
  0.09 &   3.58   &    5 &    0.19   &   0.04 &   4.01   &    8 &    0.16   \\
Co~I  &   0.55 &   1.92   &    4 &    0.18   &   0.48 &   1.92   &    3 &    0.19   &   0.24 &   1.67   &    3 &    0.11   &  
 1.03 &   1.98   &    4 &    0.21   &   0.68 &   2.12   &    3 &    0.28   \\
Ni~I  &  $-$0.04 &   2.67   &    2 &    0.06   &  $-$0.19 &   2.58   &    1 & \nodata   &  $-$0.04 &   2.72   &    3 &    0.33 
  &   0.24 &   2.52   &    2 &    0.01   &   0.15 &   2.93   &    2 &    0.04   \\
Cu~I &  $-$0.85 &  $-$0.18   &    2 &    0.06   & \nodata & \nodata   &  \nodata & \nodata   & \nodata & \nodata   &  \nodata & 
\nodata   &  $-$0.66 &  $-$0.41   &    1 & \nodata   &  $-$0.75 &  $-$0.01   &    2 &    0.01   \\
Zn~I  &   $\leq$0.84 &   $\leq$1.89 &    1 & \nodata   &   $\leq$0.94 &   $\leq$2.06 &    2 &    0.17   & 
  $\leq$0.55 &   $\leq$1.66 &    1 & \nodata   & \nodata & \nodata   &  \nodata & \nodata   & 
    $\leq$0.96 &   $\leq$2.09 &    2 &    0.12   \\
Sr~II &   0.05 &  $-$0.60   &    2 &    0.06   &  $-$1.13 &  $-$1.71   &    2 &    0.16   &  $-$1.88 &  $-$2.47   &    2 &    0.08 
  &  $\leq -$1.69 &  $\leq -$2.75 &    2 &    0.21   &  $-$1.75 &  $-$2.32   &    1 & \nodata   \\
Y~II  &   $\leq$0.33 &  $\leq -$0.98 &    1 & \nodata   &\nodata &\nodata   &  \nodata & \nodata   &  
   $\leq -$0.13 &  $\leq -$1.38 &    1 & \nodata   &   $\leq$0.26 &  $\leq -$1.46 &    2 &    0.19  
    & \nodata & \nodata   &  \nodata & \nodata   \\
Ba~II &  $-$0.85 &  $-$2.27   &    2 &    0.04   &  $-$0.62 &  $-$1.96   &    2 &    0.01   &  $-$1.19 &  $-$2.54   &    2 &    0.21 
  &  $-$0.91 &  $-$2.74   &    1 & \nodata   &  $\leq -$0.56 &  $\leq -$1.91 &    3 &    0.72   \\
Eu~II &   $\leq$1.18 &  $\leq -$1.86 &    2 &    0.18   &   $\leq$1.39 &  $\leq -$1.58 &    1 & \nodata   &  
    $\leq$0.56 &  $\leq -$2.41 &    3 &    0.18   &   $\leq$1.50 &  $\leq -$1.95 &    1 & \nodata   &
    \nodata & \nodata   &  \nodata & \nodata   \\
Dy~II & \nodata & \nodata   &  \nodata & \nodata   & \nodata & \nodata   &  \nodata & \nodata   &
   \nodata & \nodata   &  \nodata & \nodata   &   $\leq$1.56 &  $\leq -$1.30 &    2 &  \nodata   & 
   \nodata & \nodata   &  \nodata & \nodata   \\
\enddata
\tablenotetext{a}{[Fe~I/H] is given instead of [X/Fe].}
\tablenotetext{b}{A correction of +0.3~dex to [Mn/Fe] as derived from lines of the
4030~\AA\ Mn~I triplet is required, but not put in here.  See \S\ref{section_analysis}.}
\end{deluxetable}

\clearpage

\begin{deluxetable}{l | rrrr | rrrr | rrrr}
\tablenum{7}
\tabletypesize{\tiny}
\tablewidth{0pt}
\tablecaption{Abundances for the Three C-Rich EMP Stars From the HES \label{table_abundb}}
\tablehead{\colhead{Species} & 
\colhead{} & \multispan{3}{HE1012$-$1540~~~[Fe/H] $-$3.43} &
\colhead{} &\multispan{3}{HE1300+0157~~~[Fe/H] $-$3.39} &
\colhead{} &\multispan{3}{HE2323$-$0256~~~[Fe/H] $-$3.79} \\
\colhead{} & 
\colhead{[X/Fe]} & \colhead{log$\epsilon(X)$} & \colhead{No.} & \colhead{$\sigma$} &  
\colhead{[X/Fe]} & \colhead{log$\epsilon(X)$} & \colhead{No.} & \colhead{$\sigma$} &
\colhead{[X/Fe]} & \colhead{log$\epsilon(X)$} & \colhead{No.} & \colhead{$\sigma$} \\
\colhead{} & \colhead{(dex)} & \colhead{(dex)} & \colhead{Lines} & \colhead{(dex)} &
\colhead{(dex)} & \colhead{(dex)} & \colhead{Lines} & \colhead{(dex)} &
\colhead{(dex)} & \colhead{(dex)} & \colhead{Lines} & \colhead{(dex)} 
}
\startdata   
C(CH) &   2.22 &   7.38   &    1 &  \nodata   &   1.23 &   6.43   &    1 &  \nodata   &
         0.97 &   5.77   &    1 &  \nodata   \\
N(NH) &   1.25 &   5.75   &    1 &  \nodata   &   $<$0.71 &   $<$5.25 &    1 &  \nodata   & 
   2.16 &   6.30   &    2 &    0.30    \\
O(OH) &   2.25 &   7.65   &    2 &    0.25   &   1.69 &   7.13   &    2 &    0.18   &  
   1.96 &   7.20   &    1 &  \nodata     \\
Na~I  &   1.21 &   4.11   &    2 &    0.07   &  $-$0.49 &   2.44   &    2 &    0.07   & 
    1.45 &  3.98   &    2 &    0.11     \\
Mg~I  &   1.88 &   5.99   &    6 &    0.44   &   0.32 &   4.47   &    3 &    0.10   & 
  1.47 &   5.22   &    7 &    0.25       \\
Al~I  &   0.93 &   3.97   &    1 &  \nodata   &  $-$0.24 &   2.83   &    1 &  \nodata   &  
 0.48 &   3.16   &    1 &  \nodata   \\
Si~I  &   1.07 &   5.20   &    1 &  \nodata   &   0.49 &   4.64   &    1 &  \nodata   &  
 0.56 &   4.32   &    1 &  \nodata    \\
Ca~I  &   0.57 &   3.50   &    2 &    0.16   &   0.26 &   3.23   &    2 &    0.13   & 
    0.31 &   2.88   &    2 &    0.11      \\
Ca~II &  $-$0.34 &   2.59   &    1 &  \nodata   &   0.10 &   3.07   &    1 &  \nodata   & 
  0.08 &   2.65   &    1 &  \nodata       \\
Sc~II & \nodata & \nodata   &  \nodata &  \nodata   &   0.17 &  $-$0.12   &    1 &  \nodata   &  
   0.12 &  $-$0.56   &    1 &  \nodata     \\
Ti~I  &   $<$0.75 &   $<$2.31 &    1 &  \nodata   & \nodata & \nodata   &  \nodata &  \nodata   &  
   $<$0.46 &  $<$1.66 &    1 &  \nodata    \\
Ti~II &  $-$0.03 &   1.53   &    7 &    0.12   &   0.12 &   1.72   &    8 &    0.06   & 
   0.27 &   1.47   &   11 &    0.06     \\
V~II  & \nodata & \nodata   &  \nodata &  \nodata   & \nodata & \nodata   &  \nodata &  \nodata   &  
   0.09 &   0.30   &    2 &    0.03    \\
Cr~I  &  $-$0.33 &   1.91   &    2 &    0.09   &  $-$0.53 &   1.74   &    4 &    0.07   & 
   $-$0.54 &   1.34   &    5 &    0.15    \\
Mn~I\tablenotemark{a}  &  $-$0.55 &   1.41   &    2 &    0.01   &  $-$0.86 &   1.13   &    2 &    0.02   & 
  $-$0.97 &   0.63   &    2 &    0.03      \\
Mn~II &  $-$1.00 &   0.96   &    1 &  \nodata   &  $-$0.66 &   1.33   &    2 &    0.16 
   &  $-$0.43 &   1.17   &    2 &    0.01    \\
Fe~I &  $-3.43$\tablenotemark{b}  &   4.02   &   28 &    0.17   &     
    $-3.39$\tablenotemark{b}   &  4.06 & 36 &    0.19   &  
    $-3.79$\tablenotemark{b}  &   3.66   &   57 &    0.16     \\
Fe~II &  $-$0.28 &   3.75   &    5 &    0.35   &  $-$0.16 &   3.90   &    6 &    0.11   &  
 0.06 &   3.72   &    9 &    0.18     \\
Co~I  &   0.19 &   1.68   &    1 &  \nodata   &   0.39 &   1.92   &    3 &    0.08   & 
  0.42 &   1.55   &    3 &    0.18    \\
Ni~I  &  $-$0.32 &   2.50   &    2 &    0.09   &  $-$0.02 &   2.83   &    2 &    0.06   & 
  $-$0.26 &   2.20   &    2 &    0.04      \\
Cu~I &  $-$0.63 &   0.15   &    1 &  \nodata   &  $-$0.68 &   0.13   &    2 &    0.02   & 
   $-$0.78 &  $-$0.36   &    1 &  \nodata     \\
Sr~II &  $-$0.54 &  $-$1.07   &    1 &  \nodata   &  $-$1.55 &  $-$2.05   &    1 &  \nodata   & 
    0.18 &  $-$0.71   &    2 &    0.02   \\
Ba~II &  $-$0.29 &  $-$1.58   &    2 &    0.02   &  $< -$0.63 &  $< -$1.89 &    2 &    0.20   & 
   $-$0.66 &  $-$2.32   &    2 &    0.23     \\
Eu~II &  $<$1.62 &  $< -1.30$ &    1 &  \nodata   & \nodata & \nodata   &  \nodata &  \nodata   &  
   $<$0.73 &  $< -$2.55 &    1 &  \nodata    \\
Pb~I  &   $<$2.93 &   $<$1.46 &    1 &  \nodata   & \nodata & \nodata   &  \nodata &  \nodata   &
   \nodata & \nodata   &  \nodata &  \nodata   \\
\enddata
\tablenotetext{a}{A correction of +0.3~dex to [Mn/Fe] as derived from lines of the
4030~\AA\ Mn~I triplet is required, but not put in here.  See \S\ref{section_analysis}.}
\tablenotetext{b}{[Fe~I/H] is given instead of [X/Fe].}
\end{deluxetable}

\clearpage

\begin{deluxetable}{l rrrrr}
\tablenum{8}
\tablewidth{0pt}
\tablecaption{Abundance Range for Five C-normal EMP Stars From the HES \label{table_range_a}}
\tablehead{\colhead{Species [X/Fe]} & 
\colhead{Nu. stars} & \colhead{Mean [X/Fe]} & \colhead{$\sigma$} &
\colhead{Min.} & \colhead{Max.} \\
\colhead{} & \colhead{(dex)} & \colhead{(dex)} & \colhead{(dex)} & \colhead{(dex)}
 & \colhead{(dex)} }
\startdata     
${\rm{[C(CH)/Fe]}}$\tablenotemark{a} & 3 & 0.39 & 0.29 & 0.07 & 0.62 \\
${\rm{[N(NH)/Fe]}}$\tablenotemark{b} & ~~ \\
${\rm{[O(OH)/Fe]}}$\tablenotemark{c} & ~~ \\
${\rm{[Na/Fe]}}$\tablenotemark{d} & 5 & $-0.21$ & 0.36 & $-0.60$ & 0.32 \\
${\rm{[Mg/Fe]}}$ & 5 & 0.46 & 0.13 & 0.31 & 0.67 \\
${\rm{[Al/Fe]}}$\tablenotemark{e} & 5 & $-0.16$ & 0.09 & $-0.28$ & $-0.02$ \\
${\rm{[Si/Fe]}}$ & 5 & 0.21 & 0.71 & $-1.00$ & 0.81 \\
${\rm{[CaI/Fe]}}$ & 5 & 0.12 & 0.40 & $-0.56$ & 0.43 \\
${\rm{[CaII/Fe]}}$ & 3 & $-0.11$ & 0.17 & $-0.30$ & 0.00 \\
${\rm{[Sc/Fe]}}$ & 5 & 0.23 & 0.33 & $-0.08$ & 0.75 \\
${\rm{[Ti/Fe]}}$\tablenotemark{f} & 5 & 0.19 & 0.21 & $-0.17$ & 0.40 \\
${\rm{[V/Fe]}}$ & 2 & 0.32 & 0.05 & 0.28 & 0.36 \\
${\rm{[Cr/Fe]}}$ & 5 & $-0.48$ & 0.07 & $-0.54$ & $-0.38$ \\
${\rm{[MnI/Fe]}}$\tablenotemark{g} & 5 & $-0.75$ & 0.36 & $-0.99$ & $-0.10$ \\
${\rm{[MnII/Fe]}}$ & 3 & $-0.26$ & 0.39 & $-0.48$ & 0.19 \\
${\rm{[FeII/FeI]}}$ & 5 & 0.08 & 0.10 & $-0.05$ & 0.20 \\
${\rm{[Co/Fe]}}$ & 5 & 0.59 & 0.29 & 0.24 & 1.03 \\
${\rm{[Ni/Fe]}}$ &  5 & 0.02 & 0.17 & $-0.19$ & 0.24 \\
${\rm{[Cu/Fe]}}$ & 3 & $-0.75$ & 0.09 & $-0.85$ & $-0.66$ \\
${\rm{[Sr/Fe]}}$ & 4 & $-1.18$ & 0.88 & $-1.88$ & 0.05 \\
${\rm{[Ba/Fe]}}$ & 4 & $-0.89$ & 0.24 & $-1.19$ & $-0.62$ \\
\enddata
\tablenotetext{a}{Two stars only have upper limits for [C/Fe].}
\tablenotetext{b}{One detection and two upper limits for [N/Fe].}
\tablenotetext{c}{Two upper limits for [O/Fe].}
\tablenotetext{d}{Non-LTE correction of $-$0.2~dex has been applied for 
   [Na/Fe] from the Na~D lines.}
\tablenotetext{e}{Non-LTE correction of +0.6~dex has been applied for 
[Al/Fe]  from the 3950~\AA\ doublet.}
\tablenotetext{f}{From Ti~II lines}
\tablenotetext{g}{A correction of +0.3~dex to [Mn/Fe] as derived from lines of the
4030~\AA\ Mn~I triplet is required, but not put in here.  See \S\ref{section_analysis}.}
\end{deluxetable}

\begin{deluxetable}{l rrrrr}
\tablenum{9}
\tablewidth{0pt}
\tablecaption{Abundance Range for Three C-rich EMP Stars From the HES \label{table_range_b}}
\tablehead{\colhead{Species [X/Fe]} & 
\colhead{Nu. stars} & \colhead{Mean [X/Fe]} & \colhead{$\sigma$} &
\colhead{Min.} & \colhead{Max.} \\
\colhead{} & \colhead{(dex)} & \colhead{(dex)} & \colhead{(dex)} & \colhead{(dex)}
}
\startdata 
${\rm{[C/Fe]}}$ & 3 & 1.47 & 0.66 & 0.97 & 2.22 \\
${\rm{[N/Fe]}}$ & 2 & 1.37 & 0.70 & 0.64 & 2.15 \\
${\rm{[O/Fe]}}$ & 3 &  2.03 & 0.30 & 1.69 & 2.25 \\
${\rm{[Na/Fe]}}$\tablenotemark{a} & 3 & 0.72 & 1.06 & $-0.49$ & 1.45 \\
${\rm{[Mg/Fe]}}$ & 3 & 1.22 & 0.81 & 0.32 & 1.88 \\
${\rm{[Al/Fe]}}$\tablenotemark{b} & 3 & 0.39 & 0.59 & $-0.24$ & 0.93 \\
${\rm{[Si/Fe]}}$ & 3 & 0.71 & 0.32 & 0.49 & 1.08 \\
${\rm{[CaI/Fe]}}$ & 3 & 0.38 & 0.17 & 0.26 & 0.57 \\
${\rm{[CaII/Fe]}}$ & 2 & $-0.12$ & 0.32 & $-0.34$ & 0.10 \\
${\rm{[Sc/Fe]}}$ & 2 & 0.14 & 0.04 & 0.12 & 0.17 \\
${\rm{[Ti/Fe]}}$\tablenotemark{c} & 3 & 0.12 & 0.15 & $-0.03$ & 0.27 \\
${\rm{[Cr/Fe]}}$ & 3 & $-0.41$ & 0.12 & $-0.54$ & $-0.33$ \\
${\rm{[MnI/Fe]}}$\tablenotemark{d} & 3 & $-0.80$ & 0.22 & $-0.97$ & $-0.55$ \\
${\rm{[MnII/Fe]}}$ & 3 & $-0.70$ & 0.28 & $-1.00$ & $-0.43$ \\
${\rm{[FeII/FeI]}}$ & 3 & $-0.12$ & 0.17 & $-0.28$ & 0.06 \\
${\rm{[Co/Fe]}}$ & 3 & 0.33 & 0.12 & 0.19 & 0.42 \\
${\rm{[Ni/Fe]}}$ &  3 & $-0.20$ & 0.16 & $-0.32$ & $-0.02$ \\
${\rm{[Cu/Fe]}}$ & 3 & $-0.70$&  0.08 & $-0.78$ & $-0.63$ \\
${\rm{[Sr/Fe]}}$ & 3 & $-0.64$ & 0.87 & $-1.55$ & 0.17 \\
${\rm{[Ba/Fe]}}$ & 2 & $-0.52$ & 0.21 & $-0.66$ & $-0.29$ \\
\enddata
\tablenotetext{a}{Non-LTE correction of $-$0.2 dex has been applied for 
   [Na/Fe] from the Na~D lines.}
\tablenotetext{b}{Non-LTE correction of +0.6 dex has been applied for 
[Al/Fe]  from the 3950~\AA\ doublet.}
\tablenotetext{c}{From Ti~II lines}
\tablenotetext{d}{A correction of +0.3~dex to [Mn/Fe] as derived from lines of the
4030~\AA\ Mn~I triplet is required, but not put in here.  See \S\ref{section_analysis}.}
\end{deluxetable}

\clearpage

\begin{deluxetable}{l rrr}
\tablenum{10}
\tablewidth{0pt}
\tablecaption{Comparison of Detailed Abundance Analyses for HE1300+0157 
\label{table_he1300}}
\tablehead{\colhead{log[{$\epsilon$}(X)]} & 
\colhead{\cite{frebel07} } & \colhead{0Z/Frebel\tablenotemark{a}} &
\colhead{0Z} \\
\colhead{} & \colhead{(5450,3.2) (dex)} & \colhead{(5450, 3.2) (dex)} & 
\colhead{(5632,3.37) (dex)}
}
\startdata 
[Fe/H] & $-$3.73 & $-3.60$ & $-3.39$ \\ 
C(CH)  &  5.89 & 6.02 & 6.43 \\
N(NH) &  $<5.12$ & $<4.93$ & $<5.25$ \\          
O(OH)  & 6.54 & 6.78 & 7.13 \\
Na~I\tablenotemark{b} & 2.44 & 2.49 & 2.64 \\   
Mg~I & 4.10 & 4.30 & 4.47 \\ 
Al~I\tablenotemark{c} & 2.64 & 2.66 & 2.83 \\ 
Si~I & 4.50 & 4.44 & 4.65 \\    
Ca~I & 2.98 & 3.06 & 3.23 \\  
Ca~II & 2.75 & 2.96 & 3.07 \\   
Sc~II & $-0.37$ & $-0.30$ & $-0.12$ \\  
Ti~II & 1.77 & 1.58 & 1.72 \\  
Cr~I & 1.51 & 1.56 & 1.74 \\   
Mn~I\tablenotemark{d} & 0.65 & 0.91 & 1.13 \\    
Mn~II & 0.89 & 1.20 & 1.33 \\   
Fe~I & 3.72 & 3.85 & 4.06 \\ 
Fe~II & 3.57 & 3.77 & 3.90 \\ 
Co~I & 1.80 & 1.71 & 1.92 \\  
Ni~I & 2.61 & 2.61 & 2.83 \\ 
Cu~I & \nodata & $-0.10$ & 0.13 \\
Sr~II & $<-2.64$ & $-2.22$ & $-2.05$ \\
Ba~II & $<-2.56$ & $<-2.08$ & $<-1.89$ \\
\enddata
\tablenotetext{a}{The stellar parameters are those of \cite{frebel07}, but
the analysis is that of our 0Z project with our own set of \eqw.}
\tablenotetext{b}{no non-LTE correction has been applied for [Na/Fe].} 
\tablenotetext{c}{Non-LTE correction of +0.6 dex has been applied for 
[Al/Fe]  from the 3950~\AA\ doublet.}
\tablenotetext{d}{In all cases only lines from the 4030~\AA\ triplet
of Mn~I have been used.  No correction factor has been added here.
The +0.4~dex correction factor added by \cite{frebel07} has been
removed.}
\end{deluxetable}

\begin{deluxetable}{l rrr}
\tablenum{11}
\tablewidth{0pt}
\tablecaption{Comparison of \teff\ For EMP Giants
Between Our 0Z Survey and The First Stars Project 
\label{table_teff_cayrel}}
\tablehead{\colhead{Star} & \colhead{\teff (0Z)} & 
\colhead{$\Delta$\teff Hybrid\tablenotemark{a}} & 
\colhead{$\Delta$\teff [\citep{cayrel04}} \\
\colhead{}  & \colhead{(K)}  & \colhead{(K)}  & \colhead{ -- 0Z)~(K)}
}
\startdata
CD $-$38~245    &    4830     & $-30$   & $-30$ \\  
BS16477$-$003  &  4928 &  $-48$  & $-28$ \\  
HE2323$-$0256\tablenotemark{b} & 4995\tablenotemark{c}  & $-65$ &  $-95$ \\  
CS22948$-$066  &   5224   &   $-96$  &  $-124$ \\
BS16467$-$062  &   5364   &  $-66$ &  $-164$ \\    
\enddata
\tablenotetext{a}{The hybrid \teff\ uses the codes of the 0Z project,
but the reddening of the First Stars project, which is almost always
slightly higher.  In all cases, the
$V$ mag from \cite{cayrel04} is adopted. The value of
\teff(hybrid) -- \teff(0Z) is given. }
\tablenotetext{b}{Rediscovery of  CS22949$-$037 from the
HK Survey.}
\tablenotetext{c}{\teff\ for this star adopted by the 0Z project
is 4915~K as our observed $V$ mag from ANDICAM is 14.41~mag, 0.05~mag
fainter than that adopted by the First Stars project.}
\end{deluxetable}

\clearpage

%
%

\clearpage

\clearpage

\begin{figure}
\epsscale{0.9}
\plotone{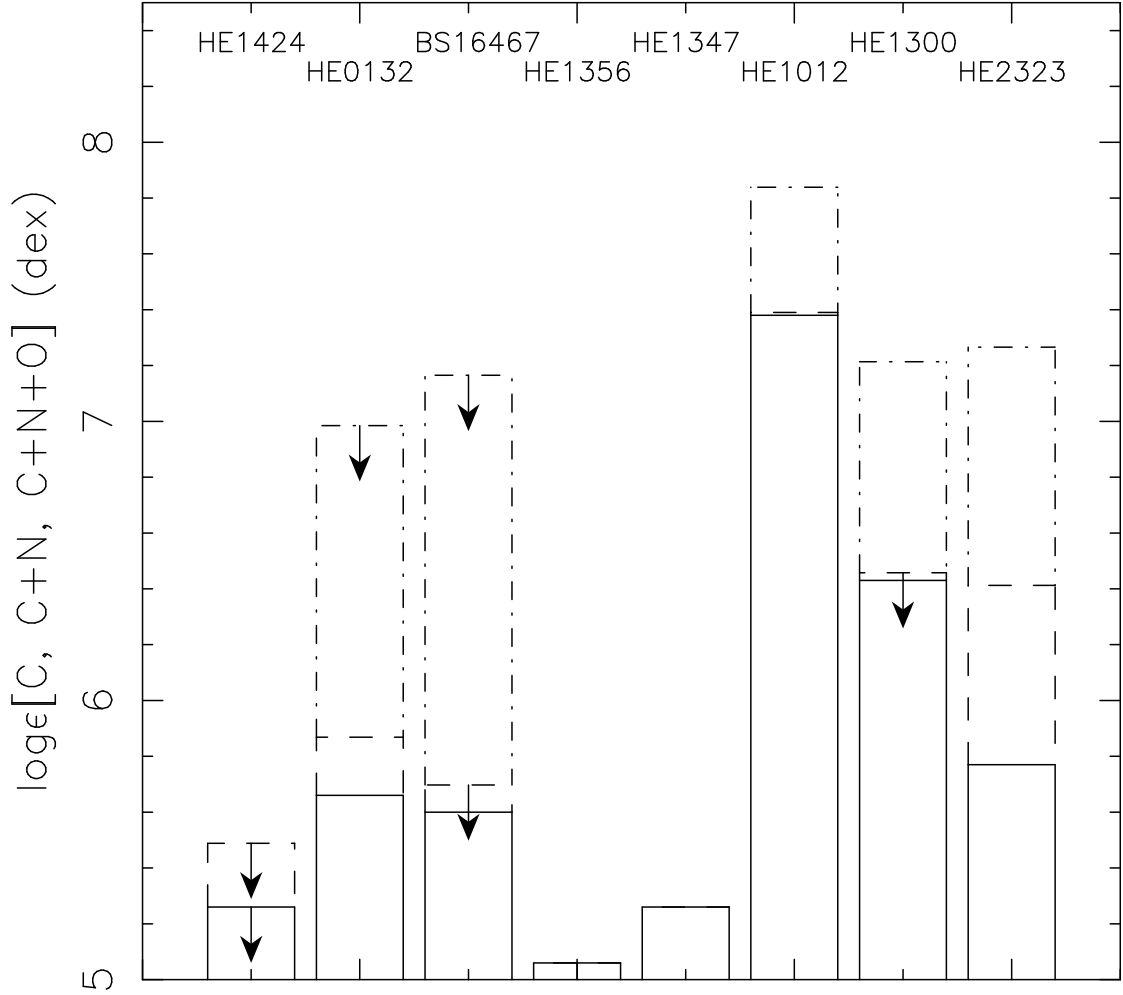}
\caption[]{Log$\epsilon$(C, C+N, C+N+O) is shown for each of the EMP stars
in the present sample.  The solid horizontal line denotes
log$\epsilon$(C), the dashed line is log$\epsilon$(C+N), while the dot-dashed
line shows
log$\epsilon$(C+N+O).  Upper limits are indicated in each case.
The 5 C-normal stars are at the left, the 3 higher C stars are at the right
side of the plot. 
\label{figure_cno}}
\end{figure}

\clearpage

\begin{figure}
\epsscale{0.9}
\plotone{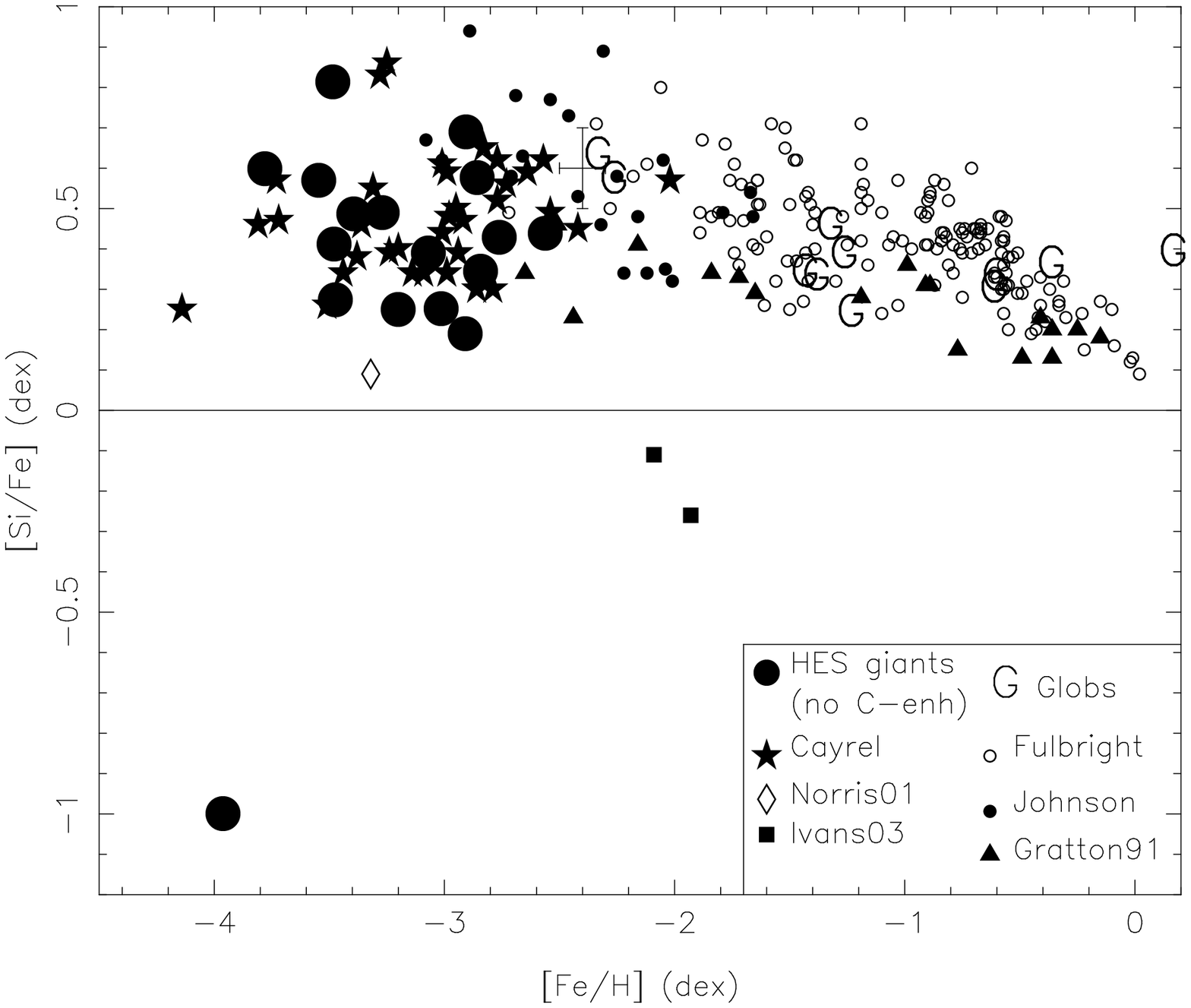}
\caption[]{[Si/Fe] is shown for all of the candidate EMP stars
with HIRES spectra analyzed by the 0Z project to date, including
the present sample.  C-rich stars are not shown. The solid horizontal line denotes
the Solar ratio.  The plot includes well studied Galactic globular clusters,
mostly from analyses by J.~Cohen and her collaborators, as well as
samples of halo field stars from the sources indicated on the symbol key
in the lower right of the figure.
Note the highly anomalous position of HE1424$-$0241, the only star
with [Si/Fe] $<< 0$~dex. 
\label{figure_si}}
\end{figure}

\begin{figure}
\epsscale{0.9}
\plotone{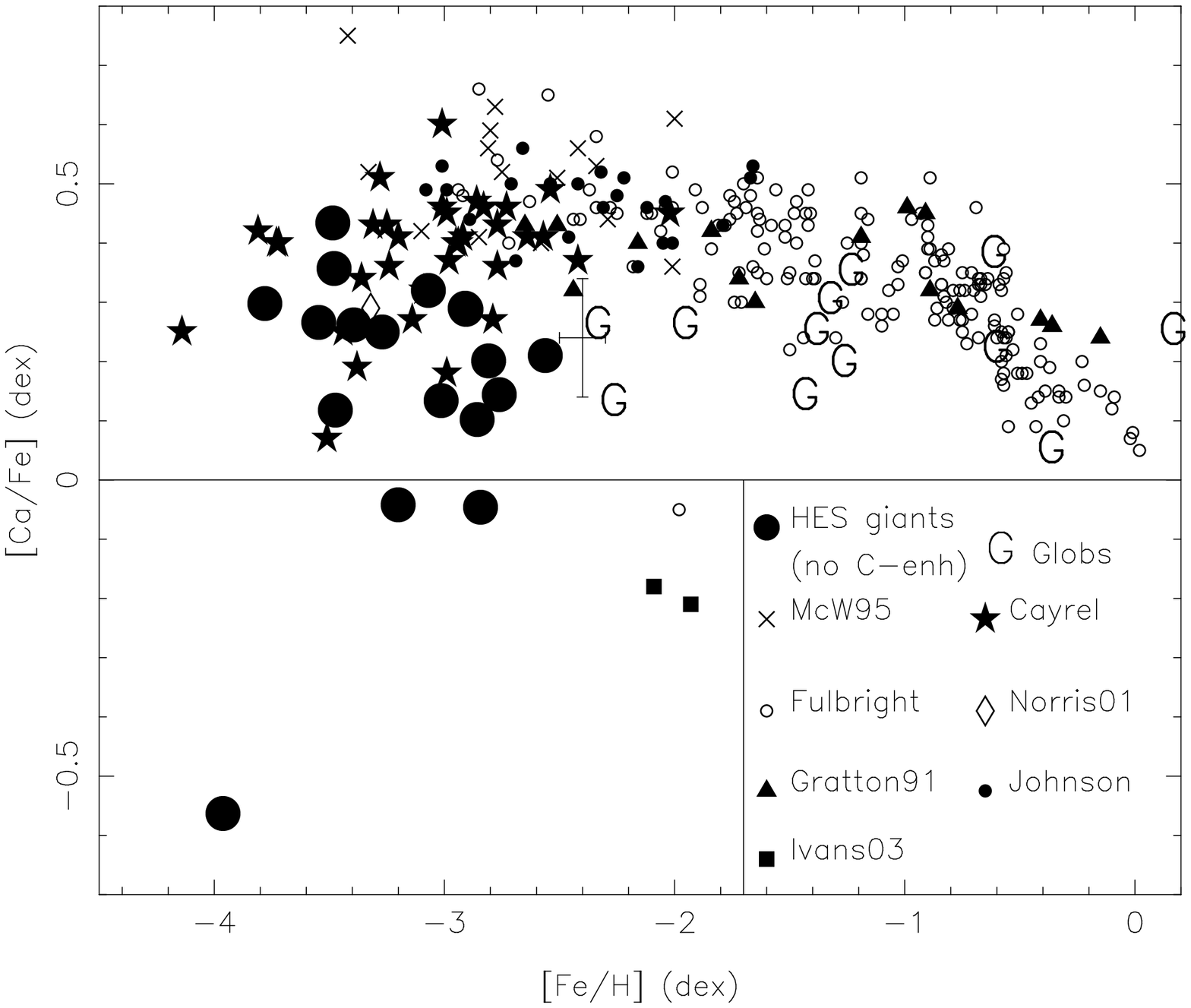}
\caption[]{[Ca/Fe] is shown for all of the candidate EMP stars
with HIRES spectra analyzed by the 0Z project to date, including
the present sample.  C-rich stars are not shown. The solid horizontal line denotes
the Solar ratio.  The plot includes well studied Galactic globular clusters,
mostly from analyses by J.~Cohen and her collaborators, as well as
samples of halo field stars from the sources indicated on the symbol key
in the lower right of the figure.
side of the plot.  Note the highly anomalous position of HE1424$-$0241,
the star with the smallest [Ca/Fe]. 
\label{figure_ca}}
\end{figure}

\begin{figure}
\epsscale{0.9}
\plotone{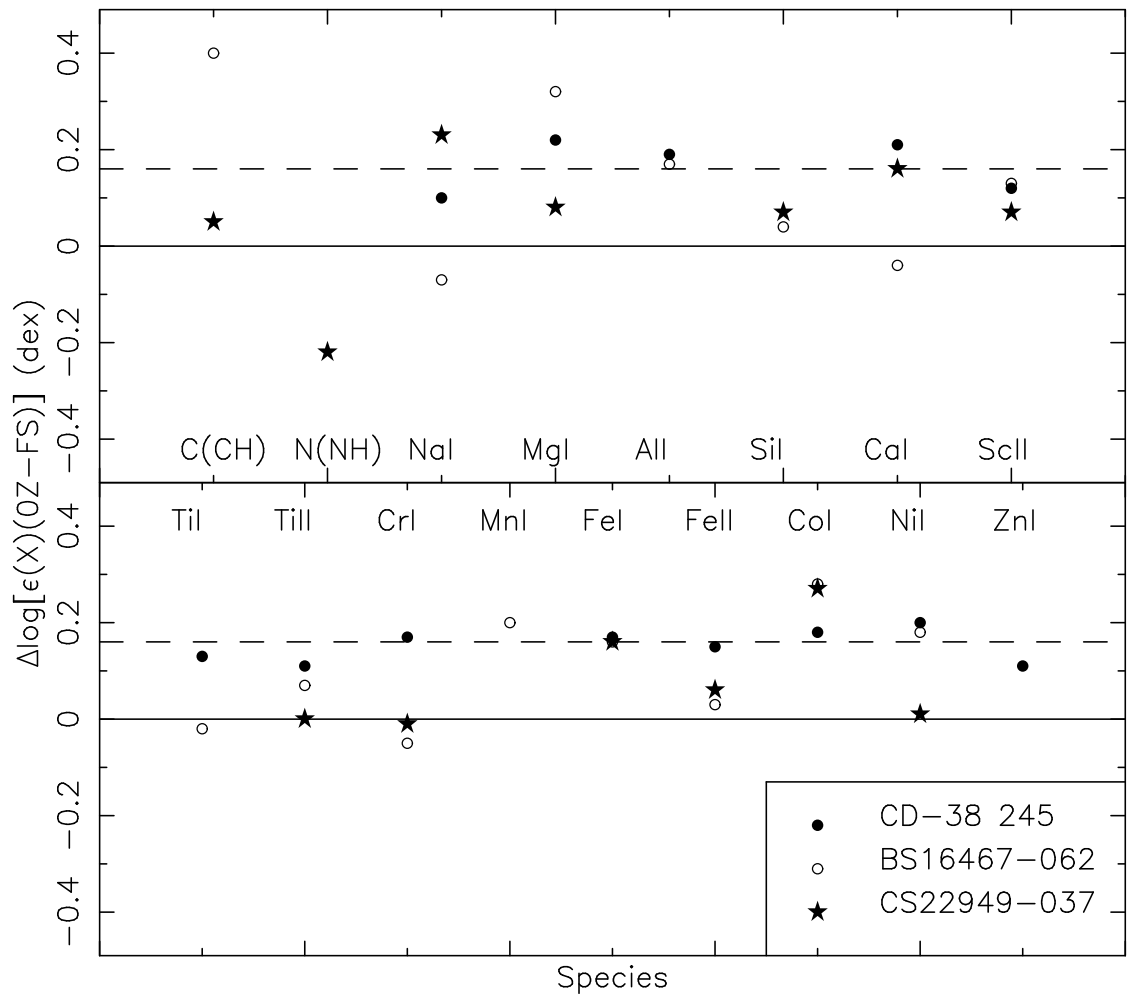}
\caption[]{The difference for log[$\epsilon$(X) between the our results
and those of the First Stars project \citep{cayrel04} for three stars.
We adopt the stellar parameters used by \cite{cayrel04} but use our
own $gf$ values and set of $W_{\lambda}$, except for CD$-$38~245, for which we have
no HIRES spectra, and for which we adopt those from the First Stars project.
The dashed line represents the mean difference for Fe~I for the three
stars.
\label{figure_cayrel_diff}}
\end{figure}


\clearpage

\begin{thebibliography}{}

\bibitem[Adelman-McCarthy \etal(2006)]{sdss_dr4}
Adelman-McCarthy, J.~K., \etal, 2006, \apjs, 162, 38

\bibitem[Allyn Smith \etal(2002)]{sdss_trans}
Allyn Smith, J. \etal, 2002, \aj, 123, 2121

\bibitem[Alonso, Arribas \& Martinez-Roger(1996)]{alonso96}
Alonso, A., Arrivas, S. \& Martinez-Roger, C., 1996, \aap, 313, 873

\bibitem[Alonso, Arribas \& Martinez-Roger(1999)]{alonso99}
Alonso, A., Arrivas, S. \& Martinez-Roger, C., 1996, \aaps, 140, 261

\bibitem[Amiot (2001)]{amiot83} Amiot C., 1983, \apjs, 52, 329

\bibitem[Anders \& Grevesse(1989)]{anders89} Anders, E. \& Grevesse, N., 1989,
Geochim. Cosmochim. Acta, 53, 197

\bibitem[Andrievsky \etal(2007)]{andrievsky07}
Andrievsky, S.~M., Spite, M., Korotin, S.~A., Spite, F.,
Bonifacio, P., Cayrel, R., Hill, V. \& Fracois, P., 2007
\aap\ (in press), 

\bibitem[Aoki \etal(2004)]{aoki04}
Aoki, W., Norris,  J.~E., Ryan, S.~G., Beers, T.~C., Christlieb, N.,
Tsangarides, S. \& Ando, H., 2004, \apj, 608, 971

\bibitem[Aoki \etal(2005)]{aoki05}
Aoki, W. \etal, 2005, \apj, 632, 611

\bibitem[Aoki \etal(2007)]{aoki07}
Aoki, W., Beers, T.~C., Christlieb, N., Norris,  J.~E., Ryan, S.~G.
\& Tsangarides, 2007, \apj, in press

\bibitem[Asplund \etal(2004)]{asplund04}
Asplund, M., Grevesse, N., Sauval, A.~J., Allende Prieto, C.
\& Kisselman, D., 2004, \aap, 417, 751

\bibitem[Asplund(2005)]{asplund_araa}
Asplund, M. 2005, \araa, 43, 481

\bibitem[Asplund \etal(2005)]{asplund05}
Asplund, M., Grevesse, N., Sauval, A.~J., Allende Prieto, C.
\& Blomme, R., 2005, \aap, 431, 693

\bibitem[Barklem \etal(2005)]{barklem05}
Barklem, P.~S. \etal, 2005, \aap, 439, 129

\bibitem[Baum\"{u}ller \& Gehren(1996)]{bau96}
Baum\"{u}ller, D. \& Gehren, T., 1996, \aap, 307, 961

\bibitem[Baum\"{u}ller \& Gehren(1997)]{bau97}
Baum\"{u}ller, D. \& Gehren, T., 1997, \aap, 325, 1088

\bibitem[Bauschlicher \& Langhoff (1987)]{bauschlicher87}
Bauschlicher, N.~B. \& Langhoff, S.~R., 1987, 
Chem. Phys. Lett. 135, 67

\bibitem[Beers, Preston \& Shectman(1985)]{beers85}
Beers, T.C.,  Preston, G.W. \& Shectman, S., 1985, \aj, 90, 2089

\bibitem[Beers, Preston \& Shectman(1992)]{beers92}
Beers, T.C.,  Preston, G.W. \& Shectman, S., 1992, \aj, 103, 1987

\bibitem[Beers \etal(1999)]{beers99}
Beers, T.~C., Rossi, S., Norris, J.~E., Ryan, S. \& Shefler, T., 1999,
\aj, 117, 981

\bibitem[Beers \& Christlieb(2005)]{beers05}
Beers, T.~C. \& Christlieb, N., 2005, \araa, 43, 531

\bibitem[Beers \etal(2007)]{beers07}
Beers, T.~C., \etal, 2007, \apjs, 168, 128

\bibitem[Bihain \etal(2004)]{bihain04}
Bihain, G., Israelian, G., Bonifacio, P. \& Molaro, P., 2004,
\aap, 423, 777

\bibitem[Biehl(1976)]{biehl76}
Biehl, D.,  1976, Diploma thesis, Kiel University

\bibitem[Bonifacio \etal(1998)]{bonifacio98}
Bonifacio, Molaro, Beers, T.C., Vladilo, 1998, \aap, 332, 672


\bibitem[Burstein \& Heiles(1982)]{burstein82}
Burstein, D \& Heiles, C., 1982, \aj, 87, 1165 

\bibitem[Busso, Gallino \& Wasserburg(1999)]{busso99}
Busso, M., Gallino, R. \& Wasserburg, G.J., 1999, \araa, 37, 239


\bibitem[Carretta \etal(2002)]{carretta02} Carretta, E.,   Gratton, R.~G.,   
Cohen,   J.~G.,  Beers,   T.~C. \&  Christlieb,   N., 2002, \aj, 124, 481

\bibitem[Castelli, Gratton \& Kurucz(1997)]{castelli}
Castelli, F., Gratton, R.~G. \& Kurucz, R.~L., 1997, \aap, 318, 841

\bibitem[Cayrel \etal(2004)]{cayrel04} Cayrel, R. \etal\, 2004, \aap, 416, 1117

\bibitem[Chieffi \& Limongi(2004)]{chieffi04}
Chieffi, N. \& Limongi, M., 2004, \apj, 608, 405


\bibitem[Christlieb(2003)]{christlieb03}
Christlieb, N., 2003, Rev. Mod. Astron. 16, 191

\bibitem[Christlieb \etal(2004a)]{christlieb04a}
Christlieb, N., Gustafsson, B., Korn, A.~J., Barklem, P.~S., Beers, T.~C.,
Bessell, M.~S., Karlsson, T. \& Mizuno-Wiedner, M., 2004, \apj, 603, 708

\bibitem[Christlieb \etal(2004b)]{christlieb04b}
Christlieb, N., \etal, 2004, \aap, 428, 1027

\bibitem[Cohen \etal(2002)]{cohen02} Cohen,   J.~G.,  Christlieb,  N., 
Beers,   T.~C.,  Gratton,   R.~G. \& Carretta,   E., 2002, \aj, 124, 470

\bibitem[Cohen \etal(2004)]{cohen04} 
Cohen, J.~G., Christlieb, N.,  McWilliam, A., Shectman, S.,
Thompson, I., Wasserburg, G.~J., Ivans, I., Dehn, Karlsson, T. \& 
Melendez, J., 2004, \apj, 612, 1107

\bibitem[Cohen \etal(2005)]{cohen05}
Cohen, J.~G. \etal, 2005, \apjl, 633, L109


\bibitem[Cohen \etal(2006)]{cohen06} 
Cohen, J.~G. \etal, 2006, \aj, 132, 137

\bibitem[Cohen \etal(2007)]{cohen07} 
Cohen, J.~G., McWilliam, A., Christlieb, N., Shectman, S., Thompson, I., 
Melendez, J.,  Wisotzki, L. \& Reimers, D.,
2007, \apjl, 659, L161

\bibitem[Collet, Asplund \& Trampedach(2006)]{collet06}
Collet, R., Asplund, M. \& Trampedach, R., 2006, \apjl, 644, L121

\bibitem[Cutri \etal(2003)]{2mass2} Cutri, R.~M. \etal, 2003,
``Explanatory Supplement to the 2MASS All-Sky Data Release,
http://www.ipac.caltech.edu/2mass/releases/allsky/doc/explsup.html

\bibitem[Ervin \& Armentrout(1987)]{ervin87}
Ervin, K.~M., \& Armentrout, P.~B., 1987, J.Chem.Phys., 86, 2659

\bibitem[Francois \etal(2003)]{francois03}
Francois et al, 2003, \aap, 403, 1105

\bibitem[Frebel \etal(2005)]{frebel05}
Frebel, A., \etal, 2005, Nature, 434, 871

\bibitem[Frebel \etal(2007)]{frebel07}
Frebel, A., \etal, 2007, \apj, 658, 545

\bibitem[Fulbright(2000)]{fulbright00}
Fulbright, J.~P., 2000, \aj, 120, 1841

\bibitem[Gillis \etal(2001)]{gillis01}
Gillis, J.~R., Goldman, A, Stark, G \& Rinsland, CP, 2001, JQSRT, 68, 225

\bibitem[Gratton \& Sneden(1991)]{gratton91}
Gratton, R. \& Sneden, C., 1991, \aap, 241, 501

\bibitem[Grevesse \& Sauval(1998)]{grevesse98}
Grevesse, N. \& Sauval, A.~J., 1998, Space Science Reviews, 85, 161

\bibitem[Gustafsson \etal(2002)]{gustaffson02}
Gustafsson, B., Edvardsson, B., Eriksson, K., Mizuno-Weidner, M.,
J{\o}rgensen,  U.~G. \& Plez, B., 2002, in ASP Conf.Ser.288,
{\it{Stellar Atmospheres Modeling}}, ed. I.~Hubeny, D.~Mihalis
\& K. Werner, (San Francisco, ASP), 331

\bibitem[Herwig(2004)]{herwig04}
Herwig, F., 2004, \apj, 605, 425

\bibitem[Holt, Scholl \& Rosner(1999)]{holt99}
Holt, R.A., Scholl, T.J. \& Rosner, S.D., 1999, MNRAS, 306, 107


\bibitem[Houdashelt, Bell \& Sweigart(2000)]{houdashelt00}
Houdashelt, M.~L., Bell, R.~A. \& Sweigart, A.~V., 2000, \aj, 119, 1448

\bibitem[Huber \& Herzberg(1979)]{huber79}
Huber, K.~P. \& Herzberg, G., 1979, {\it{Constants of Diatomic Molecules}},
(New York, Van Nostrand)

\bibitem[Ivans \etal(2003)]{ivans03}
Ivans, I.~I., Sneden, C., Renee James, C., Preston, G.~W.,
Fulbright, J.~P., Hoflich, P.~A., Carney, B.~W. \& Wheeler, J.~C.,
2003, \apj, 592, 906

\bibitem[Johnson(2002)]{johnson02}
Johnson, J., 2002, \apjs, 139, 219

\bibitem[Johnson \etal(2006)]{johnson06}
Johnson, J.~A., Herwig, F., Beers, T.~C. \& Christlieb, N., 2006,
\apj, in press

\bibitem[Kobayashi \etal(2006)]{kobayashi}
Kobayashi, C., Umeda, H., Nomoto, K., Tominaga, N.
\& Ohkubo, W., 2007, \apj, 653, 1145


\bibitem[Kurucz(1993)]{kurucz93} Kurucz, R. L., 1993, ATLAS9 Stellar 
Atmosphere Programs and 2 km/s Grid, (Kurucz CD-ROM No. 13)

\bibitem[Kurucz(1994)]{kurucz94} Kurucz, R. L., 1994, Diatomic
molecular data for opacity calculations, (Kurucz CD-ROM No. 15)

\bibitem[Landolt(1992)]{landolt92} Landolt, A.~U., 1992, \aj, 104, 340

\bibitem[Lattanzio(1992)]{lattanzio92}
Lattanzio, J.~C., 1992, Pub. Astronomical Soc. of Australia,
10, 120


\bibitem[Limongi \& Chieffi(2006)]{limongi06}
Limongi, M. \& Chieffi, A., 2006, in {\it{The Multicolored Landscape
of Compact Objects and Their Explosive Origins}}, to be published
by AIP (also available as Astro-ph/0611140)

\bibitem[Lucatello \etal(2006)]{lucatello06}
Lucatello, S., Beers, T., Christlieb, N.,
Barklem, P., Rossi, S., Marsteller, B.,
Sivarani, T. \& Lee, Y.~S.,  2006, \apjl, 652, L37

\bibitem[Luck, Kovtyuk \& Andrievsky(2006)]{luck06}
Luck, R.~E., Kovtyuk, V.~V. \& Andrievsky, S.~M., 2006, \aj, 132, 902

\bibitem[Matteucci(2007)]{matteucci07}
Matteucci, F., 2007, in {\it{Emission Line Universe}},
see Astro-ph/07040770


\bibitem[McWilliam \etal(1995a)]{mcwilliam95a}
McWilliam, A., Preston, G.~W., Sneden, C. \& Shectman, S., 1995, \aj,
109, 2736

\bibitem[McWilliam \etal(1995b)]{mcwilliam95b}
McWilliam, A., Preston, G.~W., Sneden, C. \& Searle, L., 1995, \aj,
109, 2757

\bibitem[McWilliam(1998)]{mcwilliam98}
McWilliam, A., 1998, \aj, 115, 1640

\bibitem[Mishenina \etal(2002)]{mishenina02}
Mishenina, T.~V., Kovtyukh, V.~V., Soubiran, C., Travaglio, C.
\& Busso, M., 2002, \aap, 396, 189

\bibitem[Nomoto \etal(2006)]{nomoto06}
Nomoto, K., Tominaga, N., Umeda, H., Kobayashi, C., \& Maeda, K. 2006, 
Nuclear Physics, A777, 424 (see also Astro-ph/0605725)

\bibitem[Norris, Ryan \& Beers(1997)]{norris97}
Norris, J.~E., Ryan, S.~G. \& Beers, T.C., 1997, \apjl, 489, L169

\bibitem[Norris, Ryan \& Beers   (2001)]{norris01}
Norris, J.~E., Ryan, S.~G. \& Beers, T.C., 2001, \apj, 561, 1034

\bibitem[Oke \& Gunn(1982)]{oke82}
Oke, J.~B. \& Gunn, J.~E., 1982, \pasp, 94, 586

\bibitem[Plez \& Cohen(2005)]{plez}
Plez, B. \& Cohen, J.~G., 2005, \aap, 434, 1117

\bibitem[Plez, Cohen \& Melendez(2006)]{plez2}
Plez, B., Cohen, J.~G. \& Melendez, J., 2005, in
IAU Symposium 228, {\it{ From Lithium to Uranium:
Elemental Tracers of Early Stellar Evolution}}, ed. V. Hill,
P. Francois \& F. Primas, Cambridge University Press, pg. 267

\bibitem[Prantzos, Hashimoto \& Nomoto(1990)]{prantzos90}
Prantzos, N.,  Hashimoto, M. \& Nomoto, K., 1990, \aap, 234, 211


\bibitem[Prantzos(2006)]{prantzos06}
Prantzos, N., 2006, in {\it{Nuclei in the Cosmos IX}},
CERN, Geneva, July 2006, ed. A. Mengoni et al.

\bibitem[Preston(1996)]{preston96}
Preston, G.~W., 1996, in {\it{The Formation of the Galactic Halo -- Inside
and Out}}, ed. H.~L. Morrison \& A. Sarajedini, ASP Conf. Ser. 92

\bibitem[Preston \& Sneden(2000)]{preston00}
Preston, G.~W. \& Sneden, C., 2000, \aj, 120, 1014

\bibitem[Ram\'{\i}rez \& Cohen(2003)]{ramirez03}
Ram\'{\i}rez, S.~V. \& Cohen, J.~G., 2003, \aj, 125, 224 

\bibitem[Schlegel, Finkbeiner \& Davis(1998)]{schlegel98}
Schlegel, D.~J., Finkbeiner, D.~P. \& Davis, M., 1998, \apj, 500, 525

\bibitem[Shortridge(1993)]{shortridge93}
Shortridge K. 1993, in {\it{Astronomical Data Analysis Software and
      Systems II}}, A.S.P. Conf. Ser., Vol 52, eds. R.J. Hannisch, 
      R.J.V. Brissenden, \& J. Barnes, 219

\bibitem[Simmerer \etal(2003)]{simmerer03}
Simmerer, J., Sneden, C., Ivans, I.~I., Kraft, R.~P., Shetrone, M.~A.
\& Smith, V.~v., 2003, \aj, 125, 2018
      
\bibitem[Simmerer \etal(2004)]{simmerer04}
Simmerer, J., Sneden, C., Cowan, J.~J., Collier, J., Woolf, V.~M.
\& Lawler, J.~E., 2004, \apj, 617, 1091

\bibitem[Skrutskie \etal(2006)]{2mass1}
Skrutskie, M.~F. \etal, 2006, \aj, 131, 1163


\bibitem[Sivarani \etal(2006)]{sivarani06}
Sivarani, T. \etal, 2006, \aap, 459, 125

\bibitem[Sneden(1973)]{moog} Sneden, C., 1973, Ph.D. thesis, Univ. 
of Texas

\bibitem[Sneden \etal(2003)]{sneden03}
Sneden, C., \etal, 2003, \apj, 591, 936

\bibitem[Spite \etal(2005)]{spite05}
Spite, M. \etal, 2005, \aap, 430, 655

\bibitem[Spite \etal(2006)]{spite06}
Spite, M. \etal, 2006, \aap, 455, 291

\bibitem[Takeda \etal(2003)]{takeda03}
Takeda, Y., Zhao, G., Takada-Hidai, M., Chen, Y.~Q.,
Saito, Y. \& Zhang, H.~W., 2003, Chinese Jrl Astron \& Astrophys,
3, 316

\bibitem[Tominaga, Umeda \& Nomoto(2007)]{tominaga07}
Tominaga, N., Umeda, H. \& Nomoto, K., 2007, 
\apj, 660, 516

\bibitem[Travaglio \etal(2004)]{travaglio04}
Travaglio, C., Gallino, R., Arnone, E., Cowan, J., Jordan, F.
\& Sneden, C., 2004, \apj, 601, 864

\bibitem[Umeda \& Nomoto(2002)]{umeda02}
Umeda, H. \& Nomoto, K., 2002, \apj, 565, 385

\bibitem[Vogt \etal(1994)]{vogt94} Vogt, S.~E. \etal\, 1994, SPIE, 2198, 362

\bibitem[Wallace, Hinkle \& Livingston(1998)]{wallace98}
Wallace, L., Hinkle, K. \& Livingston, W.~C.,1998,
``An Atlas of the Spectrum of the Solar Photosphere from 13,500
to 28,000 cm$^{-1}$'', N.S.O. Technical Report 98-001,
ftp://nsokp.nso.edu/pub/atlas/visatl.


\bibitem[Wisotzki \etal(2000)]{wis00} Wisotzki, L., Christlieb, N., 
Bade, N.,Beckmann, V., K\"ohler, T., Vanelle, C. \& Reimers, D., 2000, 
\aap, 358, 77

\bibitem[Woosley \& Hoffman(1992)]{woosley92}
Woosley, S.~E., \& Hoffman, R.~D. 1992, \apj, 395, 202

\bibitem[Woosley \& Weaver(1995)]{woosley95}
Woosley, S.~E. \& Weaver, T.~A., 1995, \apjs, 101, 181

\bibitem[Yi \etal(2002)]{yi01}
Yi, S., Demarque, P., Kim, Y.-C. ,  Lee, Y.-W., Ree, C.
Lejeune, Th. \&  Barnes, S., 2001, \apjs, 136, 417

\bibitem[York \etal(2000)]{york00}
York, D. \etal, 2000, \aj, 120, 1579

\end{thebibliography}
\end{document}